\documentclass[english,10pt,aps,prd,a4paper,preprintnumbers,floatfix,nofootinbib,showpacs,superscriptaddress, notitlepage]{revtex4-1}
\pdfoutput=1
\usepackage[english]{babel}
\usepackage{amsmath}
\usepackage{graphicx}
\usepackage{color}
\usepackage{mathrsfs}   %\mathscr{L}
\usepackage{amssymb}
\usepackage{hyperref}
\usepackage[normalem]{ulem} % for striking out text
\usepackage{dcolumn}% Align table columns on decimal point
\usepackage{bm}% bold math
\usepackage{graphicx}
\usepackage[sort&compress]{natbib}
\usepackage{epstopdf}
\newcommand {\be} {\begin{equation}}
\newcommand {\ee} {\end{equation}}

\definecolor{greenLinks}{rgb}{0, 0.6, 0} 
\definecolor{blueLinks}{rgb}{0, 0, 0.6}
\definecolor{redLinks}{rgb}{0.6, 0, 0}
\definecolor{tempText}{rgb}{0.55, 0.10,0.67}
\definecolor{eprintLinks}{rgb}{0.4, 0.4, 0.4}
\definecolor{journalLinks}{rgb}{0.6, 0, 0}

\newcommand{\MYhref}[3][redLinks]{\href{#2}{\color{#1}{#3}}}%
\def \nbb {${0\nu}\beta\beta$ }

\usepackage{float}

\def\21{$\mathrm{SU(2)_L \otimes U(1)_Y}$ }
\def\31{$\mathrm{SU(3)_c \otimes U(1)_Q}$ }
\def\SM{$\mathrm{SU(3)_c \otimes SU(2)_L \otimes U(1)_Y}$ }

\def\3211{$\mathrm{SU(3) \otimes SU(2)_L \otimes U(1)_R \otimes U(1)_{B-L}}$ }
\def\321{$\mathrm{SU(3) \otimes SU(2) \otimes U(1)}$ }
\def\422{$\mathrm{SU(4) \otimes SU(2) \otimes SU(2)_R}$ }
\newcommand {\ignore}[1]{}

\newcommand{\sm}{{Standard Model }}

\def\SM{$\mathrm{ SU(3)_C \otimes SU(2)_L \otimes U(1)_Y }$ }
\newcommand{\AddrAHEP}{%
  AHEP Group, Institut de F\'{i}sica Corpuscular --
  C.S.I.C./Universitat de Val\`{e}ncia, Parc Cient\'ific de Paterna.\\
 C/ Catedr\'atico Jos\'e Beltr\'an, 2 E-46980 Paterna (Valencia) - SPAIN}

\begin{document}

\title{Seesaw Dirac neutrino mass through dimension-6 operators}
\author{Salvador Centelles Chuli\'{a}}\email{salcen@ific.uv.es}
\affiliation{\AddrAHEP}
\author{Rahul Srivastava}\email{rahulsri@ific.uv.es}
\affiliation{\AddrAHEP}
\author{Jos\'{e} W. F. Valle}\email{valle@ific.uv.es}
\affiliation{\AddrAHEP}

\begin{abstract}
  \vspace{1cm} {In this paper, a followup
    of~\cite{CentellesChulia:2018gwr}, we describe the many pathways
    to generate Dirac neutrino mass through dimension-6 operators.
     By using only the Standard Model Higgs doublet in the external
    legs one gets a unique operator
    $\frac{1}{\Lambda^2} \, \bar{L}\, \bar{\Phi} \, \bar{\Phi} \, \Phi
    \, \nu_R $.
    In contrast,  the presence of new scalars implies new possible field
    contractions, which greatly increase the number of
    possibilities. Here we study in detail the simplest ones,
    involving $SU(2)_L$ singlets, doublets and triplets. The extra
    symmetries needed to ensure the Dirac nature of neutrinos can also
    be responsible for stabilizing dark matter.}
   
\end{abstract}

\maketitle

%%%%%%%%%%%%%%%%%%%%%%%%%%%%%%%%%%%%%%%%%%%%%%%%%%%%%%%%%%%%%%%%%%%%

\section{Introduction}
\label{sec:introduction}

%%%%%%%%%%%%%%%%%%%%%%%%%%%%%%%%%%%%%%%%%%%%%%%%%%%%%%%%%%%%%%%%%%%

Elucidating the nature of neutrinos constitutes a key open challenge
in particle physics.
The detection of neutrinoless double-beta decay - \nbb - would
establish the Majorana nature of at least one
neutrino~\cite{Schechter:1981bd}
So far, however, the experimental
searches~\cite{KamLAND-Zen:2016pfg,GERDA:2018,MAJORANA:2008,CUORE:2018,EXO-2018,Arnold:2016bed} for \nbb have not
borne a positive result, leaving us in the dark concerning whether 
neutrinos are their own anti-particles or not.
One should stress that, although Dirac neutrinos are not generally
expected within a gauge theoretic framework~\cite{Schechter:1980gr},
they arise in models with extra
dimensions~\cite{Chen:2015jta,Addazi:2016xuh}, where the $\nu_R$ states
are required for the consistent high energy completion of the theory.
Dirac neutrinos also emerge in conventional four dimentional gauge
theories with an adequate extra symmetry.
One appealing possibility is the quarticity symmetry, which was
originally suggested in \cite{Chulia:2016ngi, Chulia:2016giq,
  CentellesChulia:2017koy}.  This mechanism uses a  version of
$U(1)_L$ lepton number symmetry broken into its subgroup $Z_4$.

In summary, there has been recently a growing interest in Dirac
neutrinos~\cite{Heeck:2013rpa, Abbas:2013uqh, Ma:2014qra, Okada:2014vla, Ma:2015rxx,
  Ma:2015raa, Ma:2015mjd, Valle:2016kyz, Bonilla:2016zef,
  Bonilla:2016diq, Chulia:2016ngi, Chulia:2016giq, Reig:2016ewy,
  Ma:2016mwh, Wang:2016lve, Borah:2016zbd, Abbas:2016qbl,
   Yao:2017vtm, Ma:2017kgb,CentellesChulia:2017koy,
  Borah:2017leo, Wang:2017mcy, Bonilla:2017ekt, Hirsch:2017col,
  Srivastava:2017sno,
  Borah:2017dmk,Yao:2018ekp,Chen:2015jta,Addazi:2016xuh,CentellesChulia:2017sgj,
 Fonseca:2018ehk, Helo:2018bgb, Reig:2018mdk}.
The many pathways to generate Dirac neutrino mass through generalized
dimension-5 operators\textit{ a la Weinberg} have been described in
Ref.~\cite{CentellesChulia:2018gwr}.
It has been shown that the symmetry responsible for ``Diracness'' can
always be used to stabilize a WIMP Dark Matter candidate, thus
connecting the Dirac nature of neutrinos with the stability of Dark
Matter.

In this letter we take the point of view that neutrinos are Dirac
fermions, extending the results of~\cite{CentellesChulia:2018gwr} to
include the analysis of seesaw operators that lead to Dirac neutrino
mass at dimension 6 level.
Using only the Standard Model Higgs doublet in the external legs one
is lead to a unique operator
$\frac{1}{\Lambda^2} \, \bar{L}\, \bar{\Phi} \, \bar{\Phi} \, \Phi \,
\nu_R $.
However the presence of new scalar bosons beyond the standard Higgs
doublet implies many new possible field contractions.
We also notice that, also here and quite generically, the extra
symmetries needed to ensure the Dirac nature of neutrinos can also be
made responsible for stability of dark matter.

The paper is organized as follows. In
Section~\ref{sec:operator-analysis} we discuss the various dimension-6
operators that can give rise to Dirac neutrinos.
For simplicity we restrict ourselves to the simplest cases of scalar
singlets ($\chi$), doublets ($\Phi$) and triplets ($\Delta$) of
$SU(2)_L$.
We also discuss the Ultra-Violet (UV) complete theories associated to
each operator. All these completions fall under one of the five
distinct topologies, which we discuss, along with the associated
generic neutrino mass estimate.
In Section~\ref{sec:OnlySM} we explicitly construct and discuss all
the UV-complete dimension-6 models involving only \sm fields.
In Section~\ref{sec:singlet-doublet} we discuss the various
UV-completions of the dimension-6 operators involving only $SU(2)_L$
singlet $\chi$ and doublet $\Phi$ scalars.
In Section \ref{sec:singlet-doublet-triplet} we consider the possible
UV-completions of the dimension-6 seesaw operators involving all three
types of Higgs scalars, singlet $\chi$, doublet $\Phi$ and triplet
$\Delta$.
In Section \ref{sec:doublet-triplet} we discuss the UV-completion of
the dimension-6 operator involving the doublet $\Phi$ and triplet
$\Delta$ scalars.
In Section \ref{sec:summary-conclusions} we present a short summary
and discussion.

%%%%%%%%%%%%%%%%%%%%%%%%%%%%%%%%%%%%%%%%%%%%%%%%%%%%%%%%%%%%%%%%%%%%

\section{Operator Analysis}
\label{sec:operator-analysis}

%%%%%%%%%%%%%%%%%%%%%%%%%%%%%%%%%%%%%%%%%%%%%%%%%%%%%%%%%%%%%%%%%%%%

We begin our discussion by looking at the possible dimension-6
operators that can lead to naturally small Dirac neutrino masses. 
In order to cut down the number of such operators in this work we will
limit our discussion only to operators involving scalar singlet
($\chi$), doublet ($\Phi$) and triplet ($\Delta$) representations of
the weak gauge group $SU(2)_L$.
The discussion can be easily extended to other higher $SU(2)_L$
multiplets. The general form of such dimension-6 operators is given by 
\be
\frac{1}{\Lambda^2} \, \bar{L} \otimes X \otimes Y \otimes Z \otimes \nu_R
\ee
where $L = (\nu_L, e_L)^T$ is the lepton doublet, $\nu_R$ are the
right handed neutrinos which are singlet under \sm gauge group and
$X, Y,Z$ denote scalar fields which are singlets under $SU(3)_c$,
transforming under $SU(2)_L$ and carrying appropriate $U(1)_Y$
charges such that the operator is invariant under the full \sm gauge
symmetry. 
Moreover, $\Lambda$ is the cutoff scale above which the full
UV-complete theory must be taken into account. For sake of simplicity,
throughout this work we will suppress all flavor indices of the
fields. The \SM symmetry is broken by the vacuum expectation values
(vev) of the scalars. After symmetry breaking the neutrinos acquire
naturally small Dirac masses.

Invariance under \SM dictates that, if $X$ transforms as a $n$-plet
under $SU(2)_L$, then $Y \otimes Z$ must transform either as a
$n+1$-plet or a $n-1$-plet. This leaves many possible choices for the
scalars $X, Y, Z$, as we now discuss.

For example, if we take $X$ to be a singlet $\chi$, then $Y \otimes Z$
should transform as a doublet of $SU(2)_L$. Restricting ourselves only
to representations up to triplets of $SU(2)_L$ , one possibility is
$Y \otimes Z \equiv \chi \otimes \bar{\Phi}$, where $\bar{\Phi}$
denotes a scalar doublet of $SU(2)_L$ but with hypercharge opposite to
that of $\Phi$. It can be either $\Phi^\dagger$ or $\Phi^c$, depending
on the particular $SU(2)_L$ contraction.
The only other option is
$Y \otimes Z \equiv \bar{\Phi} \otimes \Delta_0$ or
$Y \otimes Z \equiv \Phi \otimes \Delta_{-2}$ where $\Delta_i$;
$i = 0, -2$ is a scalar triplet of $SU(2)_L$. Note that here,
depending on the choice of $\Phi$ or $\bar{\Phi}$, there are two
possible $U(1)_Y$ charge assignments for $\Delta_i$ i.e. $\Delta_0$
with $U(1)_Y = 0$ and $\Delta_{-2}$ with $U(1)_Y = -2$. 

Taking $X$ as a doublet $\bar{\Phi}$ or $\Phi$, then $Y \otimes Z$
must transform either as a singlet or a triplet under $SU(2)_L$
symmetry. Thus we could have $Y \otimes Z \equiv \chi \otimes \chi$
which is only allowed if $X = \bar{\Phi}$.
The other option is to have $Y \otimes Z \equiv \chi \otimes \Delta_i$;
$i = 0,-2$.
These operators are the same as already discussed for $X = \chi$ case .
Apart from these, as far as transformation under $SU(2)_L$ is
concerned, there are two new possibilities, namely
$Y \otimes Z \equiv \Delta \otimes \Delta$ and
$Y \otimes Z \equiv \bar{\Phi} \otimes \Phi$. 
The latter is the only dimension-6 operator which can be written down
with only \sm scalar fields~\footnote{ Of course Dirac neutrino masses
  always require the addition of \sm singlet right handed neutrinos
  $\nu_R$.}.
For the operator $Y \otimes Z \equiv \Delta \otimes \Delta$ there are
several possibilities depending on the hypercharge of $\Delta$, as
listed in Table \ref{Tab:op}.
If $X \sim 3$ under $SU(2)_L$ symmetry, it is easy to see that,
restricting up to triplet representations of $SU(2)_L$, no new
operator can be written. 
Each of these operators can lead to different possible $SU(2)_L$
contractions which in turn select the type of new fields needed in a
UV-complete model~\footnote{Notice that, while for operators we have
  restricted ourselves up to triplets of $SU(2)_L$, for their
  UV-completion in Table \ref{Tab:op} we have also allowed
  exchanges involving higher $SU(2)_L$ multiplets.}.
The resulting dimension-6 operators along with the number of possible
UV-completions in each of the cases are summarized in Table
\ref{Tab:op}.
\begin{table}[ht]
\begin{center}
\begin{tabular}{c c c c c || c c c c c}
  \hline \hline  
$X$        \hspace{0.5cm}    & $Y$     \hspace{0.5cm}   & $Z$           \hspace{0.5cm}  & Operator\hspace{0.5cm} &  Diagrams  &	$X$        \hspace{0.5cm}    & $Y$     \hspace{0.5cm}   & $Z$           \hspace{0.5cm}  & Operator\hspace{0.5cm} &  Diagrams  \\
\hline \hline
$\mathbf{1}$	 \hspace{0.5cm}    & $\mathbf{1}$    \hspace{0.5cm}    & $\mathbf{2}$   	 \hspace{0.5cm}   &	
$ \bar{L} \, \chi \, \chi \, \bar{\Phi} \, \nu_R$    \hspace{0.5cm}        &  	$10$        
&
$\mathbf{2}$	 \hspace{0.5cm}    & $\mathbf{2}$    \hspace{0.5cm}    & $\mathbf{2}$   	 \hspace{0.5cm}   &	
$ \bar{L} \, \bar{\Phi} \,\bar{\Phi} \, \Phi\, \nu_R$    \hspace{0.5cm}        &  	$15$       
\\
$\mathbf{1}$     \hspace{0.5cm}    & $\mathbf{2}$    \hspace{0.5cm}    & $\mathbf{3}$   	 \hspace{0.5cm}   &	
$ \bar{L} \, \chi \, \bar{\Phi} \, \Delta_0 \, \nu_R$     \hspace{0.5cm}           &  	$16$         
&
$\mathbf{1}$     \hspace{0.5cm}    & $\mathbf{1}$    \hspace{0.5cm}    & $\mathbf{2}$   	 \hspace{0.5cm}   &	
$ \bar{L} \, \bar{\chi} \,\chi \, \bar{\Phi}\, \nu_R$     \hspace{0.5cm}           &  	$15$   
\\
$\mathbf{1}$     \hspace{0.5cm}    & $\mathbf{2}$    \hspace{0.5cm}    & $\mathbf{3}$   	 \hspace{0.5cm}   &	
$ \bar{L} \, \chi \, \Phi \, \Delta_{-2}\, \nu_R$   	 \hspace{0.5cm}                   &  	$16$     
&
$\mathbf{2}$     \hspace{0.5cm}    & $\mathbf{3}$    \hspace{0.5cm}    & $\mathbf{3}$   	 \hspace{0.5cm}   &	
$ \bar{L} \, \bar{\Phi} \, \bar{\Delta}_0 \, \Delta_0 \, \nu_R$   	 \hspace{0.5cm}                   &  	$31$     
\\
$\mathbf{2}$     \hspace{0.5cm}    & $\mathbf{3}$    \hspace{0.5cm}    & $\mathbf{3}$   	 \hspace{0.5cm}   &	
$ \bar{L} \, \bar{\Phi} \, \Delta_{0}\, \Delta_0 \, \nu_R$  \hspace{0.5cm}             &  	$16$     
&
$\mathbf{2}$     \hspace{0.5cm}    & $\mathbf{3}$    \hspace{0.5cm}    & $\mathbf{3}$   	 \hspace{0.5cm}   &	
$ \bar{L} \, \bar{\Phi} \, \bar{\Delta}_{-2} \, \Delta_{-2}\, \nu_R$                \hspace{0.5cm}               &	 $26$     
\\
$\mathbf{2}$     \hspace{0.5cm}    & $\mathbf{3}$    \hspace{0.5cm}    & $\mathbf{3}$   	 \hspace{0.5cm}   &	
$ \bar{L} \, \Phi \,\Delta_0 \, \Delta_{-2}\, \nu_R$   \hspace{0.5cm}         & 	$27$	  
&
    \hspace{0.5cm}    &    \hspace{0.5cm}    &    	 \hspace{0.5cm}   &	
    \hspace{0.5cm}         & 		  
\\
  \hline
  \end{tabular}
\end{center}
\caption{ Possible $SU(2)_L$ assignments for the scalars $X$, $Y$,
  $Z$; the allowed operators and number of associated UV-complete
  models in each case. Here $\bar{\Phi}$ denotes either $\Phi^\dagger$
  or $\Phi^c$, depending on the particular $SU(2)_L$ contractions.
  Note that the hypercharge of $\bar{\Phi}$ has the opposite sign than
  the hypercharge of $\Phi$.  Similar notation is used for other
  scalar multiplets.}
 \label{Tab:op}
\end{table}

It is important to notice that the dimension-6 operators listed in
Table \ref{Tab:op} will give the leading contribution to neutrino mass
only in scenarios where other lower dimensional operators are
forbidden by some symmetry. 
Such scenarios can arise in context of many symmetries ranging from
$U(1)_X$ symmetries \cite{ Ma:2014qra, Ma:2015mjd} to abelian discrete $Z_n$ symmetries \cite{Chulia:2016ngi,  Bonilla:2016zef, Bonilla:2016diq, Reig:2016ewy,Reig:2018mdk} up to various
types of more complex flavor symmetries containing non-abelian
groups~\cite{Chulia:2016giq, CentellesChulia:2017koy, Bonilla:2017ekt}.
Keeping this in mind, Table \ref{Tab:op} is divided into two
columns. The operators in the left column are those for which the
lower dimensional operators can be forbidden by $U(1)_X$ or $Z_n$
symmetries. The operators in the right column will not be
  leading operators for neutrino mass in such case, but will require,
  for example, a soft breaking of such symmetries. Another appealing
  possibility is to have two copies of the scalars
  involved. Alternatively, one may use more involved symmetries
  involving non-abelian discrete flavor symmetries. We will further
  discuss this issue in latter sections. 

Also, notice that the number of UV-complete models for similar type of
operators e.g.  $ \bar{L} \, \chi \, \chi \, \bar{\Phi} \, \nu_R$ and
$ \bar{L} \, \bar{\chi} \, \chi \, \bar{\Phi} \, \nu_R$ are
different. This is because, while counting the number of models, we
have also taken into account the possible differences under the
symmetry forbidding lower dimensional operators. 
Apart from Hermitian conjugated (h.c.) counterparts which are not listed,
there are also other possibilities beyond the operators listed in
Table \ref{Tab:op}, where one or more hypercharge neutral fields
i.e. $\chi_0$ or $\Delta_0$, is replaced by the corresponding
$\bar{\chi}$ or $\bar{\Delta}_0$.
These possibilities are not listed here as they do not give rise to
new operators, as far as only \sm symmetries are concerned. 
However, they can be potentially differentiated by other symmetries
such as those forbidding the lower dimensional operators. We will
also briefly discuss such possibilities in latter sections whenever
they arise. 

\begin{figure}[!h]
 \centering
  \includegraphics[scale=0.25]{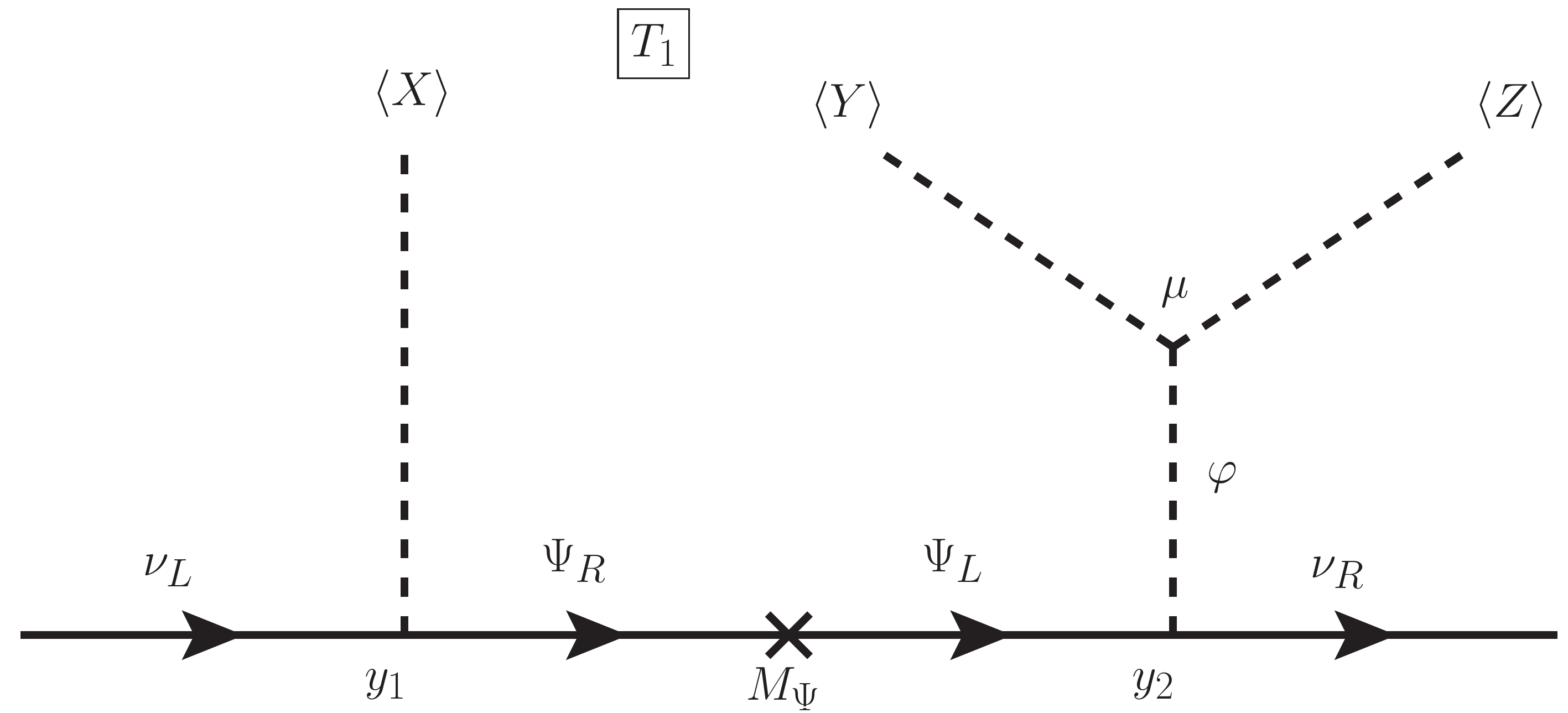} \hspace{2mm}
   \includegraphics[scale=0.25]{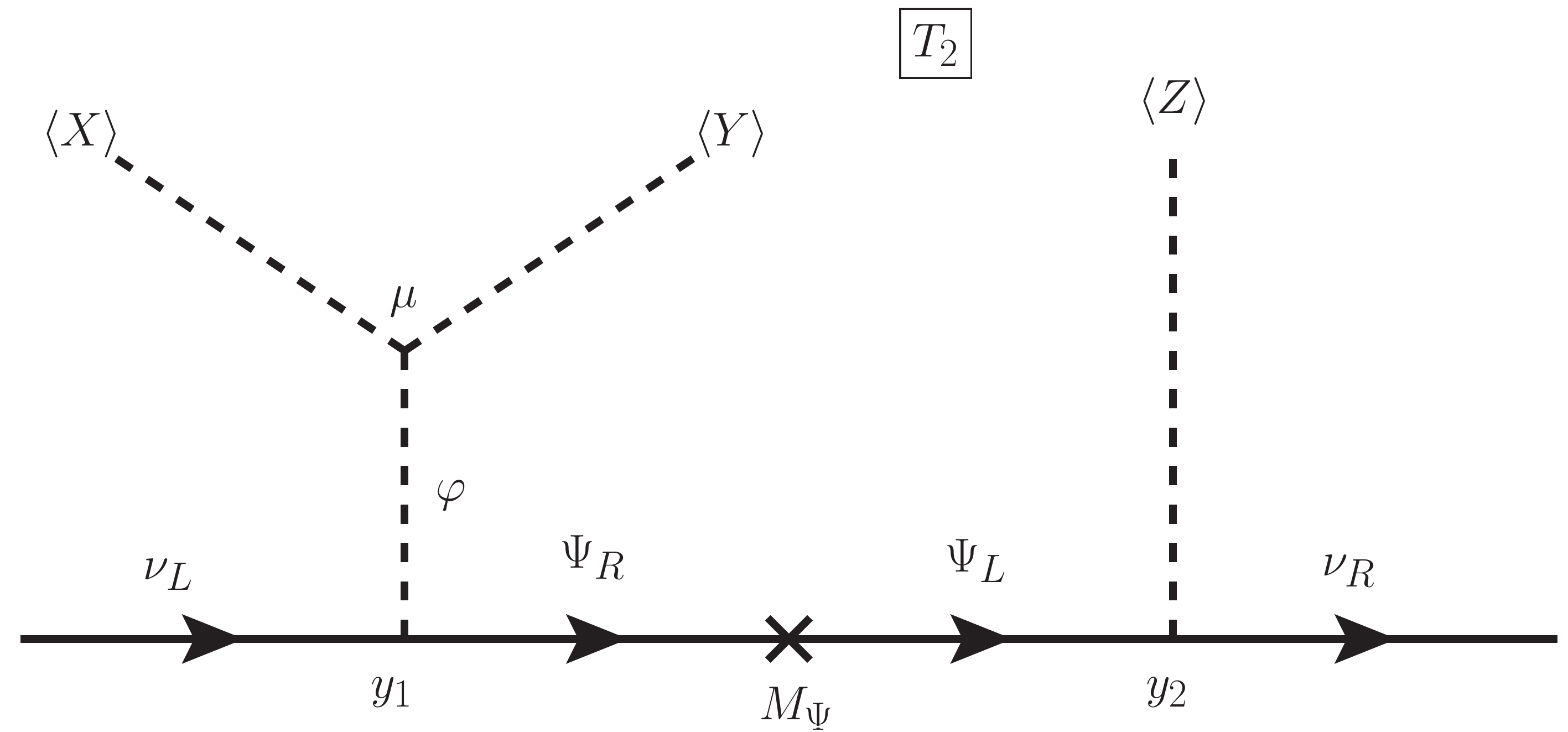} \hspace{2mm}
    \includegraphics[scale=0.25]{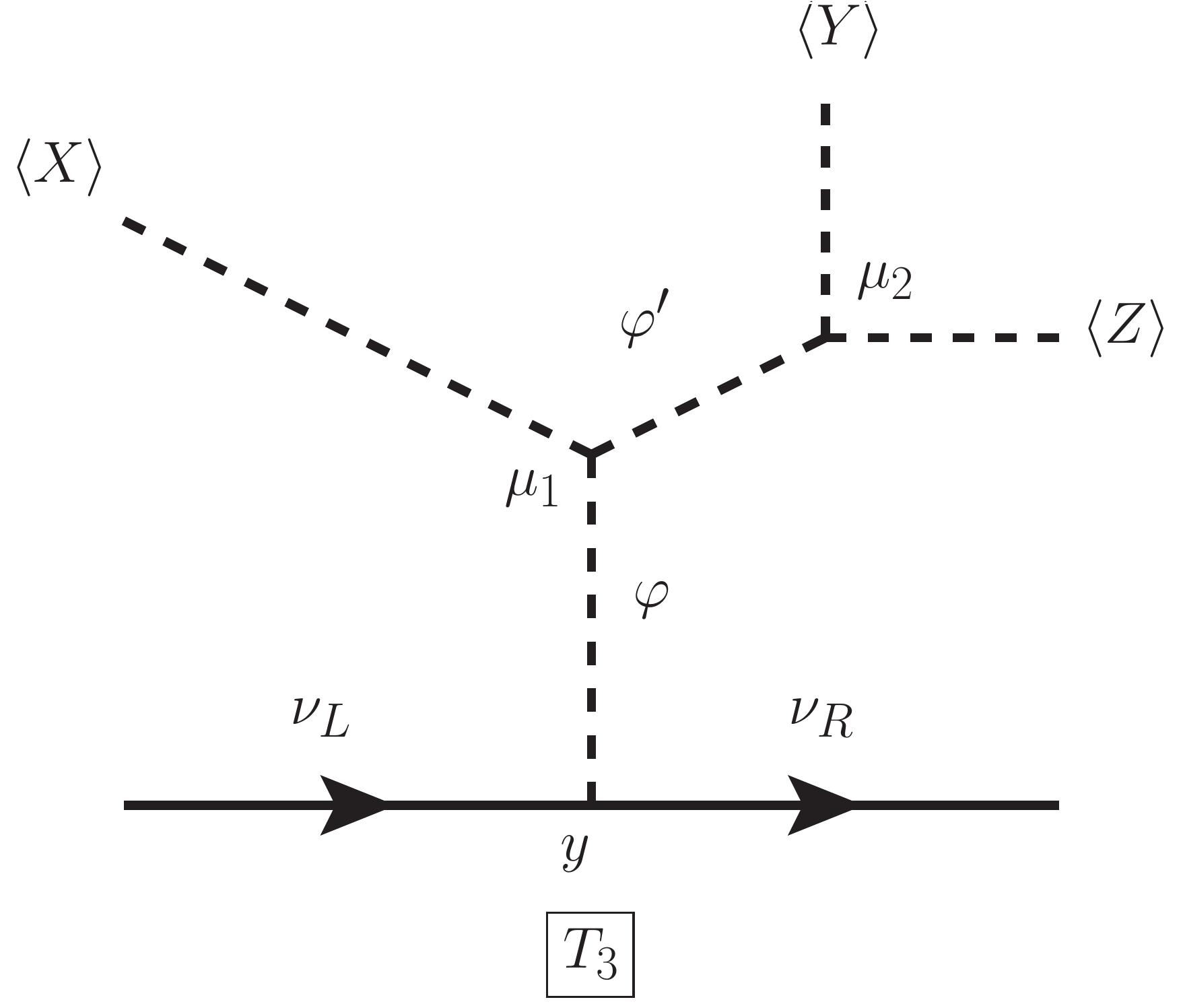} \hspace{2mm}
     \includegraphics[scale=0.25]{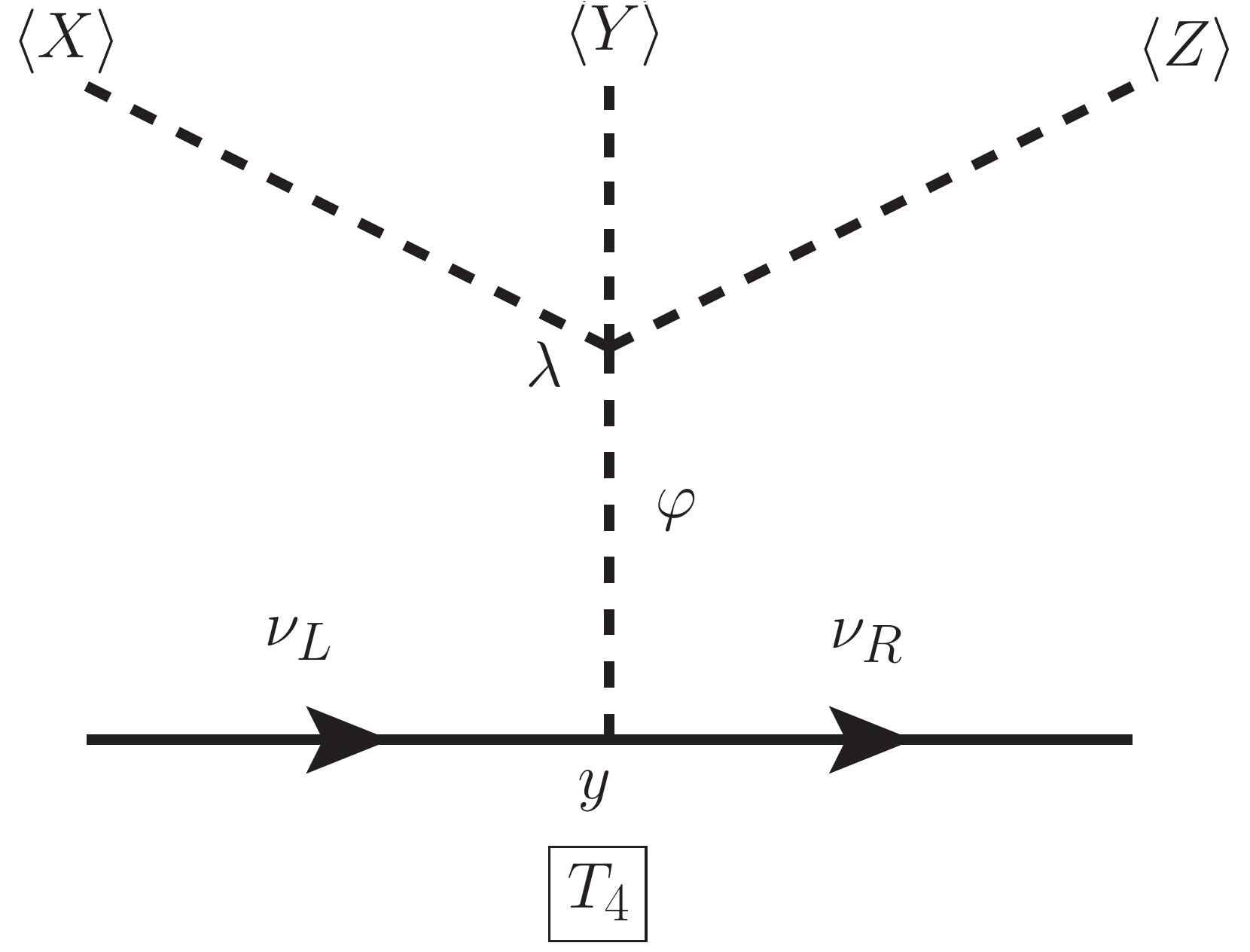} \hspace{2mm}
      \includegraphics[scale=0.25]{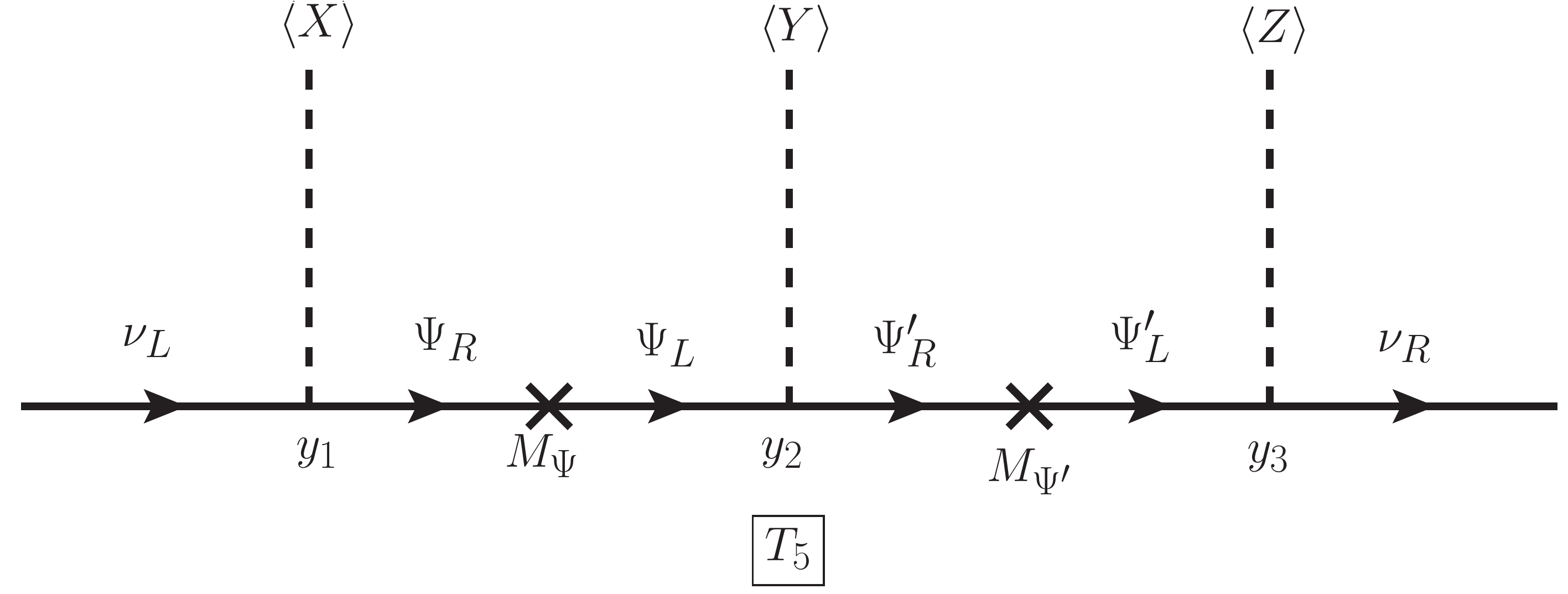}
       \caption{Feynman diagrams representing the five different topologies, $T_1$, $T_2$, $T_3$, $T_4$, $T_5$,  respectively.}
        \label{fig:topologies}
\end{figure}

We find that all possible UV-completions of the operators listed in
Table \ref{Tab:op} can be arranged into five distinct topologies for
the Feynman diagrams of neutrino mass generation.  For lack of better
names, we are calling these five topologies as $T_i$,
$i \in \{1, 2, 3, 4, 5\}$. These topologies are shown in Fig.~\ref{fig:topologies}. 

Each topology involves new ``messenger fields'' which can be either
new scalars ($\varphi$), new fermions ($\Psi$) or both. The masses of
these heavy messenger fields lying typically at or above the cutoff
scale $\Lambda$ of the dimension-6 operators.
The first topology $T_1$ involves two messenger fields, a scalar
($\varphi$) and a Dirac fermion ($\Psi$). The scalar $\varphi$ gets a
small induced vev through its trilinear coupling with $Y$ and $Z$
scalars. 
Topology $T_2$ is very similar to $T_1$ and also involves two
messenger fields, a scalar $\varphi$ and a Dirac fermion
$\Psi$. However, the small vev of $\varphi$ in $T_2$ is induced by its
trilinear coupling with $X$ and $Y$ scalars.
The third topology $T_3$ is distinct from the first two and only
involves scalar messengers $\varphi$ and $\varphi^\prime$. The scalar
$\varphi^\prime$ gets a small induced vev through its trilinear coupling
with $Y$ and $Z$ scalars. The other scalar $\varphi$ subsequently
gets a ``doubly-induced vev'' through its trilinear coupling with
$\varphi^\prime$ and $X$ scalars. 
The fourth topology $T_4$ only involves a single messenger scalar
$\varphi$ which gets an induced vev through its quartic coupling with
scalars $X$, $Y$ and $Z$. 
The final fifth topology involves only fermionic Dirac messengers
$\Psi$ and $\Psi'$, as shown in Fig.~\ref{fig:topologies}. 
The $SU(2)_L \otimes U(1)_Y$ charges of the messenger fields in all
topologies will depend on the details of the operator under
consideration and the contractions involved. We will discuss all such
possibilities in the following sections.
Each topology leads to different estimates for the associated light
neutrino mass generated in each case, neglecting the three generation
structure of the various Yukawa coupling matrices in family space.
The resulting formulas for the neutrino masses are listed in Table
\ref{Tab:mass-formula}. 
\begin{table}[ht]
\begin{center}
\begin{tabular}{c c c}
  \hline \hline  
Topology  \hspace{1cm} &   Messenger Fields   \hspace{1cm}   & Neutrino Mass Estimate                  \\
\hline \hline 
T1        \hspace{1cm} &  $\Psi$, $\varphi$   \hspace{1cm}   &  $ \frac{\mu y_1 y_2 \, v_X v_Y v_Z}{M_\psi M^2_\varphi}$   \vspace{2mm}   \\
T2        \hspace{1cm} &  $\Psi$, $\varphi$   \hspace{1cm}   &  $ \frac{\mu y_1 y_2 \, v_X v_Y v_Z}{M_\psi M^2_\varphi}$   \vspace{2mm}    \\
T3        \hspace{1cm} &  $\varphi$, $\varphi^\prime$ \hspace{1cm} &  $ \frac{\mu_1 \mu_2 y \, v_X v_Y v_Z}{M^2_{\varphi} M^2_{\varphi^\prime}}$ \vspace{2mm}  \\
T4       \hspace{1cm}  & $\varphi$            \hspace{1cm}   &  $ \frac{ y \lambda \, v_X v_Y v_Z}{M^2_\varphi}$              \vspace{2mm}  \\
T5       \hspace{1cm}  & $\Psi$, $\Psi'$      \hspace{1cm}   &  $ \frac{ y_1 y_2 y_3 \, v_X v_Y v_Z}{M_{\psi} M_{\psi'}}$        \vspace{2mm}                \\
  \hline
  \end{tabular}
\end{center}
\caption{Possible topologies and messengers leading to light neutrino
  masses, and the associated estimates for each topology.  }
 \label{Tab:mass-formula}
\end{table}
 
Having discussed the various possible dimension-6 operators for Dirac
neutrino mass generation and the various topologies involved in the UV-
complete-models, in the following sections we discuss the various
operators and topologies in more detail. 
To clarify the notation used in upcoming sections we list, as an
illustration, all possible $SU(2)$ contractions of $\Phi$ and
$\bar{\Phi}$ explicitly in Table \ref{contractions}. 
\begin{table}[ht]
\begin{center}
\begin{tabular}{c c c c}
  \hline \hline  
Field 1         \hspace{1cm}    & Field 2 \hspace{1cm}   & Implicit contraction & Explicit contraction \\
\hline   \hline 
$\Phi$      \hspace{1cm}    & $\Phi$             \hspace{1cm}    &  $\underbrace{\Phi \otimes \Phi}_{1}$             &    $\Phi^\alpha \epsilon_{\alpha \beta} \Phi^\beta = 0 $              \\
$\Phi$       \hspace{1cm}    & $\bar{\Phi}$       \hspace{1cm}    &  $\underbrace{\Phi \otimes \bar{\Phi}}_{1}$       &    $\Phi^\alpha \delta_\alpha^\beta \Phi^*_\beta $                    \\
$\Phi$        \hspace{1cm}    & $\Phi$             \hspace{1cm}    &  $\underbrace{\Phi \otimes \Phi}_{3}$             &    $\Phi^\alpha (\tau^a)_{\alpha}^{\,\,\,\,\beta} \epsilon_{\beta \sigma} \Phi^\sigma$                                           \\
$\Phi$         \hspace{1cm}    & $\bar{\Phi}$       \hspace{1cm}    &  $\underbrace{\Phi \otimes \bar{\Phi}}_{3}$       &    $\Phi^\alpha (\tau^a)_\alpha^{\,\,\,\,\beta} \Phi^*_\beta$                         \\
$\bar{\Phi}$    \hspace{1cm}    & $\bar{\Phi}$       \hspace{1cm}    &  $\underbrace{\bar{\Phi} \otimes \bar{\Phi}}_{1}$ &    $\Phi^*_\alpha \epsilon^{\alpha \beta} \Phi^*_\beta = 0$           \\
$\bar{\Phi}$     \hspace{1cm}    & $\bar{\Phi}$       \hspace{1cm}    &  $\underbrace{\bar{\Phi} \otimes \bar{\Phi}}_{3}$ &    $\Phi^*_\alpha \epsilon^{\alpha \sigma} (\tau^a)_{\sigma}^{\,\,\,\,\beta} \Phi^*_\beta$                                       \\
  \hline
  \end{tabular}
\end{center}
\caption{The possible $SU(2)_L$ contractions of $\Phi$ and $\bar{\Phi}$, where $\tau^a = \tau^+, \tau ^3, \tau^-$ are the three $SU(2)_L$ generators.}
 \label{contractions}
\end{table}

Similar notation will be used for contractions of other field
multiplets in upcoming sections. For the sake of brevity we will not
write them explicitly, though the contractions involved should be
clear from the context. 
The UV-complete models arising from the operators listed in Table
\ref{Tab:op} will all involve certain new bosonic and/or fermionic
messengers. These fields will be heavy, with masses close to the
cutoff scale $\Lambda$. 
All relevant messenger fields will be singlet under $SU(3)_C$, while
their transformation under $SU(2)_L \otimes U(1)_Y$ will vary.
For quick reference we list all such messenger fields, their Lorentz
transformation and $SU(2)_L$ and $U(1)_Y$ charges are given in the
Table. \ref{tab:mes-fields}.
 Everywhere, except for Standard Model fields, the subscript
  denotes the hypercharge. To make the notation lighter this is not
  done for Standard Model fields. In contrast to $\chi_0$, $\Phi$ and
  $\Delta_i$, which are the external scalars that acquire vevs, we
  denote as $\chi^\prime_0$ the messenger singlets which develop only
  an induced vev. The corresponding doublets will be denoted as
  $\sigma_1 \equiv \Phi^\prime$ and the triplets are denoted as
  $\Delta_i^\prime$. The electric charge conservation implies
  that all the components appearing in the diagrams must be
  electrically neutral.
\begin{table}[ht]
\begin{center}
\begin{tabular}{c c c c}
  \hline \hline  
Messenger Field                       \hspace{1cm}    & Lorentz   \hspace{1cm}    &  $SU(2)_L$    \hspace{1cm}   &   $U(1)_Y$ \\
\hline   \hline 
$\chi'_0$                      \hspace{1cm}    & Scalar                   \hspace{1cm}    &  $1$          \hspace{1cm}   &   $0$              \\
$N_{L,0}, N_{R, 0}$             \hspace{1cm}    & Fermion                  \hspace{1cm}    &  $1$          \hspace{1cm}   &   $0$              \\
$\sigma_1$                       \hspace{1cm}    & Scalar                   \hspace{1cm}    &  $2$          \hspace{1cm}   &   $1$              \\
$E_{L, 1}, E_{R, 1}$              \hspace{1cm}    & Fermion                  \hspace{1cm}    &  $2$          \hspace{1cm}   &   $1$              \\
$E_{L, -1}, E_{R, -1}$             \hspace{1cm}    & Fermion                  \hspace{1cm}    &  $2$          \hspace{1cm}   &   $-1$              \\
$\Delta^\prime_0$                   \hspace{1cm}    & Scalar                   \hspace{1cm}    &  $3$          \hspace{1cm}   &   $0$              \\
$\Delta^\prime_2$                    \hspace{1cm}    & Scalar                   \hspace{1cm}    &  $3$          \hspace{1cm}   &   $2$              \\
$\Delta^\prime_{-2}$                  \hspace{1cm}    & Scalar                   \hspace{1cm}    &  $3$          \hspace{1cm}   &   $-2$             \\
$\Sigma_{L, 0}, \Sigma_{R, 0}$         \hspace{1cm}    & Fermion                  \hspace{1cm}    &  $3$          \hspace{1cm}   &   $0$              \\
$\Sigma_{L, 2}, \Sigma_{R, 2}$          \hspace{1cm}    & Fermion                  \hspace{1cm}    &  $3$          \hspace{1cm}   &   $2$              \\
$\Sigma_{L, -2}, \Sigma_{R, -2}$         \hspace{1cm}    & Fermion                  \hspace{1cm}    &  $3$          \hspace{1cm}   &   $-2$             \\
$\Xi_1$                                   \hspace{1cm}    & Scalar                   \hspace{1cm}    &  $4$          \hspace{1cm}   &   $1$              \\
$Q_{L, -1}, Q_{R,-1}$      		   \hspace{1cm}    & Fermion                  \hspace{1cm}    &  $4$          \hspace{1cm}   &   $-1$             \\
  \hline
  \end{tabular}
\end{center}
\caption{Messengers transform under $SU(2)_L$ and $U(1)_Y$ as given, they are color singlets and we are using the convention
  $Q = T_3 + \frac{Y}{2}$. See text for the explanation of the notation used.}
 \label{tab:mes-fields}
\end{table} 

Note that certain messenger fields e.g $\Delta^\prime_2$ and $\Delta^\prime_{-2}$
are related with each other, for example
$\bar{\Delta}^\prime_2 \equiv \Delta^\prime_{-2}$. We have chosen to give them
different symbols to avoid any confusion and also for aesthetics
reasons.
The notation and transformation properties of these messenger fields
as listed in Table. \ref{tab:mes-fields} will be used throughout the
rest of the paper. 

%%%%%%%%%%%%%%%%%%%%%%%%%%%%%%%%%%%%%%%%%%%%%%%%%%%%%%%%%%%%%%%%%%%%%%%

\section{Operator Involving only the Standard Model Doublet} 
\label{sec:OnlySM}

%%%%%%%%%%%%%%%%%%%%%%%%%%%%%%%%%%%%%%%%%%%%%%%%%%%%%%%%%%%%%%%%%%%%%%%
 
We begin our discussion with the operator involving only \sm scalar
doublet $\Phi$ and discuss the various possible UV-complete models for
this case. 
As has been argued in \cite{CentellesChulia:2018gwr}, for Dirac
neutrinos, after the Yukawa term, the lowest dimensional operator
involving only the \sm scalar doublet appears at dimension-6 and is
given by
\begin{equation}
 \frac{1}{\Lambda^2} \bar{L}\otimes \bar{\Phi} \otimes \bar{\Phi} \otimes \Phi \otimes \nu_R
 \label{op-sm}
\end{equation}
where $L$ and $\Phi$ denote the lepton and Higgs doublets, $\nu_R$ is
the right-handed neutrino field and $\Lambda$ represents the cutoff
scale.
Above $\Lambda$ the Ultra-Violet (UV) complete theory is at play,
involving new ``messenger'' fields, whose masses lie close to the
scale $\Lambda$. Recently, this operator has also been studied in
\cite{Yao:2017vtm} and our results agree with those obtained in that
work. 

Before starting our systematic classification of the UV-complete
seesaw models emerging from this operator, we stress that, in order
for this operator to give the leading contribution to Dirac neutrino
masses, the lower dimensional Yukawa term $\bar{L} \bar{\Phi} \nu_R$
should be forbidden. 
This can happen in many scenarios involving flavor
symmetries~\cite{Chulia:2016ngi,Chulia:2016giq,CentellesChulia:2017koy}
and/or additional $U(1)_{B-L}$ symmetries with unconventional charges
for $\nu_R$~\cite{Ma:2014qra,Ma:2015mjd}.
If this dimension-4 operator is forbidden by a simple $U(1)$
or $Z_n$ symmetry, then it is easy to see that the dimension-6
operator will also be forbidden.
However, the dimension-4 operator can be forbidden in other ways.
As a first possibility, one could have a softly broken symmetry, such
as $Z_3$~\cite{Yao:2017vtm}.
Alternatively, one can add a new Higgs doublet $\Phi_2$ transforming
as the \sm Higgs under the gauge group.
One can show that, in such a two-doublet Higgs model (2HDM) the
imposition of a $Z_3$ symmetry, under with the two Higgses transform
non-trivially, is sufficient. The diagrams and discussion of this
section will be identical for the 2HDM case, except that a label
$\Phi_1$ or $\Phi_2$ should be added instead of simply $\Phi$.
Finally, one may invoke non-abelian discrete symmetries such as
$S_4$~\cite{Yao:2017vtm}.  

The operator in \eqref{op-sm} can lead to several different
UV-complete seesaw models, depending on the field contractions
involved.
There are fifteen inequivalent ways of contracting these fields, each
of which will require different types of messenger fields for
UV-completion.
Out of these, one is similar to type I Dirac seesaw but with
induced vev for the singlet scalar.
Three of them are type-II-like, with induced vevs for the messenger
scalars.
Five of them are analogous to the three type-III Dirac seesaws
discussed in \cite{CentellesChulia:2018gwr}, with induced vevs for
either the singlet or the triplet, while the other six are new
diagrams.
We now look at these possibilities one by one. 

%%%%%%%%%%%%%%%%%%%%%%%%%%%%%%%%%%%%%%%%%%%%%%%%%%%%%%%%%%%%%%%%%%%%%%%
\subsection{Type I seesaw mechanism with induced vev}
%%%%%%%%%%%%%%%%%%%%%%%%%%%%%%%%%%%%%%%%%%%%%%%%%%%%%%%%%%%%%%%%%%%%%%%

\begin{figure}[!h]
 \centering
  \includegraphics[scale=0.25]{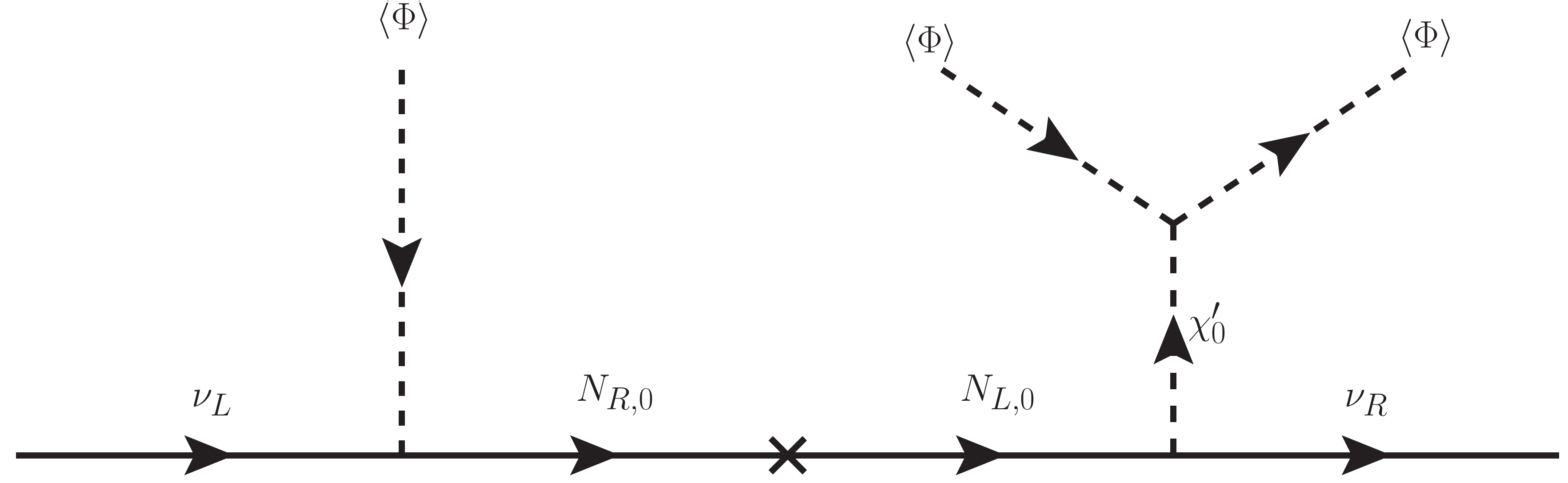}
  \caption{Feynman diagram representing the Dirac Type-I seesaw with
    an induced vev for $\chi^\prime_0$.}
    \label{DT1}
 \end{figure}

One of the simplest contractions for the dimension-6 operator of
\eqref{op-sm} is as follows:
\be
\underbrace{\underbrace{\underbrace{\bar{L} \otimes \bar{\Phi}}_1 \otimes \underbrace{\bar{\Phi} \otimes \Phi}_1}_{1} \otimes \underbrace{\nu_R}_{1}}_{\textnormal{Type I with induced vev  Fig.~\ref{DT1}}}
\label{DT1-op}
\ee

In~\eqref{DT1-op} the under-brace denotes a $SU(2)_L$ contraction of
the fields involved, whereas the number given under it denotes the
transformation of the contracted fields under $SU(2)_L$ (note that the
other possible contraction in which $\bar{\Phi} \otimes \bar{\Phi}$
goes to a singlet is simply $0$).
Although not made explicit, we take it for granted that the global
contraction leading to a UV-complete model where the neutrino mass is
generated by the diagram shown in Fig.~\ref{DT1} should always be an
$SU(2)_L$ singlet.
 
 The diagram in Fig.~\ref{DT1} belongs to the topology $T_1$ and
 involves two messenger fields, a vector-like neutral fermion $N_0$ and
 a scalar $\chi^\prime_0$, both of which are singlet under the \SM gauge
 group. 
 As listed in Table \ref{Tab:mass-formula}, the light neutrino mass is
 doubly suppressed first by the mass of the fermion $N_0$ and also by
 the small induced vev for $\chi^\prime_0$. 
 In contrast to the type I Dirac seesaw diagram of
 Ref.~\cite{CentellesChulia:2018gwr}, here the messenger field $\chi^\prime_0$
 required for the UV-completion gets a small induced vev via its cubic
 coupling with the Standard Model Higgs doublet.

%%%%%%%%%%%%%%%%%%%%%%%%%%%%%%%%%%%%%%%%%%%%%%%%%%%%%%%%%%%%%%%%%%
\subsection{Type II seesaw mechanism with induced vev}
%%%%%%%%%%%%%%%%%%%%%%%%%%%%%%%%%%%%%%%%%%%%%%%%%%%%%%%%%%%%%%%%%

\begin{figure}[!h]
 \centering
  \includegraphics[scale=0.3]{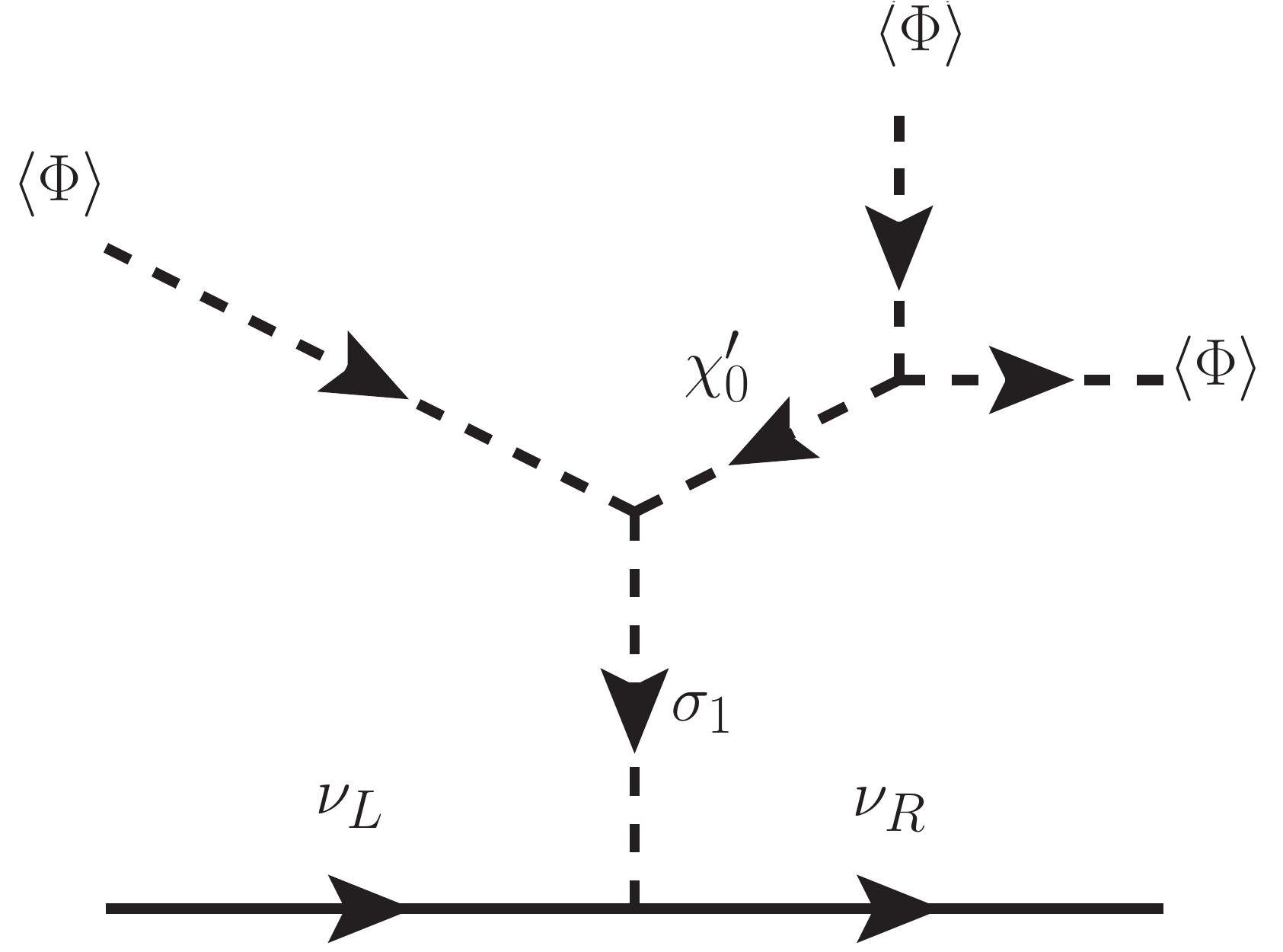}, \hspace{2mm}
   \includegraphics[scale=0.3]{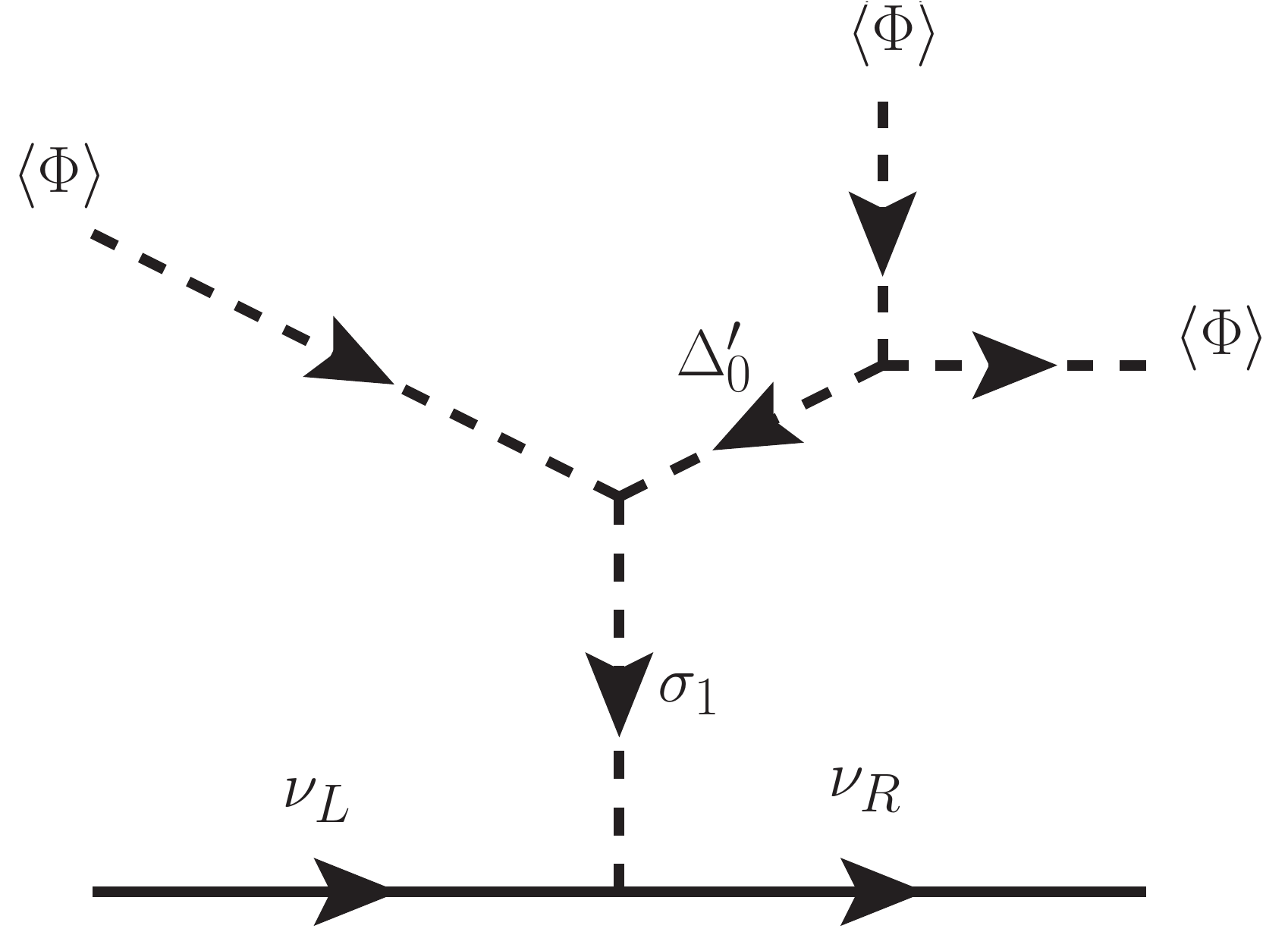}, \hspace{2mm}
    \includegraphics[scale=0.3]{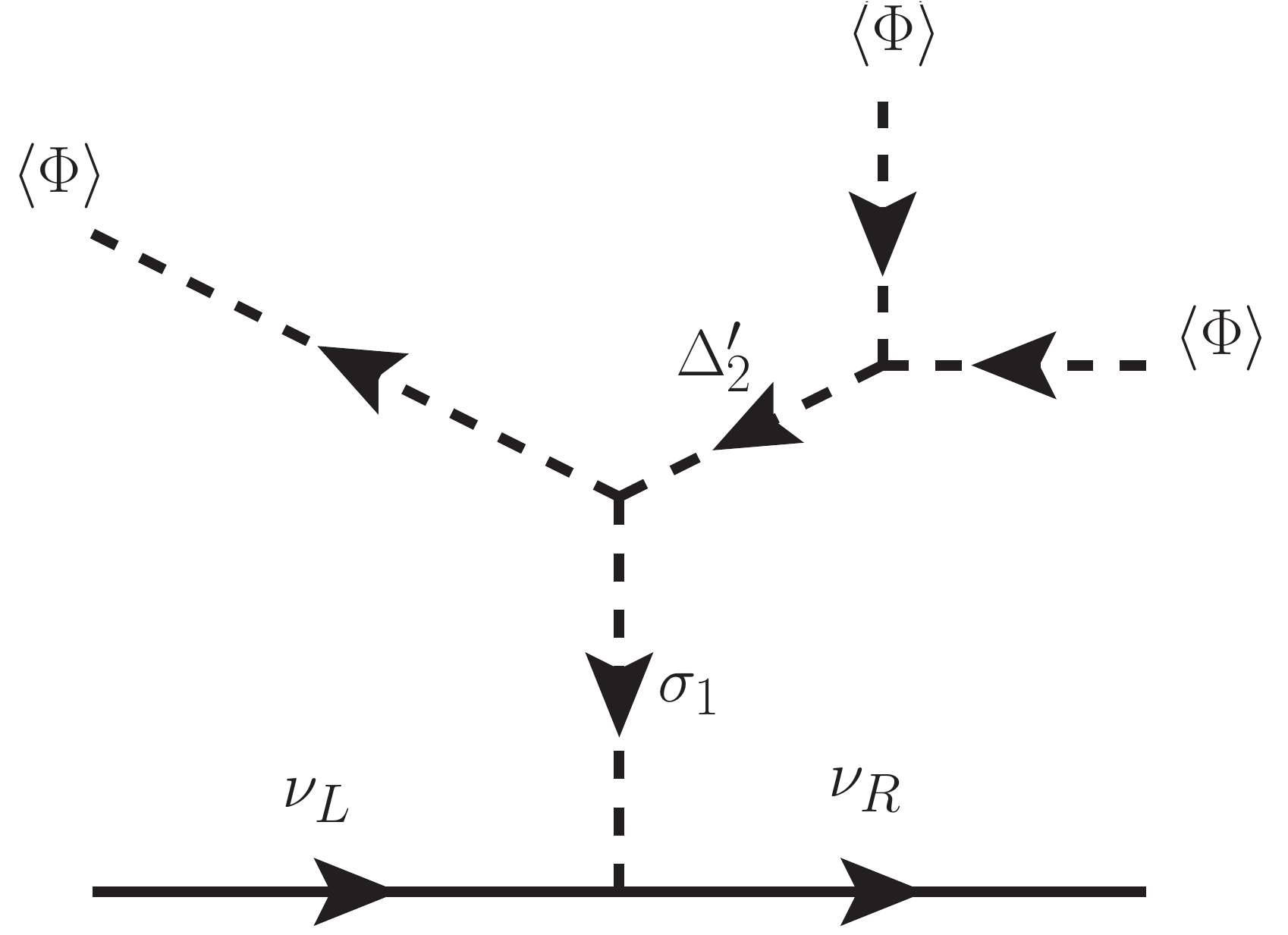}, \hspace{2mm}
     \caption{Feynman diagrams representing the three realizations of the Dirac Type-II seesaw with an induced vev for $\chi^\prime_0$ or $\Delta^\prime_i$.}
    \label{DT2}
 \end{figure}

The three possibilities for this case are shown in \eqref{DT2-op}. 
\be
\underbrace{\underbrace{\underbrace{\bar{L} \otimes \nu_R}_2 \otimes \underbrace{\bar{\Phi} \otimes \Phi}_1}_{2} \otimes \underbrace{\bar{\Phi}}_{2}}_{\textnormal{Type II with induced vev  Fig.~\ref{DT2}}}, \hspace{0.5cm}
\underbrace{\underbrace{\underbrace{\bar{L} \otimes \nu_R}_2 \otimes \underbrace{\bar{\Phi} \otimes \Phi}_3}_{2} \otimes \underbrace{\bar{\Phi}}_{2}}_{\textnormal{Type II with induced vev   Fig.~\ref{DT2}}} , \hspace{0.5cm}
\underbrace{\underbrace{\underbrace{\bar{L} \otimes \nu_R}_2 \otimes \underbrace{\bar{\Phi} \otimes  \bar{\Phi}_3}_{2}} \otimes \underbrace{\Phi}_{2}}_{\textnormal{Type II with induced vev Fig.~\ref{DT2}}}
\label{DT2-op}
\ee

These three contraction possibilities lead to three different
UV-completions, as illustrated in Fig.~\ref{DT2}.
 All diagrams in Fig.~\ref{DT2} belong to the $T_3$ topology, and
 require two scalar messengers.  The diagram on the left requires a
 $SU(2)_L$ singlet $\chi^\prime_0$ and a new doublet $\sigma_1$ (different from
 the \sm Higgs doublet) with $U(1)_Y = 1$.
 The middle one requires an $SU(2)_L$ triplet $\Delta^\prime_0$ (with
 $U(1)_Y = 0$) and an $SU(2)_L$ doublet $\sigma_1$ (with $U(1)_Y = 1$)
 scalar messengers.  
 The third diagram is identical to the second, exchanging
 $\Phi \leftrightarrow \bar{\Phi}$ in two external legs. Note that the
 hypercharges of the intermediate fields $\Delta^\prime_{0}$ and $\Delta^\prime_{2}$
 are different so that, although the UV-completions share the same
 topology, the underlying models are different. 
 The associated light neutrino mass estimate is given in Table
 \ref{Tab:mass-formula}. 
%

%%%%%%%%%%%%%%%%%%%%%%%%%%%%%%%%%%%%%%%%%%%%%%%%%%%%%%%%%%%%%%%%%%%%%%%%%
\subsection{Type III seesaw mechanism with induced vevs}
%%%%%%%%%%%%%%%%%%%%%%%%%%%%%%%%%%%%%%%%%%%%%%%%%%%%%%%%%%%%%%%%%%%%%%%%%

\begin{figure}[H]
 \centering
  \includegraphics[scale=0.24]{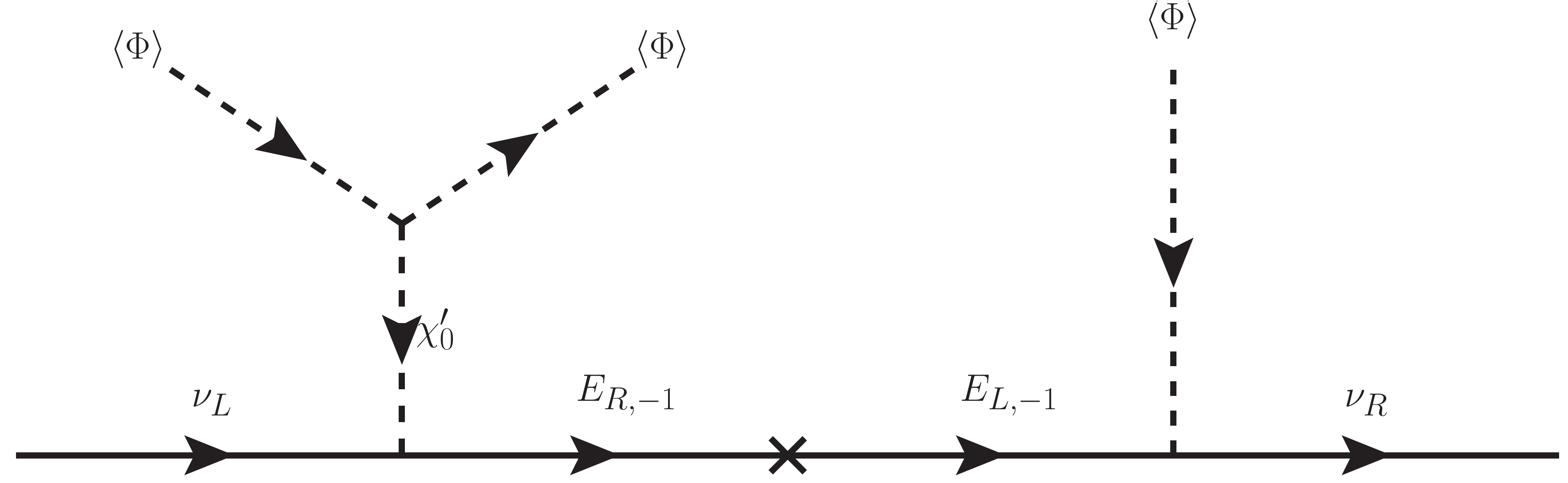}, \hspace{2mm}
   \includegraphics[scale=0.24]{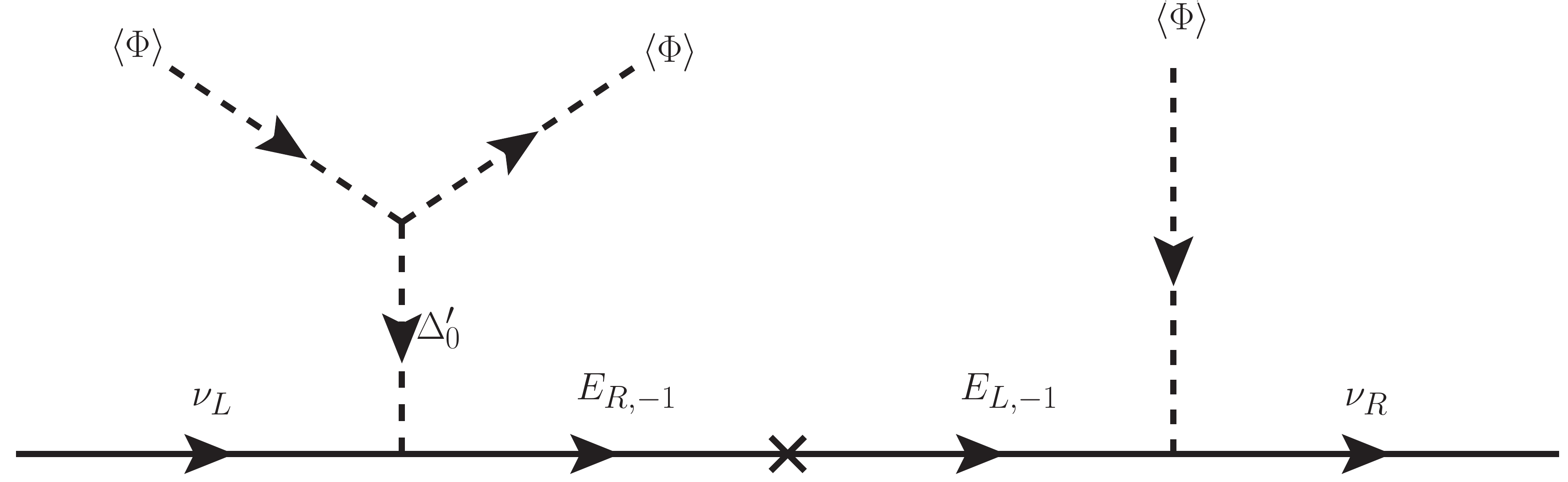}, \hspace{2mm}
    \includegraphics[scale=0.24]{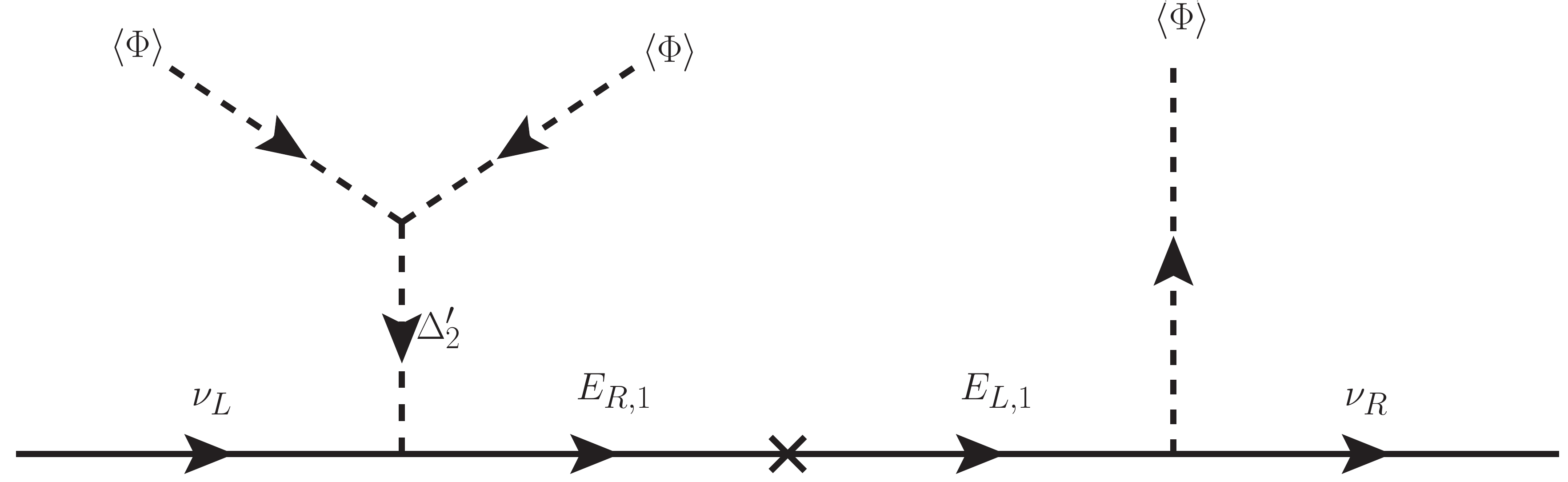}, \hspace{2mm}
     \includegraphics[scale=0.24]{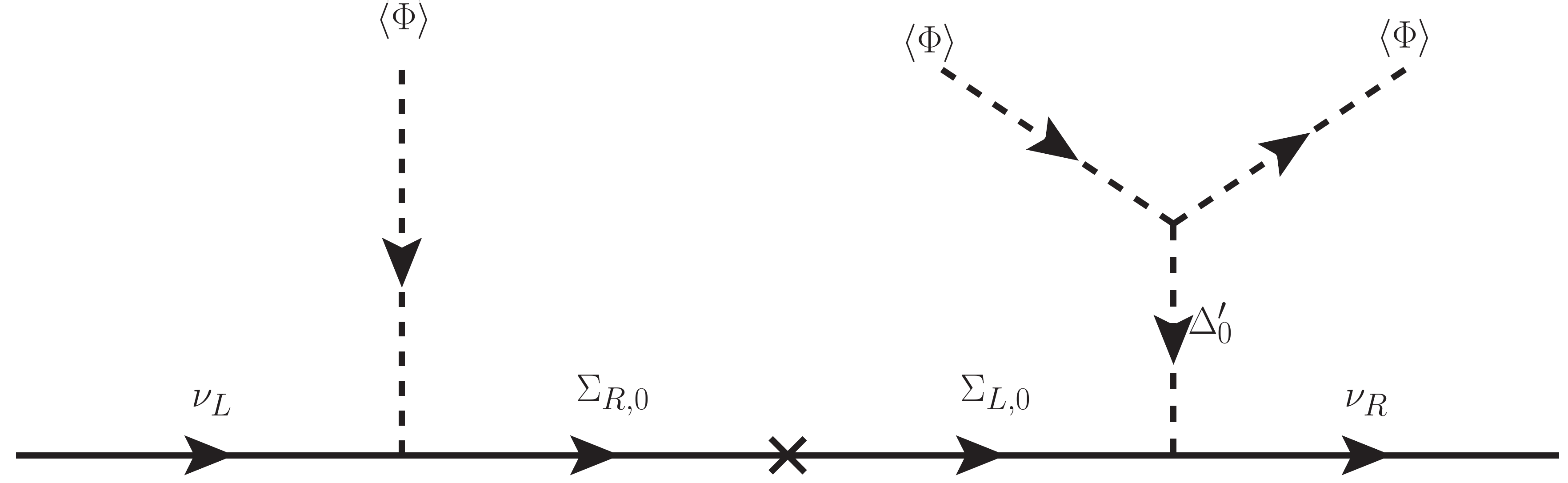}, \hspace{2mm}
      \includegraphics[scale=0.24]{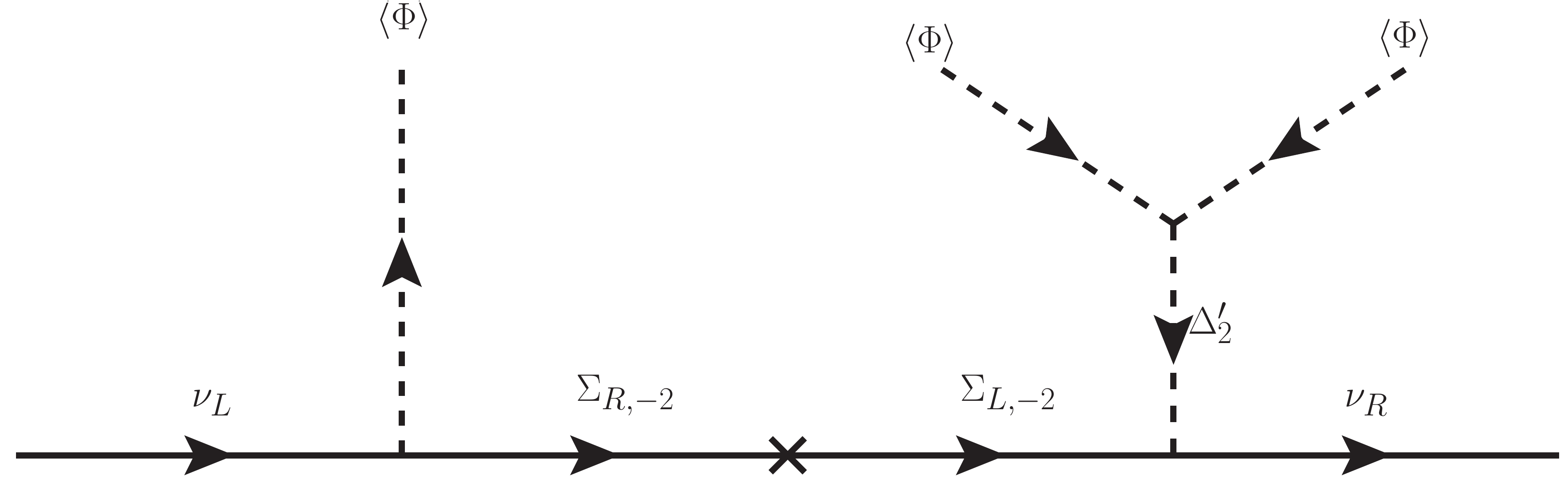}
       \caption{Feynman diagram representing the five realizations of the Dirac Type-III seesaw with an induced vev for $\chi^\prime_0$ or $\Delta^\prime_i$.}
    \label{DT3}
 \end{figure}

The operator of \eqref{op-sm} also leads to five distinct type-III
like seesaw possibilities with induced vevs~\footnote{We denote all
  diagrams with $T_1$ or $T_2$ topologies as type-III seesaw-like if
  they involve fermions transforming non-trivially under $SU(2)_L$.}.
The various possible contractions leading to such possibilities as
shown in \eqref{DT32-op}. 
\be
\underbrace{\underbrace{\underbrace{\bar{L}}_2 \otimes \underbrace{\bar{\Phi} \otimes \Phi}_1}_2 \otimes \underbrace{\bar{\Phi} \otimes \nu_R}_{2}}_{\textnormal{Type III with induced vev Fig.~\ref{DT3}}} , \hspace{0.5cm}
\underbrace{\underbrace{\underbrace{\bar{L}}_2 \otimes \underbrace{\bar{\Phi} \otimes \Phi}_3}_2 \otimes \underbrace{\bar{\Phi} \otimes \nu_R}_{2}}_{\textnormal{Type III with induced vev Fig.~\ref{DT3}}} , \hspace{0.5cm}
\underbrace{\underbrace{\underbrace{\bar{L}}_2 \otimes \underbrace{\bar{\Phi} \otimes \bar{\Phi}}_3}_2 \otimes \underbrace{{\Phi} \otimes \nu_R}_{2}}_{\textnormal{Type III with induced vev Fig.~\ref{DT3}}} \nonumber
\label{DT3-op}
\ee
\be
\underbrace{\underbrace{\underbrace{\bar{L} \otimes \bar{\Phi}}_3 \otimes \underbrace{\bar{\Phi} \otimes \Phi}_3}_{1} \otimes \underbrace{\nu_R}_{1}}_{\textnormal{Type III with induced vev Fig.~\ref{DT3}}}, \hspace{0.5cm}
\underbrace{\underbrace{\underbrace{\bar{L} \otimes {\Phi}}_3 \otimes \underbrace{\bar{\Phi} \otimes \bar{\Phi}}_3}_{1} \otimes \underbrace{\nu_R}_{1}}_{\textnormal{Type III with induced vev Fig.~\ref{DT3}}}
\label{DT32-op}
\ee

The UV-completions of each of these possible contractions involve
different messenger fields, leading to five inequivalent models. The
neutrino mass generation in these models is shown diagrammatically in
Figure \ref{DT3}. 
 The first diagram in Figure \ref{DT3} involves, as messenger fields,
 scalar singlet $\chi^\prime_0$ and vector-like fermions $E_{-1}$
 transforming as an $SU(2)_L$ doublet with $U(1)_Y = -1$. 
 The second diagram involves hypercharge-less $SU(2)_L$ triplet
 scalars $\Delta^\prime_0$ and the same $SU(2)_L$ doublet vector-like
 fermion $E_{-1}$ as messenger fields. 
 The third diagram is identical to the second one, but with exchange
 $\Phi \leftrightarrow \bar{\Phi}$ in the external legs. This leads to
 a different hypercharge $U(1)_Y = 2$ for the intermediate scalar
 triplet $\Delta^\prime_{2}$, as well as for the $SU(2)_L$ doublet vector
 fermion $E_1$, with $U(1)_Y = 1$. 
 The fourth and fifth diagrams again are related to each other by
 exchanging $\Phi \leftrightarrow \bar{\Phi}$ in two external
 legs. They involve, as messenger fields, $SU(2)_L$ triplet scalars
 $\Delta^\prime_i$; $i = 0, 2$ together with vector-like $SU(2)_L$ triplet
 fermions $\Sigma_i$; $i = 0, -2$.  The hypercharges of $\Delta_i^\prime$ are
 $U(1)_Y = 0, 2$ and of $\Sigma_i$ are $U(1)_Y = 0, -2$
 respectively. 
 The first three diagrams belong to $T_2$ topology, while the fourth
 and fifth diagrams have the topology $T_1$ and the associated light
 neutrino masses for $T_1$ and $T_2$ are given in Table
 \ref{Tab:mass-formula}. 
 Notice that, in contrast to the type III like Dirac seesaw diagrams
 discussed in~\cite{CentellesChulia:2018gwr}, here the $\chi^\prime_0$ and
 $\Delta^\prime_i$ both get induced vevs from their cubic interaction terms
 with the Standard Model Higgs doublet.
 
%%%%%%%%%%%%%%%%%%%%%%%%%%%%%%%%%%%%%%%%%%%%%%%%%%%%%%%%%%%%%%%%%

\subsection{New diagrams}

%%%%%%%%%%%%%%%%%%%%%%%%%%%%%%%%%%%%%%%%%%%%%%%%%%%%%%%%%%%%%%%%%

\begin{figure}[!h]
 \centering
  \includegraphics[scale=0.30]{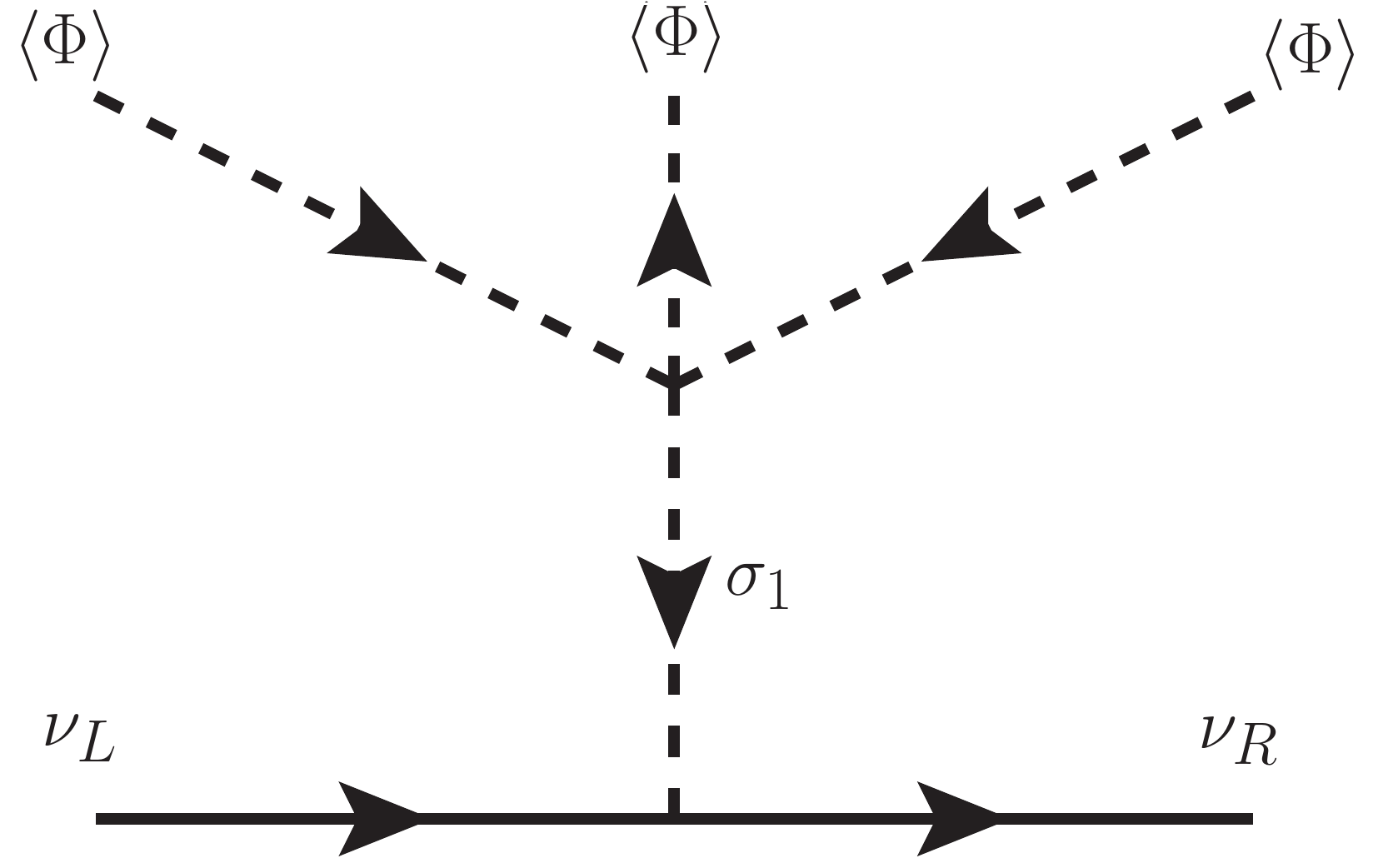}
  \caption{Feynman diagram representing the new possible UV completion
    belonging to topology $T_4$.}
    \label{DT2new}
 \end{figure}

Apart from the above diagrams, there are also six new ones which have
no dimension-5 analogues listed in Ref.~\cite{CentellesChulia:2018gwr}.
The first of these possibilities arise from the field contraction shown in \eqref{DT2new-op}.   

\be
\underbrace{\underbrace{\bar{L} \otimes \nu_R}_2 \otimes \underbrace{\bar{\Phi} \otimes \Phi\otimes \bar{\Phi}}_2}_{\textnormal{Fig. \ref{DT2new}}}, \hspace{0.5cm}
\label{DT2new-op}
\ee
This particular contraction of the operators leads to a UV-complete
model where the neutrino mass arises from the Feynman diagram shown in
Fig.~\ref{DT2new} involving a single scalar messenger field $\sigma_1$
transforming as  $SU(2)_L$ doublet with $U(1)_Y = 1$.
 The field $\sigma_1$ gets a small induced vev through its quartic
 coupling with the \sm Higgs doublet. 
 This diagram belongs to the $T_4$ topology and the resulting light
 neutrino mass estimate is given in Table \ref{Tab:mass-formula}.

\begin{figure}[!h]
 \centering
  \includegraphics[scale=0.24]{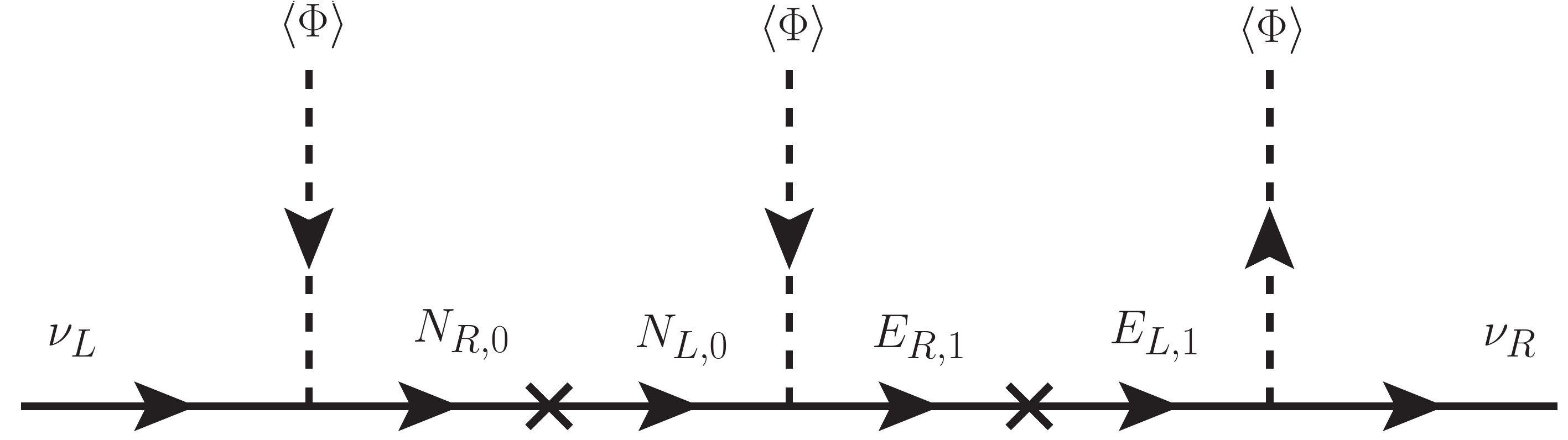}, \hspace{2mm}
  \includegraphics[scale=0.24]{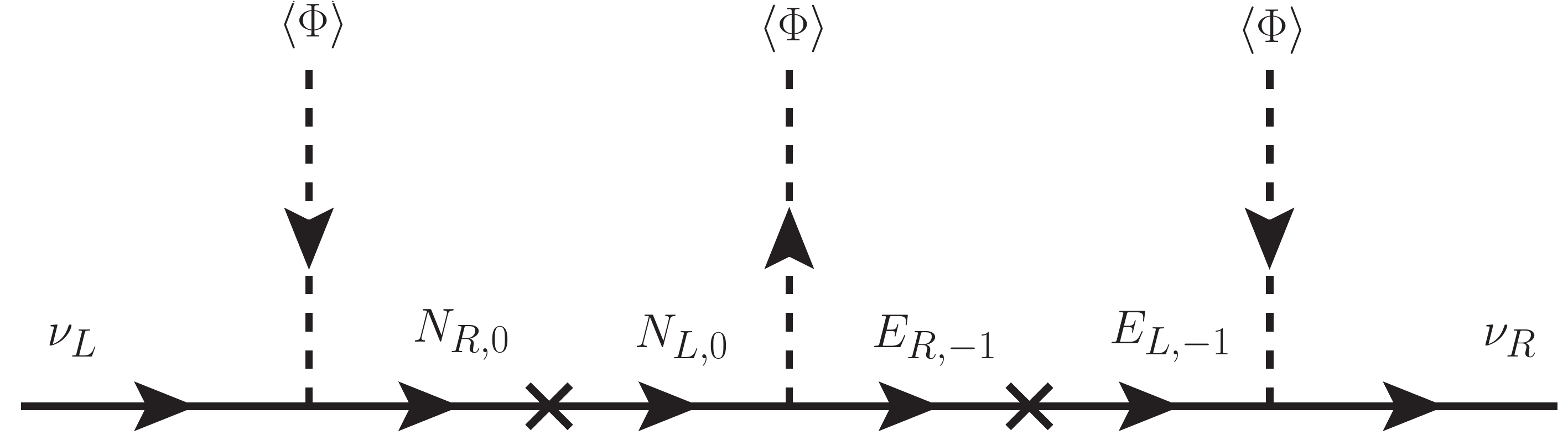}, \\
      \includegraphics[scale=0.24]{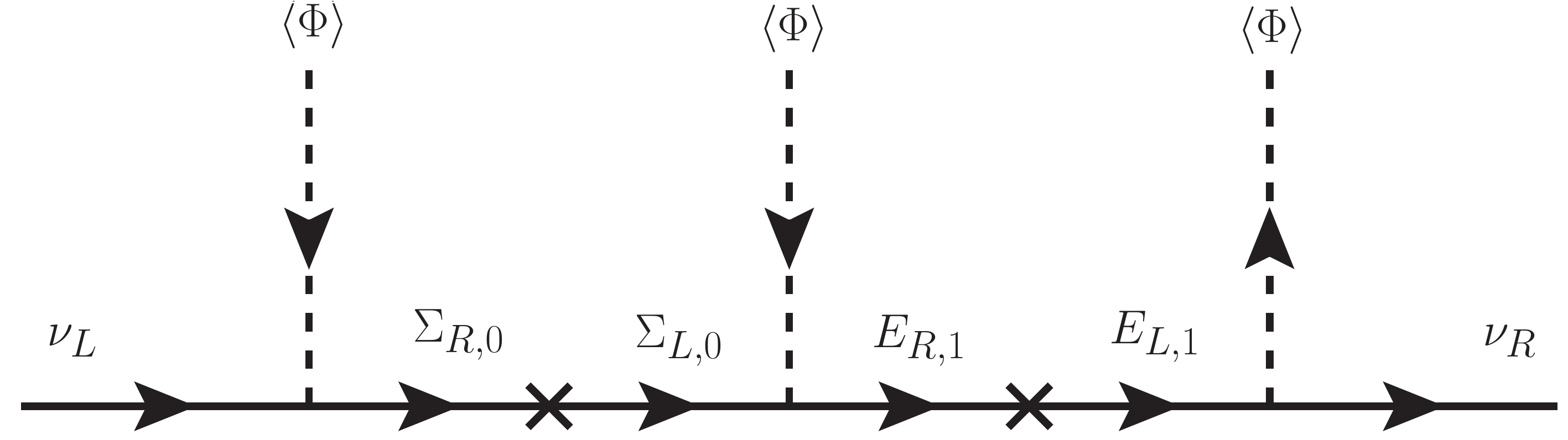}, \hspace{2mm}
      \includegraphics[scale=0.24]{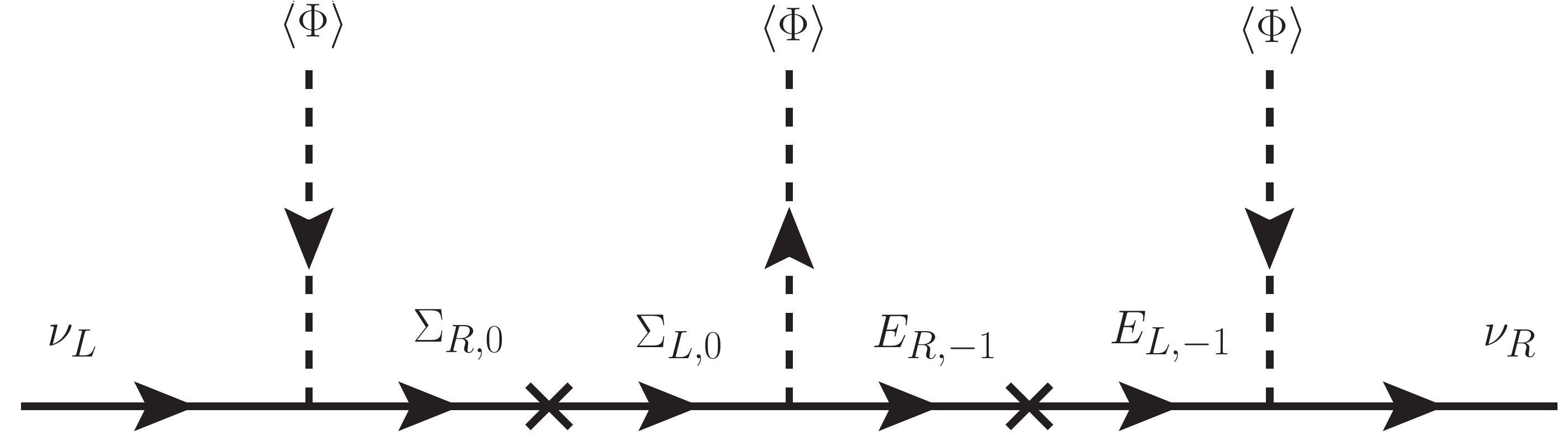}, \hspace{2mm}
      \includegraphics[scale=0.24]{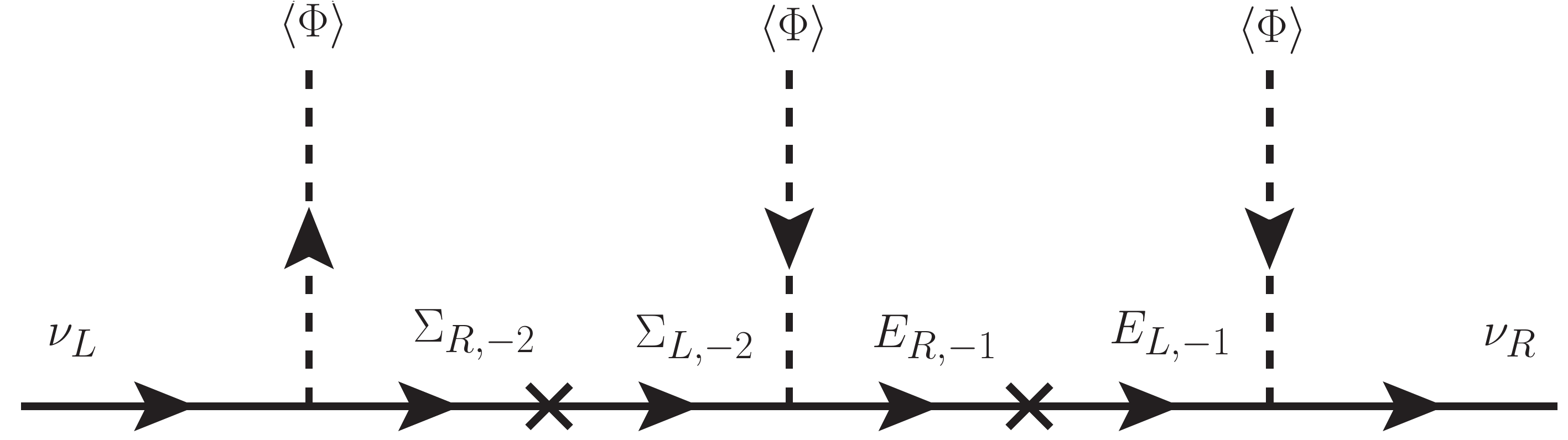}
      \caption{Feynman diagram representing the five realizations of
        the topology $T_5$ diagrams.}
    \label{DN1}
 \end{figure}

 Finally, there are five other field contractions of the dimension-6
 operator, as shown in \eqref{DN1-op} and \eqref{DN1-op2}. 
\be
\underbrace{\underbrace{\underbrace{\bar{L} \otimes \bar{\Phi}}_1 \otimes \underbrace{\bar{\Phi}}_2}_{2} \otimes \underbrace{\Phi \otimes \nu_R}_{2}}_{\textnormal{Fig. \ref{DN1}}}, \hspace{0.5cm}
\underbrace{\underbrace{\underbrace{\bar{L} \otimes \bar{\Phi}}_1 \otimes \underbrace{\Phi}_2}_{2} \otimes \underbrace{\bar{\Phi} \otimes \nu_R}_{2}}_{\textnormal{Fig. \ref{DN1}}}
\label{DN1-op}
\ee
\be
\underbrace{\underbrace{\underbrace{\bar{L} \otimes \bar{\Phi}}_3 \otimes \underbrace{\bar{\Phi}}_2}_{2} \otimes \underbrace{\Phi \otimes \nu_R}_{2}}_{\textnormal{Fig. \ref{DN1}}}, \hspace{0.5cm}
\underbrace{\underbrace{\underbrace{\bar{L} \otimes \bar{\Phi}}_3 \otimes \underbrace{\Phi}_2}_{2} \otimes \underbrace{\bar{\Phi} \otimes \nu_R}_{2}}_{\textnormal{Fig. \ref{DN1}}}, \hspace{0.5cm}
\underbrace{\underbrace{\underbrace{\bar{L} \otimes \Phi}_3 \otimes \underbrace{\bar{\Phi}}_2}_{2} \otimes \underbrace{\bar{\Phi} \otimes \nu_R}_{2}}_{\textnormal{Fig. \ref{DN1}}}
\label{DN1-op2}
\ee
Notice that the UV-completions of these five field contractions lead
to the neutrino mass generation through topology $T_5$~\footnote{This
  is analogous to the topologies characterizing the inverse (or
  double) seesaw
  mechanism~\cite{Mohapatra:1986bd,GonzalezGarcia:1988rw} of Majorana
  neutrino mass generation.}, as shown diagrammatically in
Fig.~\ref{DN1}.

All these UV complete models only involve fermionic messengers. 
One sees that two different types of fermionic messengers are
needed. 
In the first and second diagrams the $N_0$ is a vector-like gauge singlet
fermion, whereas the vector fermions $E_1$ or $E_{-1}$ are $SU(2)_L$
doublets. 
In the first diagram the field $E_1$ carries a hypercharge $U(1)_Y = 1$
while in the second diagram $E_{-1}$ carries a hypercharge $U(1)_Y = -1$.
The last three diagrams in Fig.~\ref{DN1} also involve two types of
fermionic messengers the vector-like fermions $E_{1}$ or $E_{-1}$ are
$SU(2)_L$ doublets, while the vector-like fermions $\Sigma_ i$ transform
as triplet under the $SU(2)_L$ symmetry. 
In the third diagram $\Sigma_0$ carries no hypercharge while $E_{1}$ has
$U(1)_Y = 1$. In the fourth diagram $\Sigma_0$ carries no hypercharge,
but $E_{-1}$ has $U(1)_Y = -1$. 
In the fifth diagram $\Sigma_{-2}$ has hypercharge $U(1)_Y = -2$,
while $E_{-1}$ again has $U(1)_Y = -1$.
The light neutrino mass expected for all these diagrams is the same as
that given in Table \ref{Tab:mass-formula} for $T_5$ topology.

%%%%%%%%%%%%%%%%%%%%%%%%%%%%%%%%%%%%%%%%%%%%%%%%%%%%%%%%%%%%%%%%%%%

\section{Operators Involving Only Singlet ($\chi$) and Doublet ($\Phi$)}
\label{sec:singlet-doublet}

%%%%%%%%%%%%%%%%%%%%%%%%%%%%%%%%%%%%%%%%%%%%%%%%%%%%%%%%%%%%%%%%%%%

Having discussed the dimension-6 operator involving only the \sm Higgs
doublet, we now move on to discuss other dimension-6 operators listed
in Table \ref{Tab:op} and their UV completions.
We start our discussion with the relatively simpler operator
$\bar{L} \otimes \chi_0 \otimes \chi_0 \otimes \bar{\Phi} \otimes \nu_R$
which, apart from the \sm Higgs doublet $\Phi$, also has a vev
carrying singlet $\chi_0$.
It is clear that the $\chi_0$ should not carry any $U(1)_Y$ charge,
otherwise its vev will spontaneously break the electric charge
conservation. 

If $\chi_0$ is a complex field then, apart from h.c. of above operator,
one can also write down two other operators namely
$\bar{L} \otimes \bar{\chi_0} \otimes \bar{\chi_0} \otimes \bar{\Phi}
\otimes \nu_R$ and
$\bar{L} \otimes \bar{\chi_0} \otimes \chi_0 \otimes \bar{\Phi} \otimes
\nu_R$.
In this section we will focus on the
$\bar{L} \otimes \chi_0 \otimes \chi_0 \otimes \bar{\Phi} \otimes \nu_R$
case, since the other operator contractions and UV-completions are
very similar and can be treated analogously. 
The operator
$\bar{L} \otimes \chi_0 \otimes \chi_0 \otimes \bar{\Phi} \otimes \nu_R$
will give the leading contribution to neutrino masses only if similar
operators of equal or lower dimensionality are forbidden by some
symmetry. 
Thus lower dimensional operators allowed by $SU(2)_L \otimes U(1)_Y$
gauge symmetry, such as $\bar{L} \bar{\Phi} \nu_R$,
$\bar{L} \bar{\Phi} \chi_0 \nu_R$ and
$\bar{L} \bar{\Phi} \bar{\chi_0} \nu_R$, should be forbidden. 
It is also desirable that this operator provides the sole contribution
to neutrino masses, avoiding other dimension-6 operators such as
$\bar{L} \bar{\Phi} \bar{\Phi} \Phi \nu_R$.
A consistent scenario can arise in many ways with different
symmetries. For example one of the simplest symmetries can be a $Z_4$
symmetry (distinct from quarticity symmetry) under which the fields
transform as
\begin{eqnarray} 
\bar{L} \otimes \nu_R \sim z^2, \hspace{2mm}
\chi_0 \sim z, \hspace{2mm}
 \Phi \sim 1, 
\label{op1-charges}
\end{eqnarray}

Note that under these charge assignments the operator
$\bar{L} \otimes \bar{\chi_0} \otimes \chi_0 \otimes \bar{\Phi} \otimes
\nu_R$
is forbidden though both
$\bar{L} \otimes \chi_0 \otimes \chi_0 \otimes \bar{\Phi} \otimes \nu_R$
and
$\bar{L} \otimes \bar{\chi_0} \otimes \bar{\chi_0} \otimes \bar{\Phi}
\otimes \nu_R$ are allowed.
Hence, in principle they can simultaneously contribute to neutrino
mass generation, as long as they have consistent UV-completions.  Here
we will only discuss the first operator, though the other may be
present in some cases.
Moreover, note that the masses of charged leptons and quarks can be
generated through the usual Yukawa terms, i.e. $\bar{L} \Phi l_R$,
$\bar{Q} \Phi d_R$ and $\bar{Q} \bar{\Phi} u_R$ can be trivially
allowed by this symmetry with appropriate $Z_4$ charges of $Q$, $l_R$,
$u_R$ and $d_R$.   

\begin{figure}[!h] 
\centering
 \includegraphics[scale=0.25]{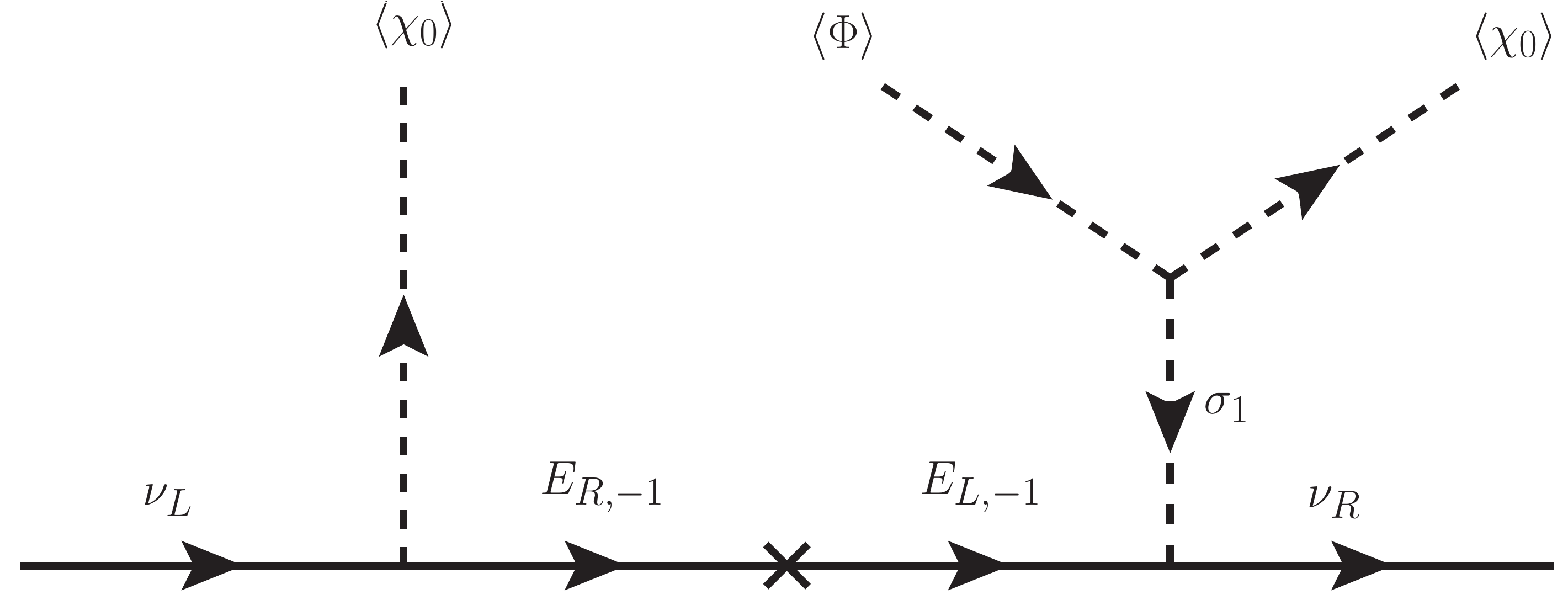}, \hspace{2mm}
  \includegraphics[scale=0.25]{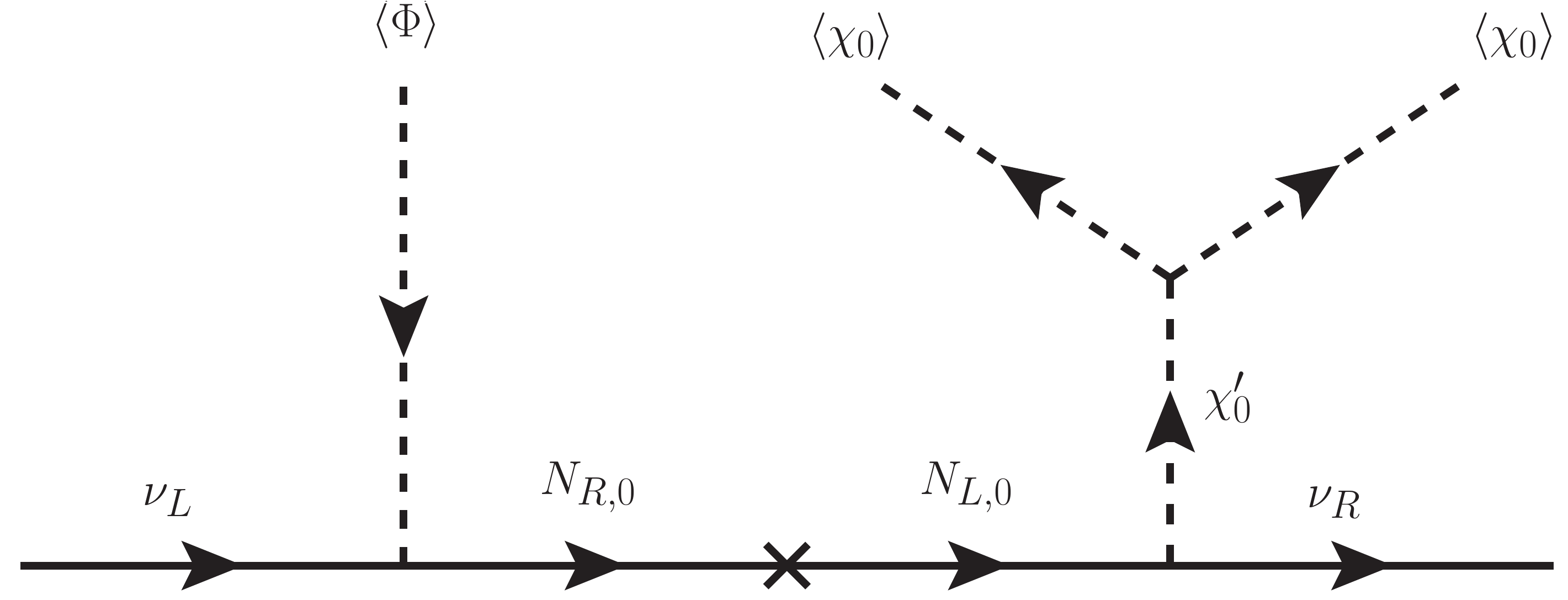}, \hspace{2mm}
   \includegraphics[scale=0.25]{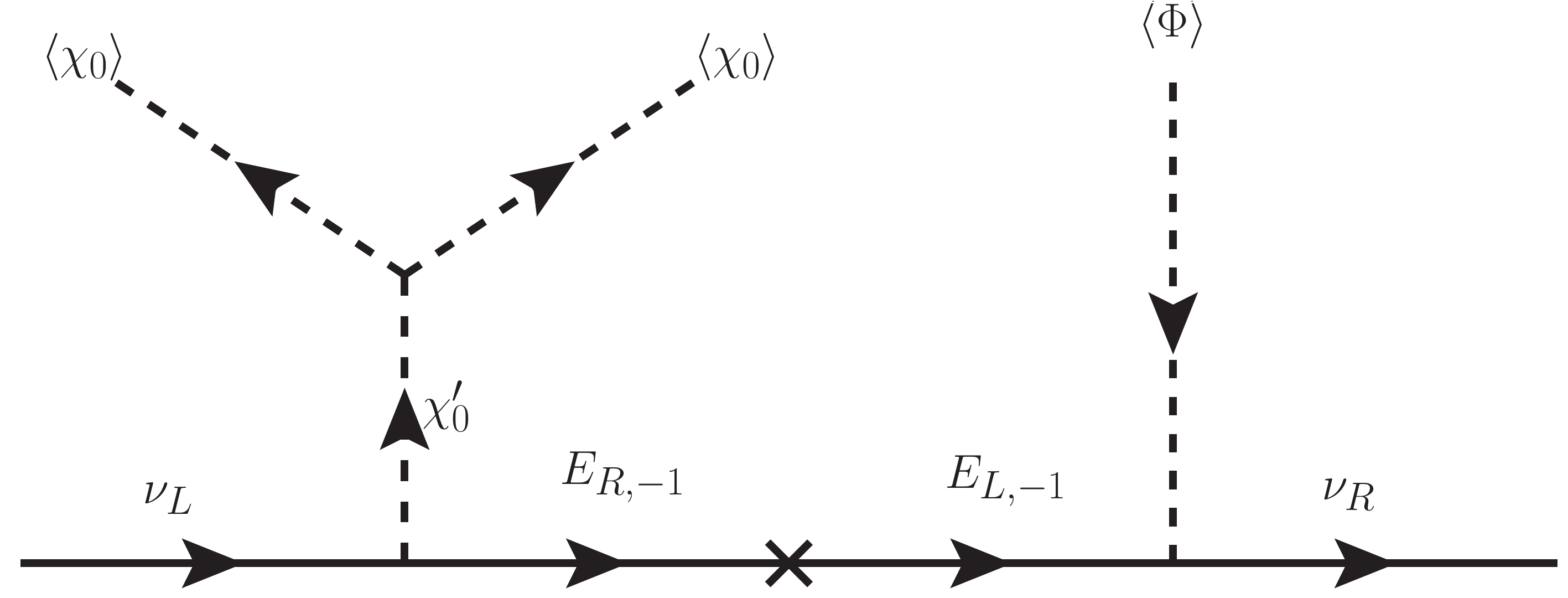}, \hspace{2mm}
    \includegraphics[scale=0.25]{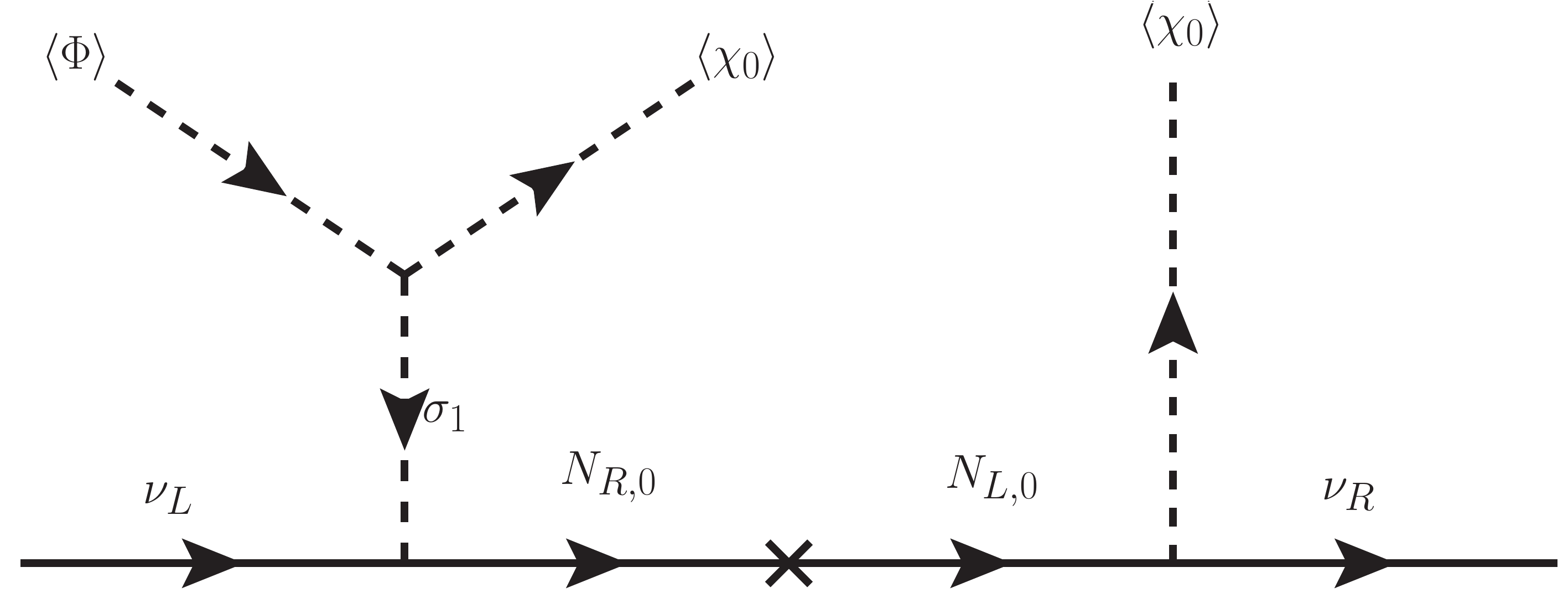}
    \caption{Diagrams showing the $T_1$ and $T_2$ topologies of the
      operator $\bar{L} \chi_0 \chi_0 \bar{\Phi} \nu_R$}
   \label{o1t1t2}
 \end{figure}

Moving on to the operator contractions and UV completions, there are
ten different ways of contracting this operator. Each of the
topologies $T_1$, $T_2$ and $T_3$ appears twice, while one diagram
belongs to the $T_4$ topology and the three others belong to the $T_5$
topology.
The operator contractions which lead to diagrams with $T_1$ or
$T_2$ topologies are listed in \eqref{ssd-t1-t2}, and the
corresponding diagrams are shown in Fig.~\ref{o1t1t2}.
\be
\underbrace{\underbrace{\bar{L} \otimes \chi_0}_2 \otimes \underbrace{\bar{\Phi} \otimes \chi_0}_2}_{1} \otimes \underbrace{\nu_R}_{1}, \hspace{0.5cm}
\underbrace{\underbrace{\bar{L} \otimes \bar{\Phi}}_1 \otimes \underbrace{\chi_0 \otimes \chi_0}_1}_{1} \otimes \underbrace{\nu_R}_{1}, \hspace{0.5cm}
\underbrace{\underbrace{\bar{L}}_2 \otimes \underbrace{\chi_0 \otimes \chi_0}_1}_{2} \otimes \underbrace{\bar{\Phi} \otimes \nu_R}_{2}, \hspace{0.5cm}
\underbrace{\underbrace{\bar{L}}_2 \otimes \underbrace{\bar{\Phi} \otimes \chi_0}_2}_{1} \otimes \underbrace{\chi_0 \otimes \nu_R}_{1}
\label{ssd-t1-t2}
\ee

 As before, the UV completion of these diagrams will involve new
 messenger fields. In all of the cases shown in Fig. \ref{o1t1t2}
 there is a scalar ($\chi_0'$ or $\sigma_1$) and a vector-like lepton
 ($N_{0}$ or $E_{i}$) messenger involved. 
 Notice that $\chi'_0$ cannot be identified with $\chi_0$ as in that case
 lower dimension-5 operators would be allowed. Thus, they must carry
 different charges under the symmetry forbidding lower dimensional
 operators (e.g $Z_4$ mentioned above). \\[-.2cm] 
 
\begin{figure}[!h]
\centering
 \includegraphics[scale=0.25]{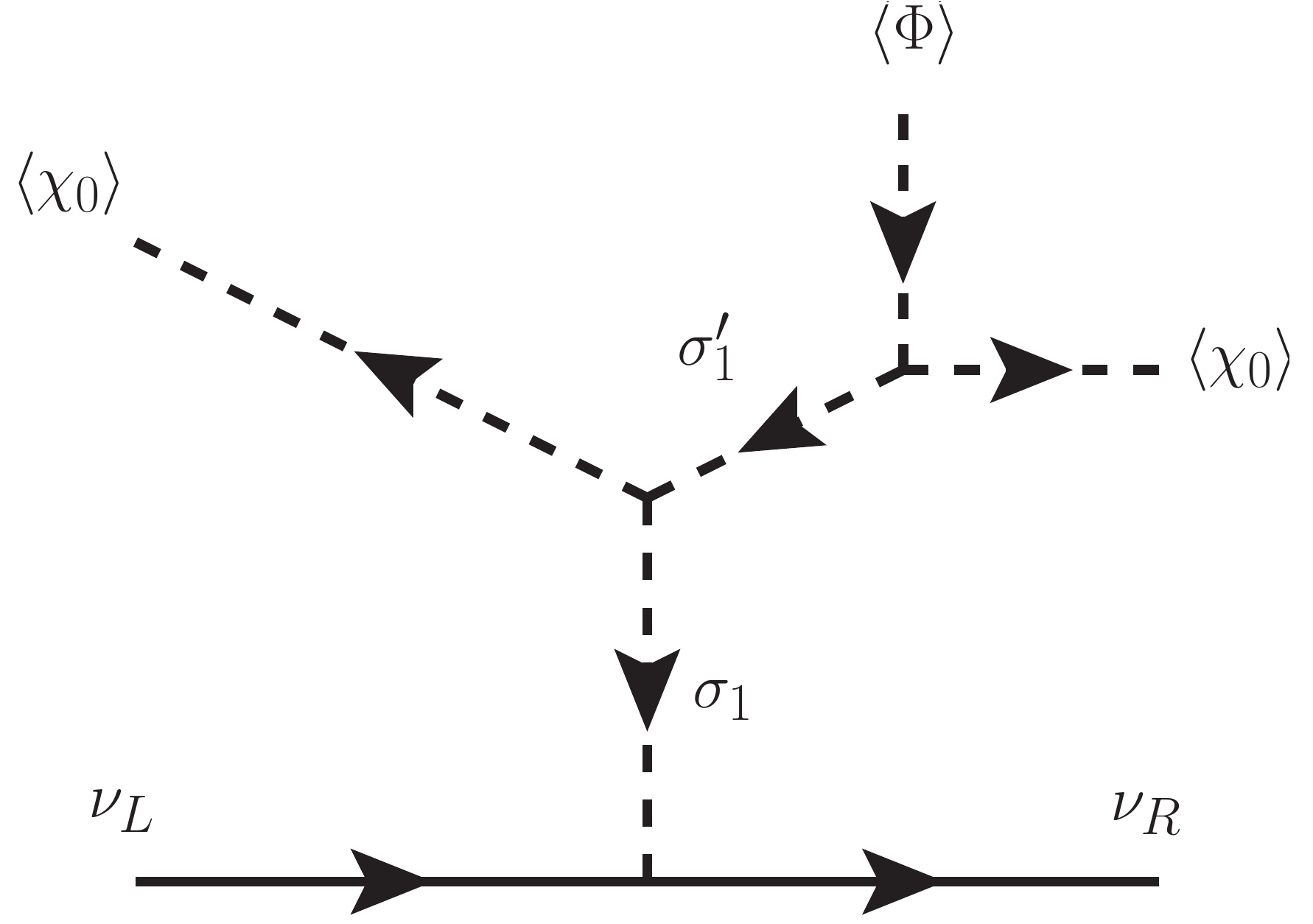}, \hspace{2mm}
  \includegraphics[scale=0.25]{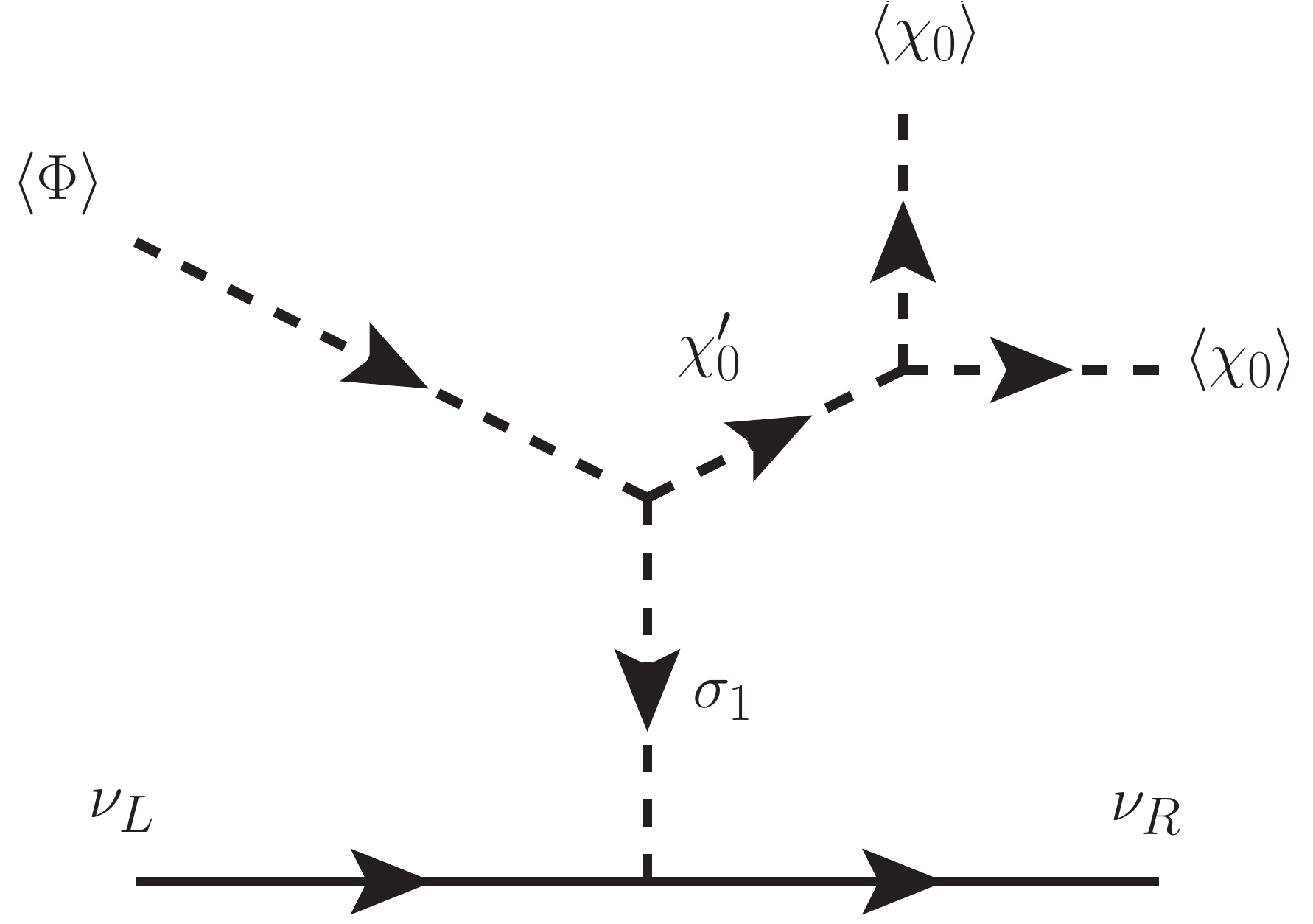}, \hspace{2mm}
   \includegraphics[scale=0.25]{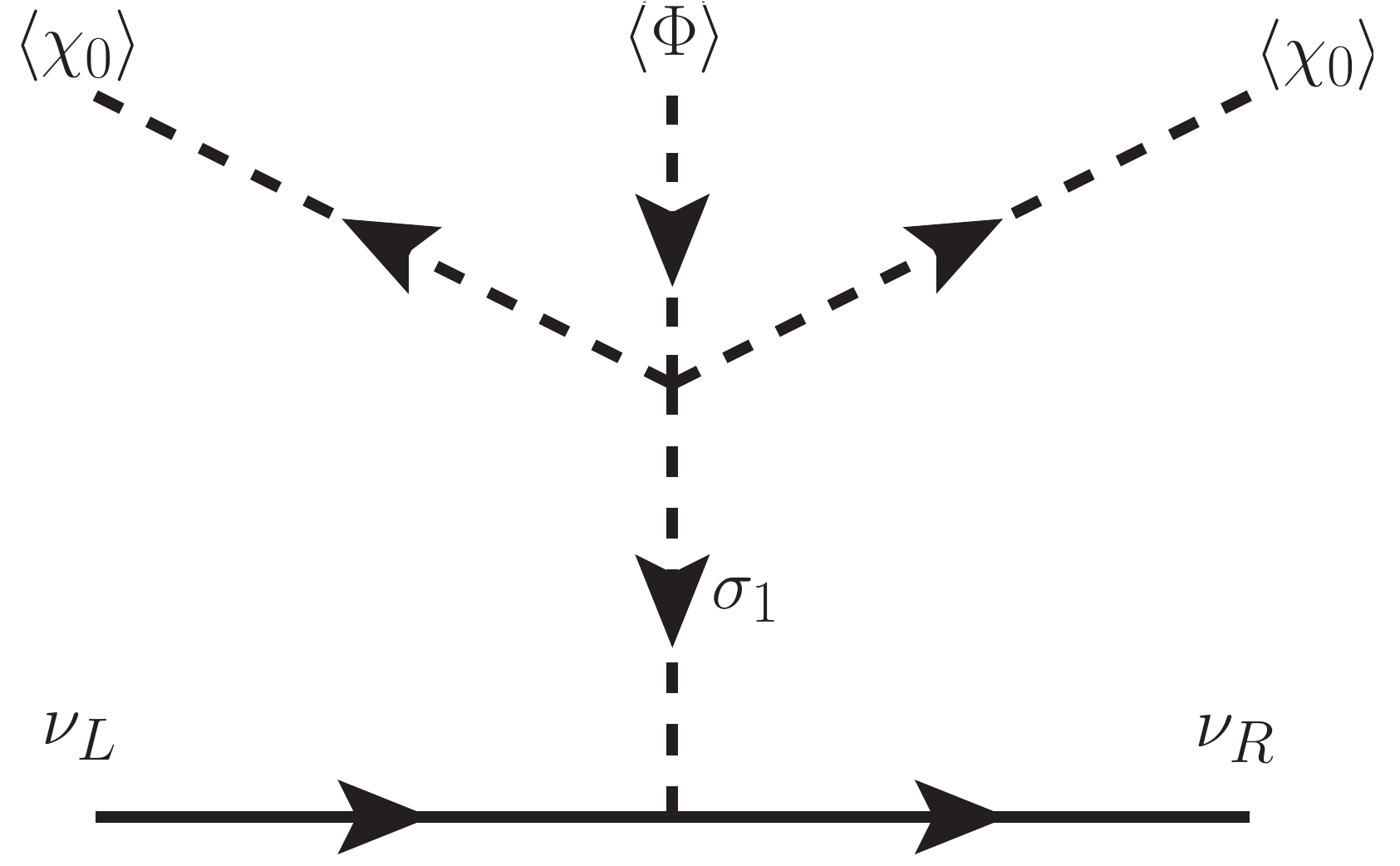}
   \caption{Diagrams showing the $T_3$ and $T_4$ topologies of the operator $\bar{L} \chi_0  \chi_0  \bar{\Phi} \nu_R$}
    \label{o1t3t4}
 \end{figure}

 UV-completion lead to other possible contractions with the $T_3$ and
 $T_4$ topology, as shown in Fig.~\ref{o1t3t4}. 
\be
\underbrace{\underbrace{\bar{L} \otimes \nu_R}_2 \otimes \underbrace{\chi_0}_1}_{2} \otimes \underbrace{\chi_0 \otimes \bar{\Phi}}_{2}, \hspace{0.5cm}
\underbrace{\underbrace{\bar{L} \otimes \nu_R}_2 \otimes \underbrace{\bar{\Phi}}_2}_{1} \otimes \underbrace{\chi_0 \otimes \chi_0}_{1}, \hspace{0.5cm}
\underbrace{\underbrace{\bar{L} \otimes \nu_R}_2 \otimes \underbrace{\bar{\Phi} \otimes \chi_0 \otimes \chi_0}_2}_{1}
\ee
 
 All the three diagrams in Fig. \ref{o1t3t4} involve only scalar
 messenger fields which can be either $\chi'_0$ or $\sigma_1$. The last diagram was realized in the Diracon model
 of Ref.~\cite{Bonilla:2016zef}.
 Notice that in the first diagram the messengers $\sigma_1$ and
 $\sigma_1^\prime$ must be different fields, since they will transform
 differently under the symmetry forbidding lower dimensional operators
 (e.g $Z_4$ of \eqref{op1-charges}). 
 Likewise, $\chi'_0$ has to be distinct from $\chi_0$ and $\sigma_1$ and $\sigma_1^\prime$ must be different from $\Phi$.

\begin{figure}[!h]
\centering
 \includegraphics[scale=0.25]{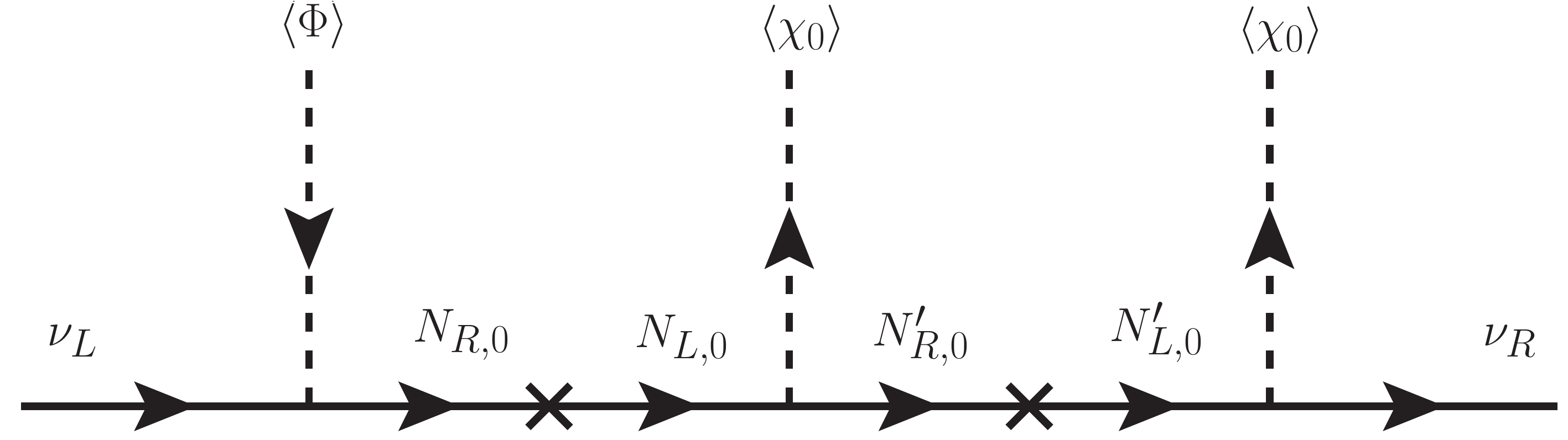}, \hspace{2mm}
  \includegraphics[scale=0.25]{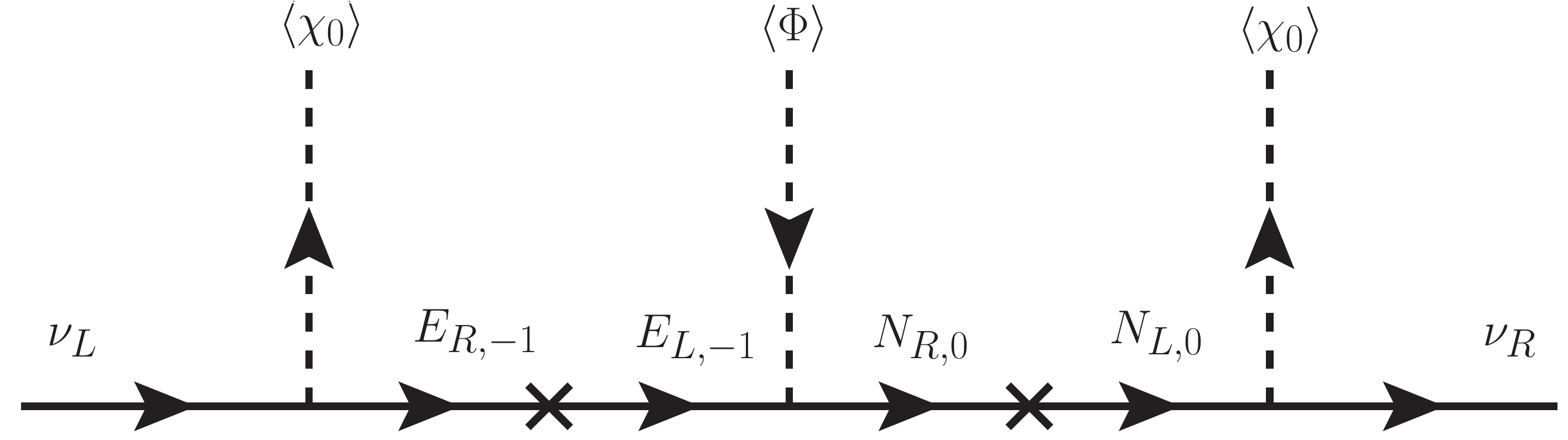}, \hspace{2mm}
   \includegraphics[scale=0.25]{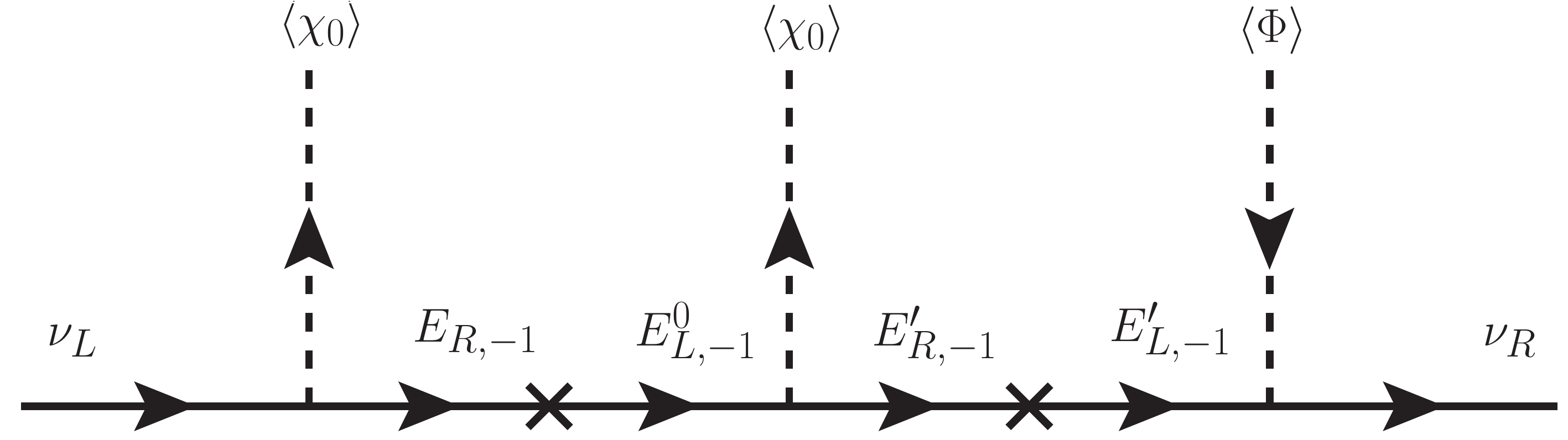}
   \caption{Diagrams showing the $T_5$ topology of the operator
     $\bar{L} \chi_0 \chi_0 \bar{\Phi} \nu_R$}
    \label{o1t5}
 \end{figure}

   Finally three other possible contractions lead to the $T_5$-type
   UV-completion. These diagrams are shown in Fig.~\ref{o1t5}.
\be
\underbrace{\underbrace{\bar{L} \otimes \bar{\Phi}}_1 \otimes \underbrace{\chi_0}_1}_{1} \otimes \underbrace{\chi_0 \otimes \nu_R}_{1}, \hspace{0.5cm}
\underbrace{\underbrace{\bar{L} \otimes \chi_0}_2 \otimes \underbrace{\bar{\Phi}}_2}_{1} \otimes \underbrace{\chi_0 \otimes \nu_R}_{1}, \hspace{0.5cm}
\underbrace{\underbrace{\bar{L} \otimes \chi_0}_2 \otimes \underbrace{\chi_0}_1}_{2} \otimes \underbrace{\bar{\Phi} \otimes \nu_R}_{2}
\ee
 
 All of the diagrams in Fig.~\ref{o1t5} involve only fermionic
 messengers $N_{L,R}$ or $E_{L,R}$.
 Again, note that the fields $N_0$ and $N_0^\prime$ in the first diagram and
 the fields $E_{-1}$ and $E^\prime_{-1}$ in the third diagram must be different
 fields, since they transform differently under the symmetry
 forbidding lower dimensional operators (e.g $Z_4$ of
 \eqref{op1-charges}).

 Before closing the section, let us briefly discuss the operator
 $\bar{L} \otimes \bar{\chi_0} \otimes \chi_0 \otimes \bar{\Phi} \otimes
 \nu_R$.  In addition to diagrams analogous to those above,
 there will be five other diagrams, as noted in Table~\ref{Tab:op}. 
 These new diagrams appear because, under the symmetry forbidding
 lower dimensional operators, $\chi_0$ and $\bar{\chi_0}$ transform
 differently so that the exchange $\chi_0 \leftrightarrow \bar{\chi_0}$
 leads to distinct UV completions.
 Finally, also for this operator, lower dimensional operators cannot
 be forbidden by simple $U(1)$ or $Z_n$ symmetries. More involved
 symmetries similar to those discussed in Section \ref{sec:OnlySM}
 would be needed.
 Alternatively, similar to the discussion in Section \ref{sec:OnlySM},
 one may introduce two different singlet scalar fields $\chi^ {(1)}_0$ and
 $\chi^{(2)}_0$ transforming differently under a $Z_n$ symmetry.

%%%%%%%%%%%%%%%%%%%%%%%%%%%%%%%%%%%%%%%%%%%%%%%%%%%%%%%%%%%%%%%%%

\section{Operators Involving Singlet ($\chi$), Doublet ($\Phi$) and Triplet ($\Delta$)}
\label{sec:singlet-doublet-triplet}

%%%%%%%%%%%%%%%%%%%%%%%%%%%%%%%%%%%%%%%%%%%%%%%%%%%%%%%%%%%%%%%%%

For this class of operators, there are two $U(1)_Y$ possibilities for
scalar field $\Delta$. 
One is the operator
$\bar{L} \otimes \chi_0 \otimes \Phi \otimes \Delta_{-2} \otimes \nu_R$, in which $\Delta_{-2} \sim -2$
under $U(1)_Y$, while the other possibility is
$\bar{L} \otimes \chi_0 \otimes \bar{\Phi} \otimes \Delta_0 \otimes
\nu_R$ with $\Delta_0 \sim 0$ under $U(1)_Y$.
Apart from hermitian conjugation, one can also write down several
other operators by replacing one or more of $\chi_0, \Phi, \Delta_i$ by
$\bar{\chi_0}, \bar{\Phi}, \bar{\Delta}_i$ respectively.
Since the operator contractions and UV completion of these operators
are all very similar, in this section we will primarily focus on
$\bar{L} \otimes\chi_0 \otimes \bar{\Phi} \otimes \Delta_{0} \otimes \nu_R$
operator.  
We will also comment on the changes required for the operator
$\bar{L} \otimes \chi_0 \otimes \Phi \otimes \Delta_{-2} \otimes
\nu_R$.
The discussion here will equally apply to the other operators which
can be treated analogously. 

As before, in order to ensure that this operator gives the
  leading contribution to neutrino masses we need extra symmetries.
For the case of the operator $\bar{L} \chi_0 \bar{\Phi} \Delta_0 \nu_R$,
due to the zero hypercharge of $\Delta_0$ there are many operators
to forbid, so that $Z_4$ will not be enough, although a $Z_6$ can work
with the charge assignments shown in \ref{op22-charges}.
\begin{eqnarray} 
\bar{L} \otimes \nu_R \sim \lambda^3, \hspace{2mm}
\Phi \sim 1, \hspace{2mm}
\chi_0 \sim \lambda^2, \hspace{2mm}
\Delta_0 \sim \lambda
\label{op22-charges}
\end{eqnarray}
where $\lambda^6=1$. 
Note that these charge assignments forbid the
  $SU(2)_L \otimes U(1)_Y$ allowed operators
  $\bar{L} \bar{\Phi} \nu_R$,
  $\bar{L} \bar{\Phi}\bar{\Phi}\Phi \nu_R$,
  $\bar{L} \chi_0 \chi_0 \bar{\Phi} \nu_R$,
  $\bar{L} \chi_0^\dagger \chi_0 \bar{\Phi} \nu_R$,
  $\bar{L} \chi_0^\dagger \chi_0^\dagger \bar{\Phi} \nu_R$,
  $\bar{L} \Delta \Delta \bar{\Phi} \nu_R$,
  $\bar{L} \Delta^\dagger \Delta \bar{\Phi} \nu_R$,
  $\bar{L} \Delta^\dagger \Delta^\dagger \bar{\Phi} \nu_R$,
  $\bar{L} \chi_0 \bar{\Phi} \nu_R$,
  $\bar{L} \chi_0^\dagger \bar{\Phi} \nu_R$ and
  $\bar{L} \Delta \Phi \nu_R$.

For the other $U(1)_Y$ allowed operator $\bar{L} \chi_0 \Phi \Delta_{-2} \nu_R$, the minimal symmetry that works is another $Z_4$, under which the charges
of particles are given in Eq.~\ref{op21-charges}. 
\begin{eqnarray} 
\bar{L} \otimes \nu_R \sim z, \hspace{2mm}
\Phi \sim 1, \hspace{2mm}
\chi_0 \sim z^2, \hspace{2mm}
\Delta_{-2} \sim z
\label{op21-charges}
\end{eqnarray}
This charge assignment will forbid operators such as
$\bar{L} \bar{\Phi} \nu_R$, $\bar{L} \bar{\Phi}\bar{\Phi}\Phi \nu_R$,
$\bar{L} \chi_0 \chi_0 \bar{\Phi} \nu_R$,
$\bar{L} \chi_0^\dagger \chi_0 \bar{\Phi} \nu_R$,
$\bar{L} \chi_0^\dagger \chi_0^\dagger \bar{\Phi} \nu_R$,
$\bar{L} \Delta^\dagger \Delta \bar{\Phi} \nu_R$,
$\bar{L} \chi_0 \bar{\Phi} \nu_R$ and
$\bar{L} \chi_0^\dagger \bar{\Phi} \nu_R$ and
$\bar{L} \Delta \Phi \nu_R$.

Note that Yukawa terms for other \sm fermions, i.e.
$\bar{L} \Phi l_R$, $\bar{Q} \Phi d_R$ and $\bar{Q} \bar{\Phi} u_R$
can be trivially allowed by these symmetries with appropriate $Z_4$ or $Z_6$
charges of $\bar{Q}$, $l_R$, $u_R$ and $d_R$]. 

  It is easy to see that with this charge assignment, all unwanted
  operators will be forbidden. 
  We stress that the above two symmetries are given just for
  illustration. The unwanted operators can be forbidden in many
  other ways.

  As mentioned before, for sake of brevity we will only explicitly
  discuss the case of
  $\bar{L} \otimes\chi_0 \otimes \bar{\Phi} \otimes \Delta_{0} \otimes
  \nu_R$ operator.  
  The diagrams and topologies of the
  $\bar{L} \otimes\chi_0 \otimes \Phi \otimes \Delta_{-2} \otimes \nu_R$
  operator will be quite similar and can be obtained from the
  $\Delta_{0}$ case by just changing the direction of the arrow
    and the $U(1)_Y$ charges of the intermediate messenger fields.
  The diagrams for the other operators mentioned before can also be
  obtained in a similar manner.

  Moving on to the possible operator contractions and UV-completions,
  here we have now sixteen possibilities.  
  Three diagrams in each topologies $T_1$, $T_2$ and $T_3$, plus a
  diagram in $T_4$ and the remaining six diagrams in $T_5$ topology. 

\begin{figure}[!h] 
\centering
 \includegraphics[scale=0.25]{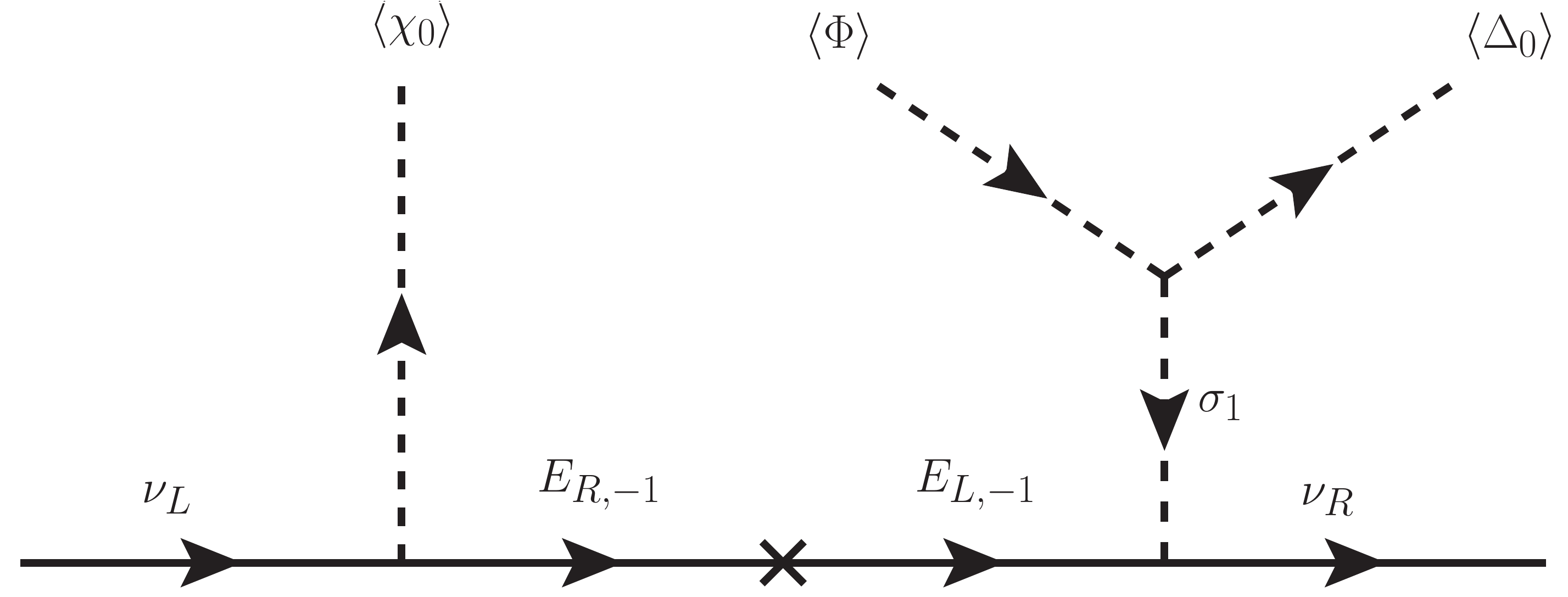}, \hspace{2mm}
  \includegraphics[scale=0.25]{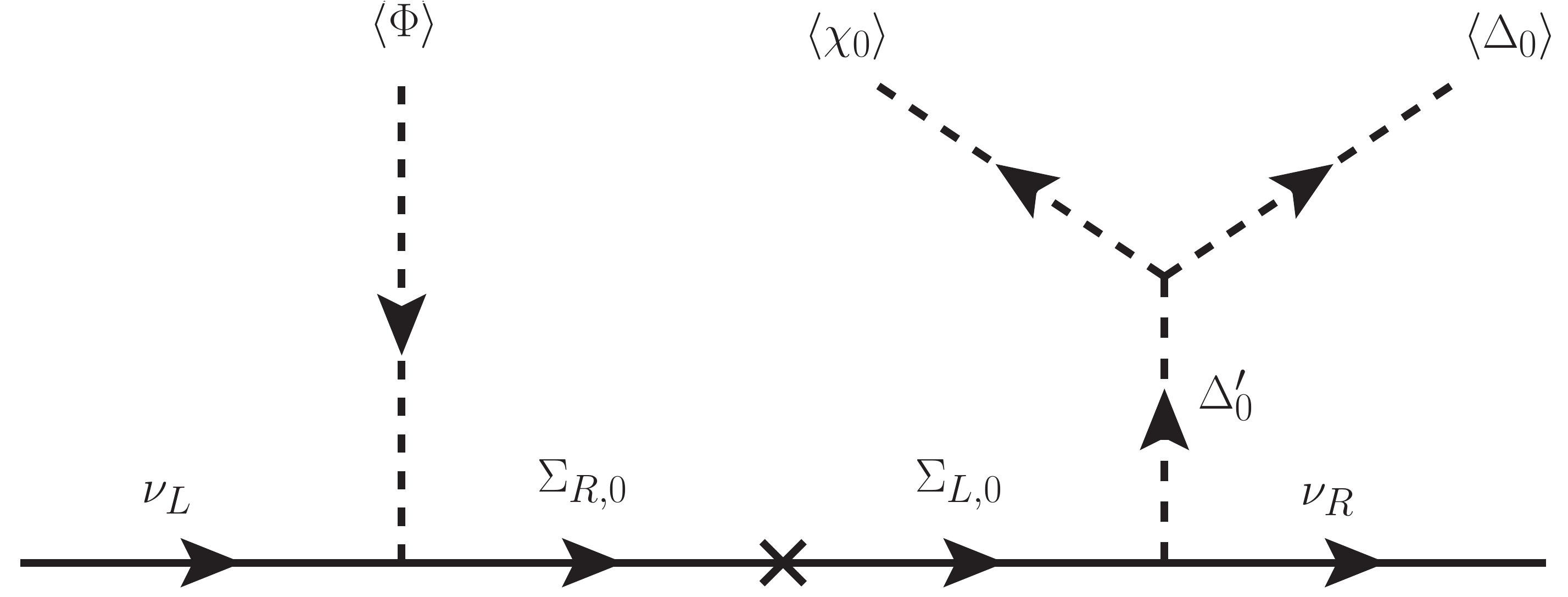}, \hspace{2mm}
   \includegraphics[scale=0.25]{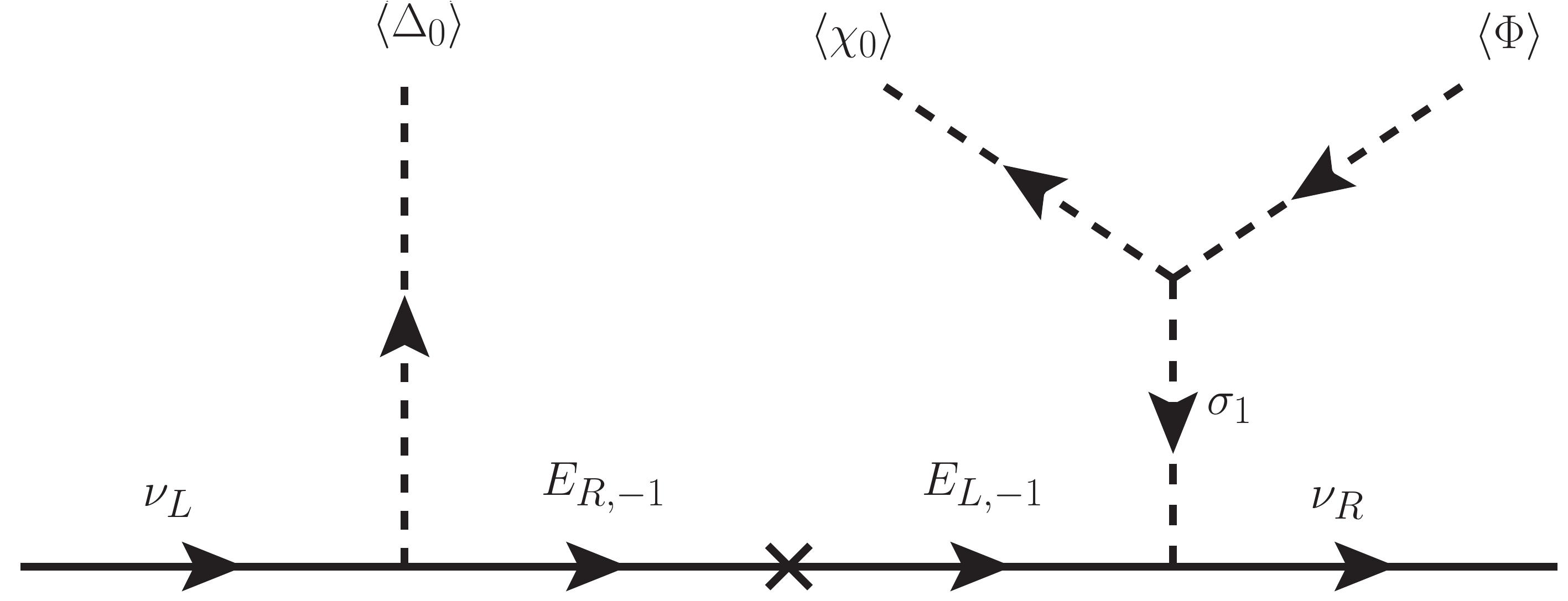}
    \caption{Diagrams associated to the $T_1$ topology of the operator 
    $\bar{L} \chi_0  \bar{\Phi}  \Delta_{0} \nu_R$.  
      Note that for the other choice of $U(1)_Y$, i.e. the operator
      $\bar{L} \chi_0 \Phi \Delta_{-2} \nu_R$, the only difference
      in the diagrams will be flipping the direction of the arrow of
      the external $\Phi$ and $\Delta_0$. }
     \label{o2t1}
 \end{figure}

  The three operator contractions which lead to diagrams with $T_1$
  topology are given in \eqref{123T1} and their the UV complete diagrams
  are shown in Fig.~\ref{o2t1}.
\be
 \underbrace{\underbrace{\bar{L} \otimes \chi_0}_2 \otimes \underbrace{\bar{\Phi} \otimes \Delta_0}_2}_{1} \otimes \underbrace{\nu_R}_{1}, \hspace{0.5cm}
  \underbrace{\underbrace{\bar{L} \otimes \bar{\Phi}}_3 \otimes \underbrace{\chi_0 \otimes \Delta_0}_3}_{1} \otimes \underbrace{\nu_R}_{1}, \hspace{0.5cm}
   \underbrace{\underbrace{\bar{L} \otimes \Delta_0}_2 \otimes \underbrace{\chi_0 \otimes \bar{\Phi}}_2}_{1} \otimes \underbrace{\nu_R}_{1}
   \label{123T1}
\ee
 
 Note that the first and third diagrams in Fig.~\ref{o2t1} have the
 same field content and therefore coexist in the same model having
 this particle content unless a symmetry like the example symmetry in \eqref{op22-charges} forbids one of them.
 
\begin{figure}[!h] 
\centering
 \includegraphics[scale=0.25]{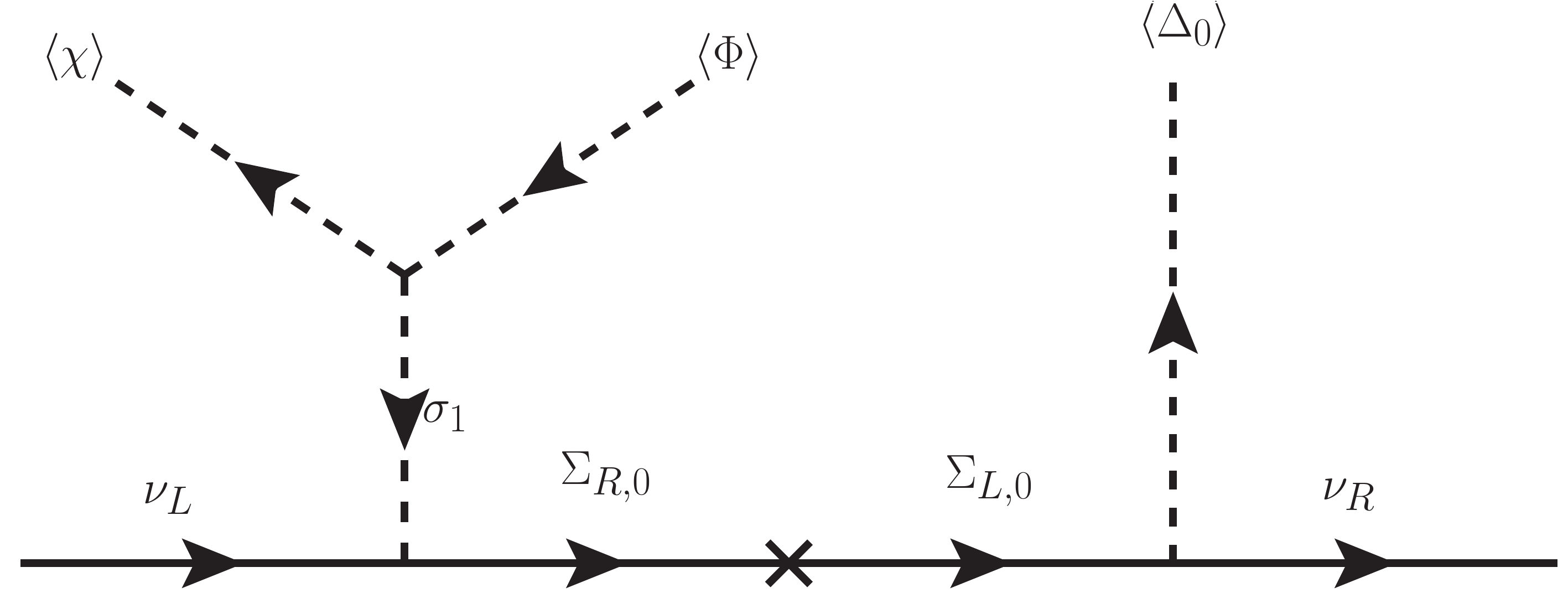}, \hspace{2mm}
  \includegraphics[scale=0.25]{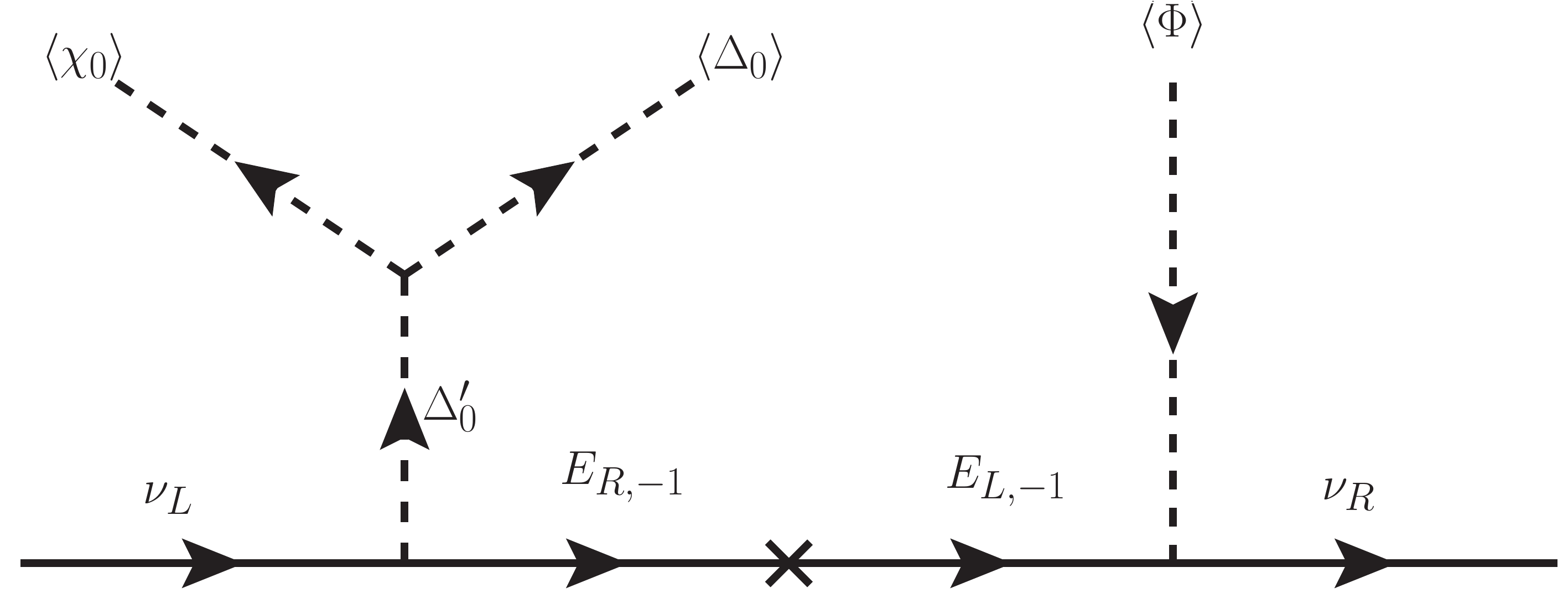}, \hspace{2mm}
   \includegraphics[scale=0.25]{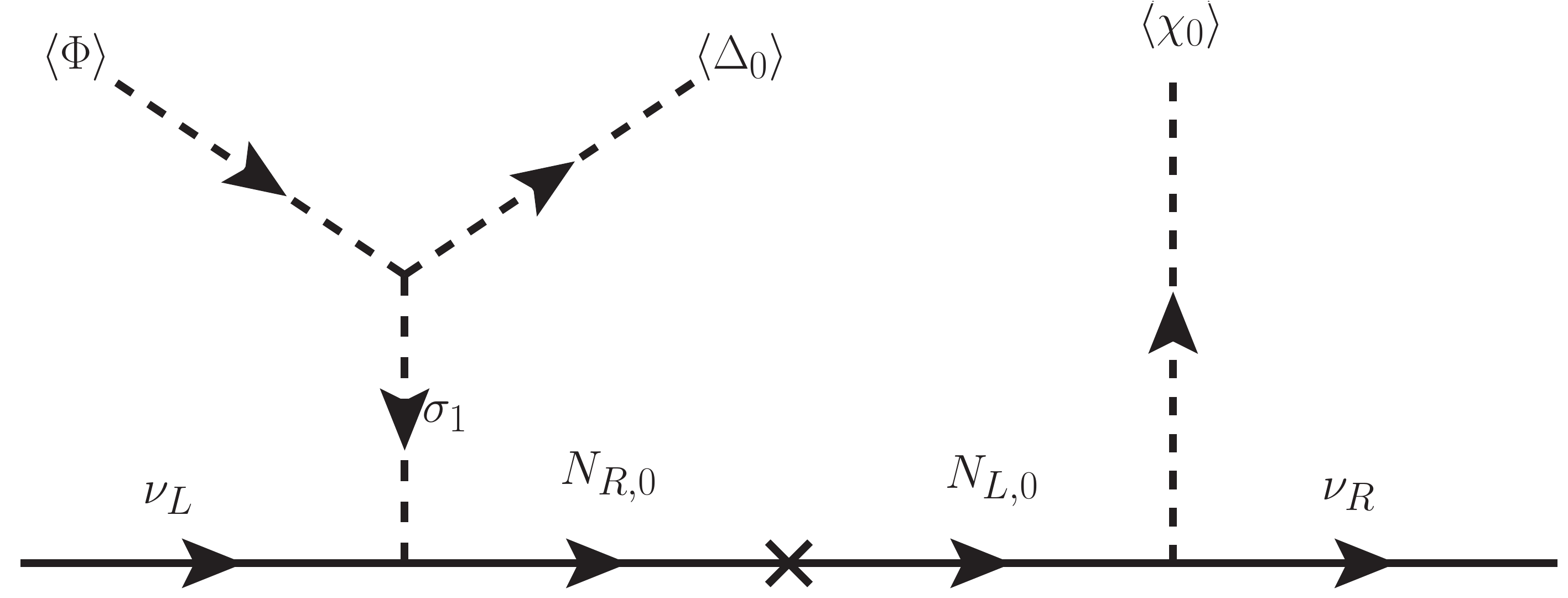}
    \caption{Diagrams showing the $T_2$ topology of the operator $\bar{L} \chi_0  \bar{\Phi}  \Delta_{0} \nu_R$.
      Note that the diagrams for operator
      $\bar{L} \chi_0 \Phi \Delta_{-2} \nu_R$ can be obtained from
      these by flipping the direction of the arrow of the external
      $\Phi$ and $\Delta_0$ fields. }
     \label{o2t2}
 \end{figure}

 The three operator contractions \eqref{123T2} and the corresponding
 diagrams for the $T_2$ topology are shown in Fig.~\ref{o2t2}.
\be
 \underbrace{\underbrace{\bar{L}}_2 \otimes \underbrace{\chi_0 \otimes \bar{\Phi}}_2}_{3} \otimes \underbrace{\Delta_0 \otimes \nu_R}_{3}, \hspace{0.5cm}
  \underbrace{\underbrace{\bar{L}}_2 \otimes \underbrace{\chi_0 \otimes \Delta_0}_3}_{2} \otimes \underbrace{\bar{\Phi} \otimes \nu_R}_{2}, \hspace{0.5cm}
   \underbrace{\underbrace{\bar{L}}_2 \otimes \underbrace{\bar{\Phi} \otimes \Delta_0}_2}_{1} \otimes \underbrace{\chi_0 \otimes \nu_R}_{1}
\label{123T2}
\ee
 
\begin{figure}[!h] 
\centering
 \includegraphics[scale=0.3]{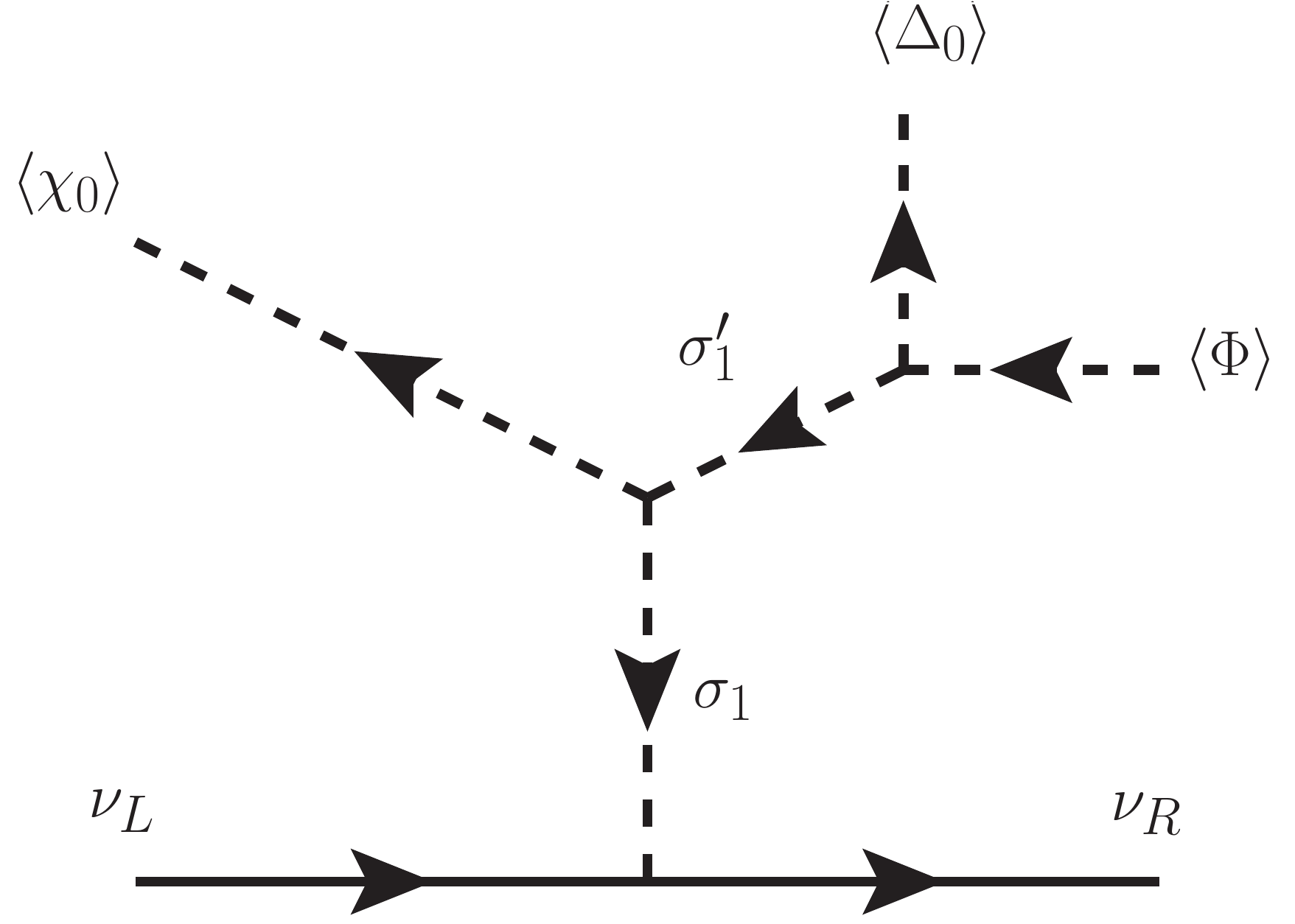}, \hspace{2mm}
  \includegraphics[scale=0.3]{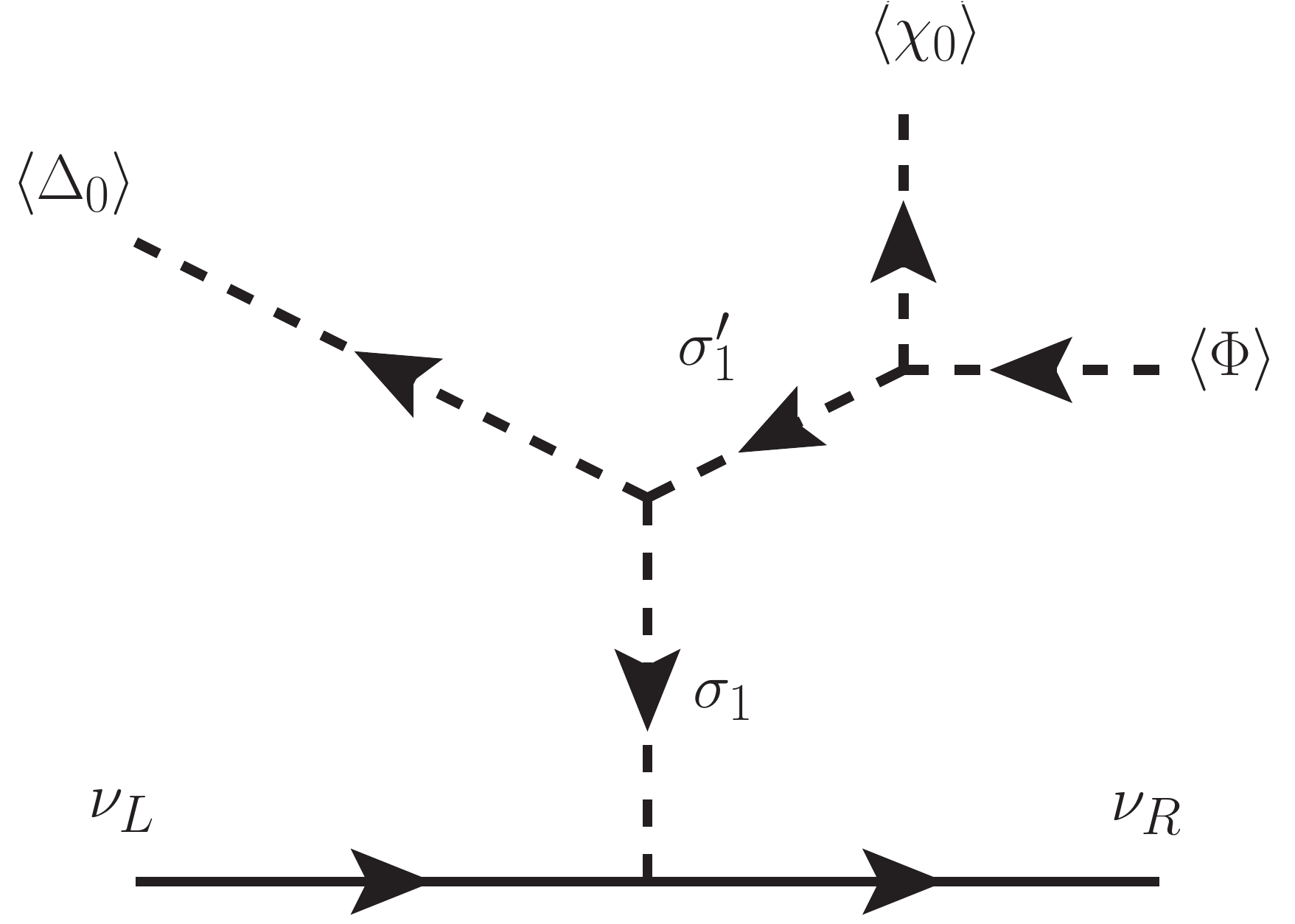}, \hspace{2mm}\\
   \includegraphics[scale=0.3]{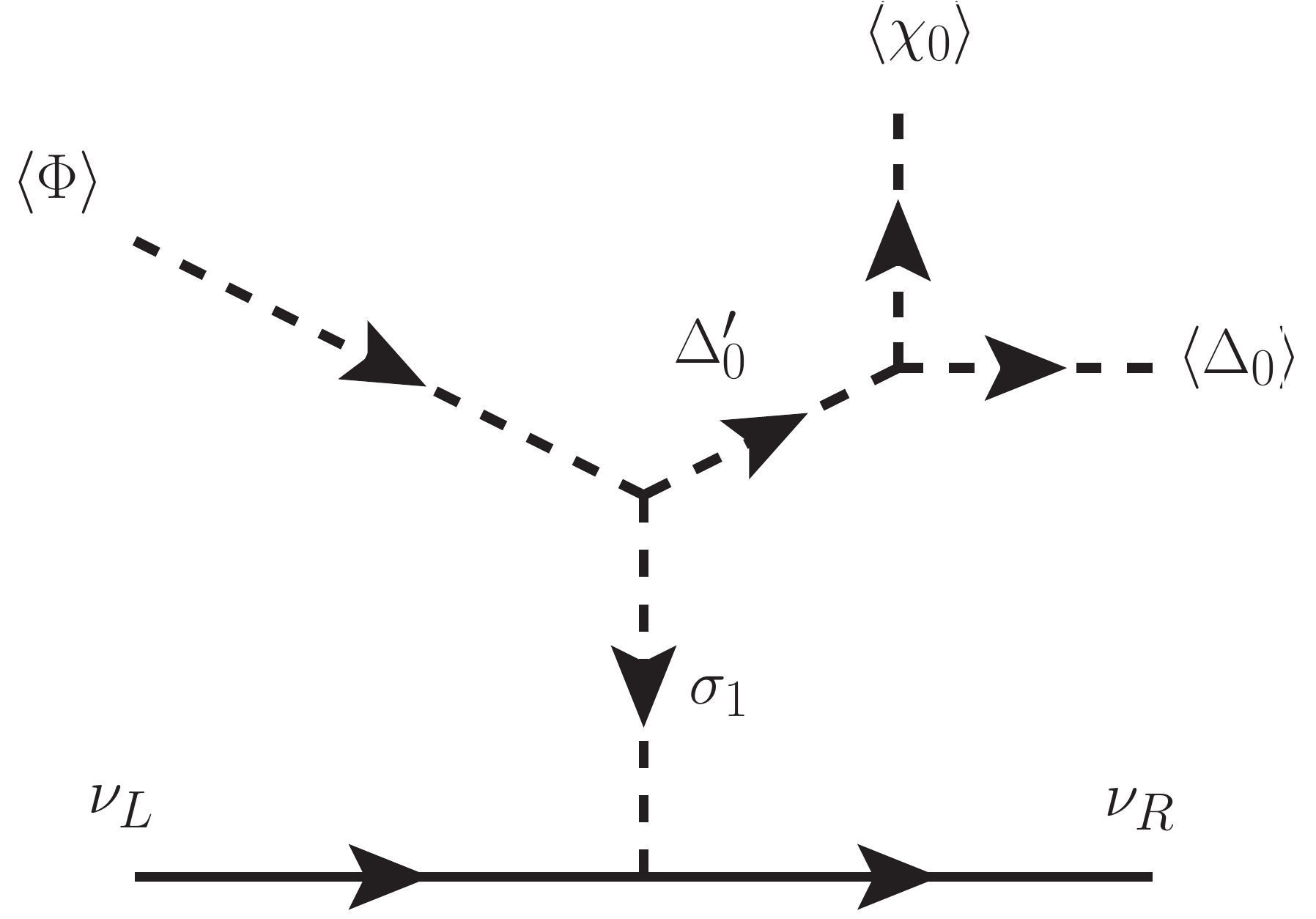}, \hspace{2mm}
    \includegraphics[scale=0.3]{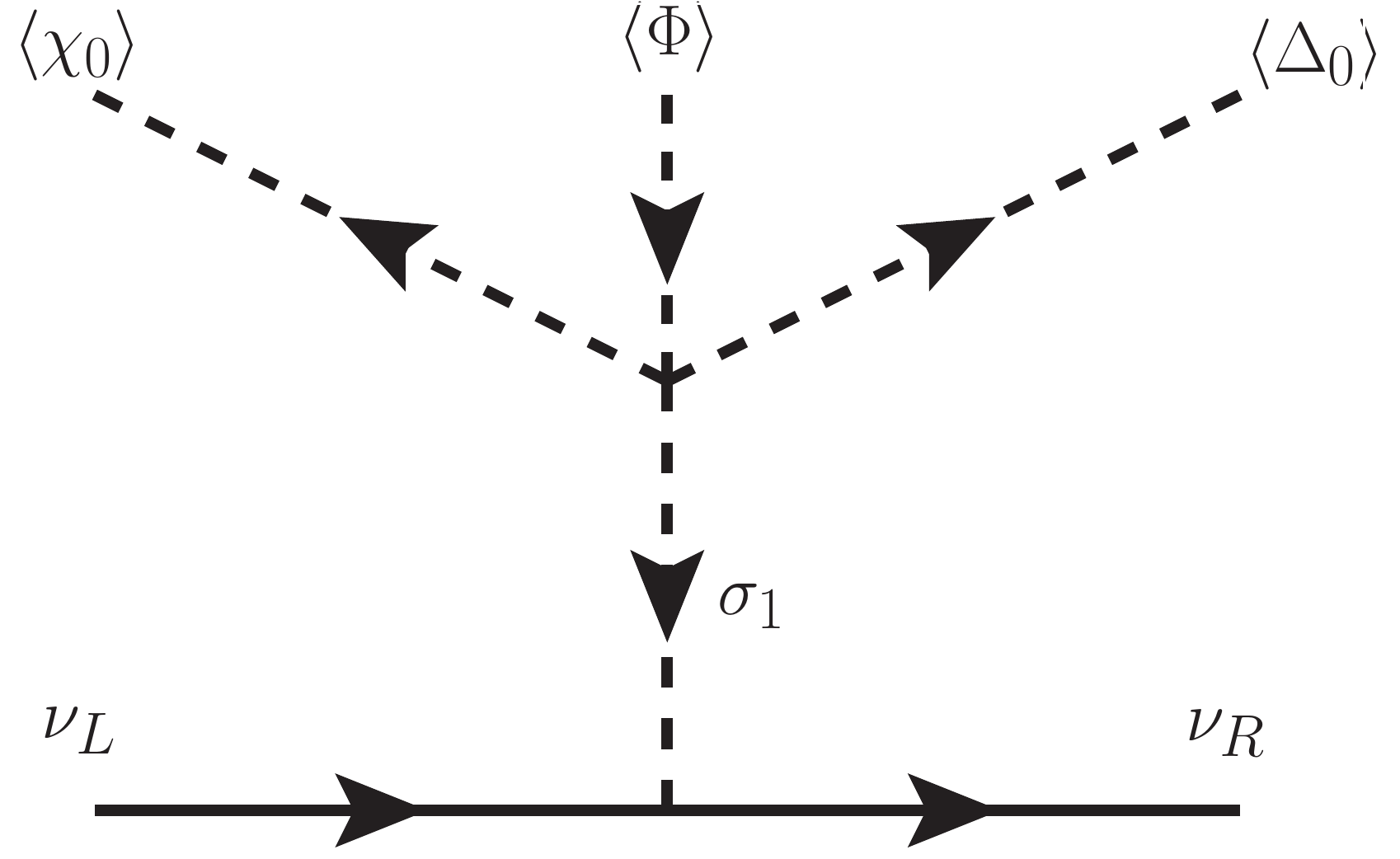}
     \caption{Diagrams showing the $T_3$ and $T_4$ topologies of the operator $\bar{L} \chi_0  \bar{\Phi}  \Delta_{0} \nu_R$. 
       Note that the diagrams for operator
       $\bar{L} \chi_0 {\Phi} \Delta_{-2} \nu_R$ can be obtained from
       these by flipping the direction of the arrow of the external
       $\Phi$ and $\Delta_0$ fields.}
   \label{o2t3t4}
 \end{figure}

 The contractions \eqref{123t3t4} and the UV-completions that lead to
 topologies $T_3$ or $T_4$ are shown in Fig.~\ref{o2t3t4}.  
\be
 \underbrace{\underbrace{\bar{L} \otimes \nu_R}_2 \otimes \underbrace{\chi_0}_1}_{2} \otimes \underbrace{\bar{\Phi} \otimes \Delta_0}_{2}, \hspace{0.5cm}
  \underbrace{\underbrace{\bar{L} \otimes \nu_R}_2 \otimes \underbrace{\Delta_0}_3}_{2} \otimes \underbrace{\chi_0 \otimes \bar{\Phi}}_{2}, \hspace{0.5cm}
   \underbrace{\underbrace{\bar{L} \otimes \nu_R}_2 \otimes \underbrace{\bar{\Phi}}_2}_{3} \otimes \underbrace{\chi_0 \otimes \Delta_0}_{3}, \hspace{0.5cm}
    \underbrace{\bar{L} \otimes \nu_R}_2 \otimes \underbrace{\chi_0 \otimes \bar{\Phi} \otimes \Delta_0}_2
\label{123t3t4}
\ee

 Again, we note that in Fig.~\ref{o2t3t4} the first and second
 diagrams have the same field content. Therefore, in a typical model
 both diagrams will contribute to neutrino masses, unless one of them
 is explicitly forbidden by some symmetry. 

\begin{figure}[!h] 
\centering
 \includegraphics[scale=0.25]{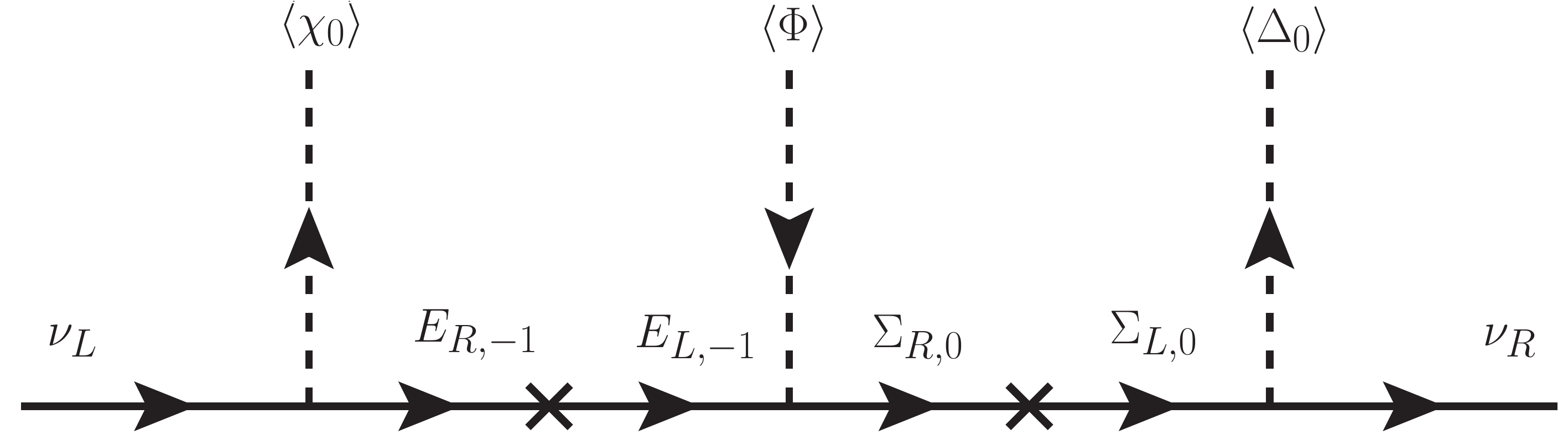}, \hspace{2mm}
  \includegraphics[scale=0.25]{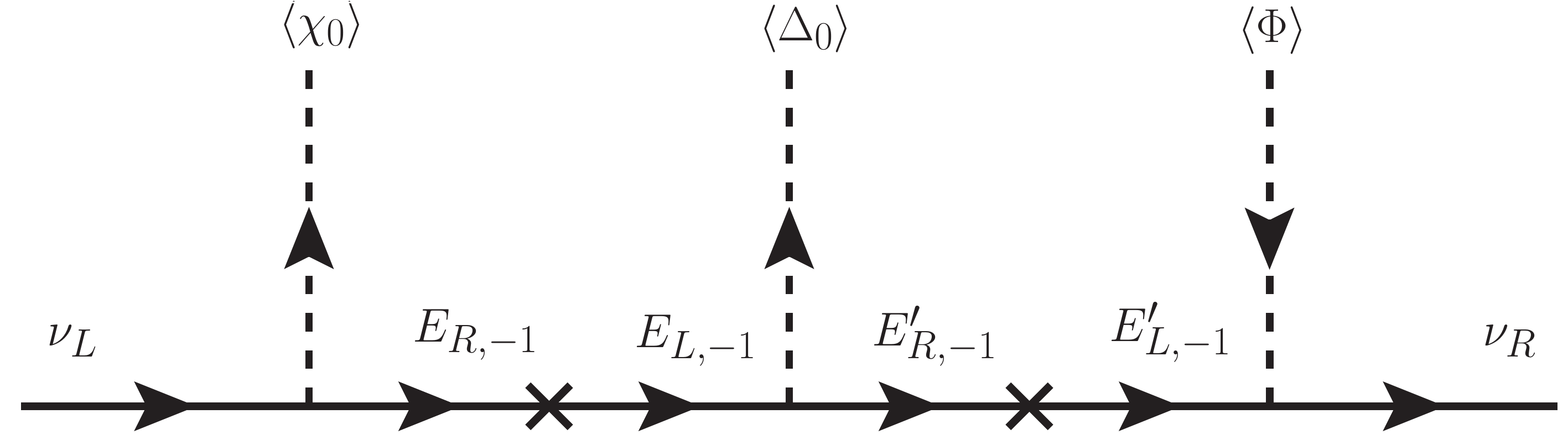}, \hspace{2mm}
   \includegraphics[scale=0.25]{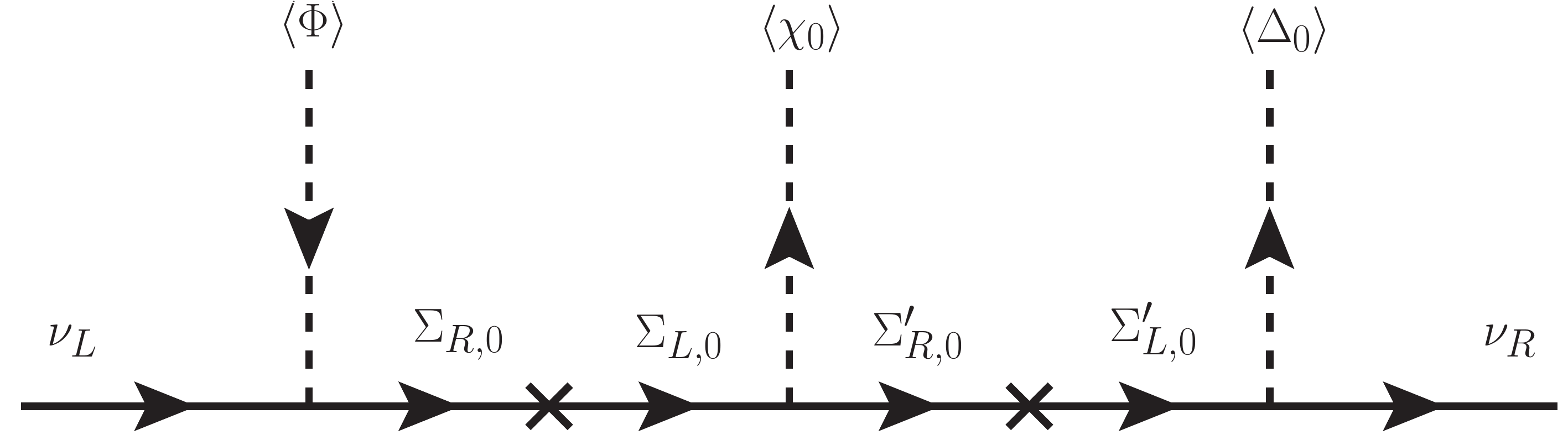}, \hspace{2mm}
    \includegraphics[scale=0.25]{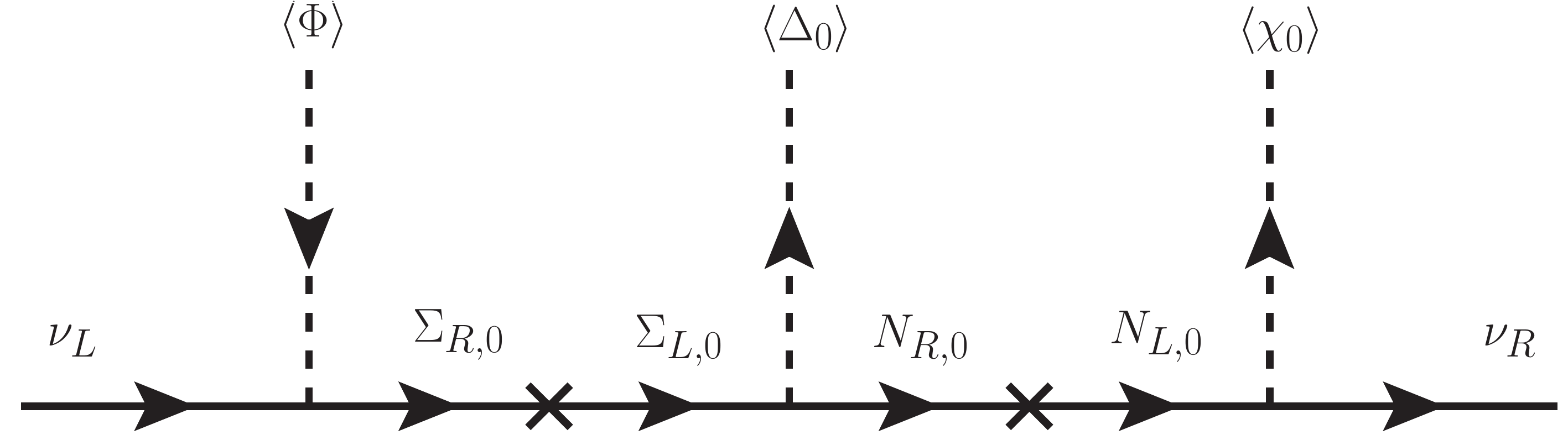}, \hspace{2mm}
     \includegraphics[scale=0.25]{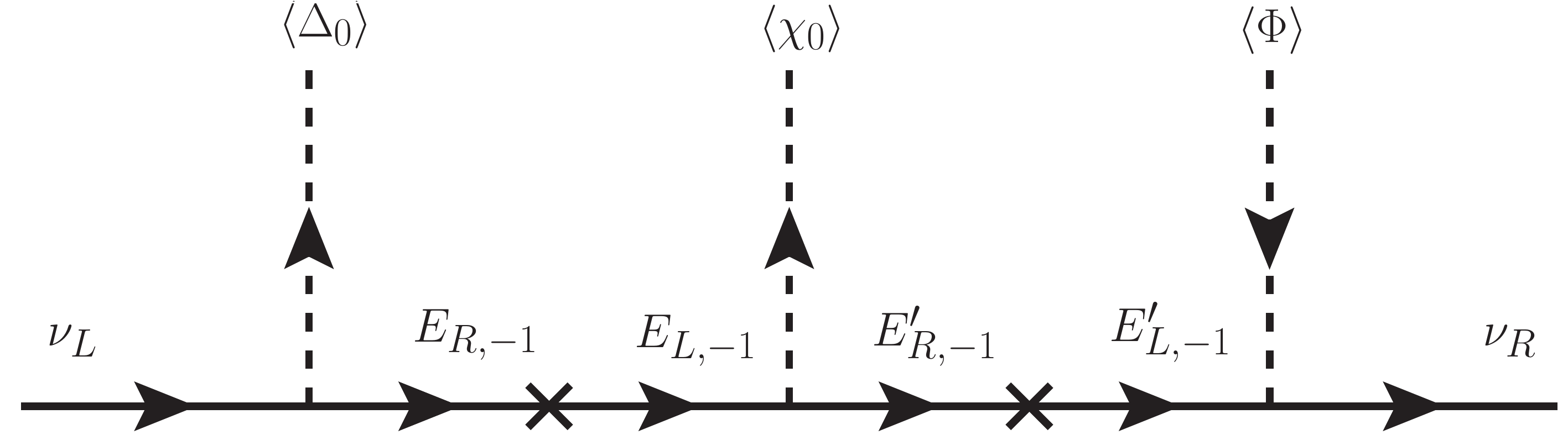}, \hspace{2mm}
      \includegraphics[scale=0.25]{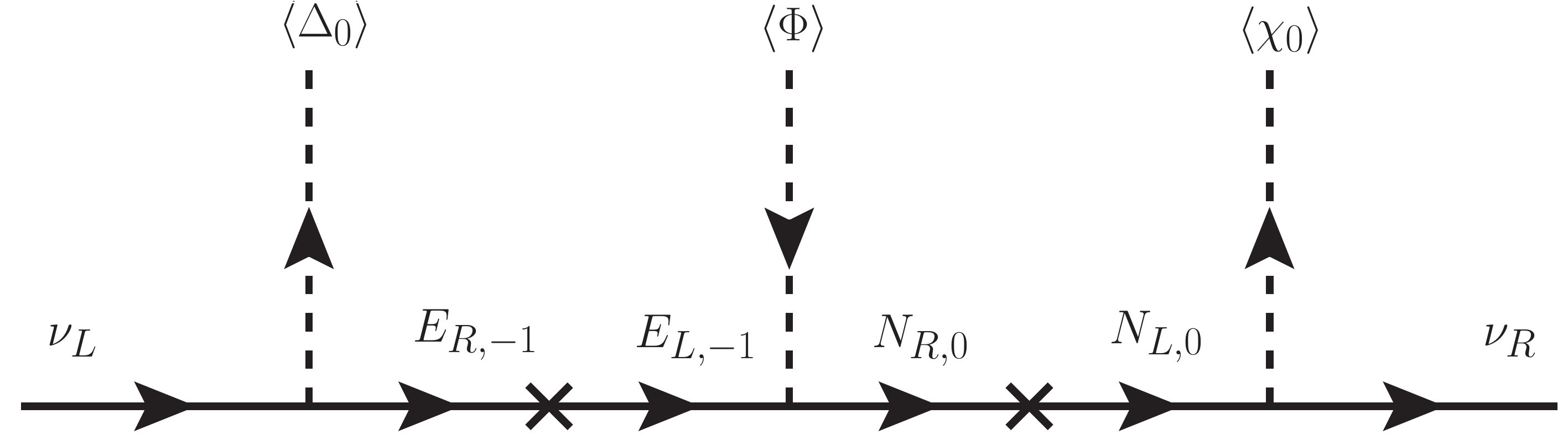}
\caption{Diagrams showing the $T_5$ topology of the operator 
$\bar{L} \chi_0  \bar{\Phi}  \Delta_{0} \nu_R$.
  Note that the diagrams for operator
  $\bar{L} \chi_0 \bar{\Phi} \Delta_{-2} \nu_R$ can be obtained from these
  by flipping the direction of the arrow of the external $\Phi$ and
  $\Delta_0$ fields.}
 \label{o2t5}
  \end{figure}

 Finally, there are six possible operator contractions leading to the
 $T_5$ topology, as shown in Fig.~\ref{o2t5}.
\be
 \underbrace{\underbrace{\bar{L} \otimes \chi_0}_2 \otimes \underbrace{\bar{\Phi}}_2}_{3} \otimes \underbrace{\Delta_0 \otimes \nu_R}_{3} , \hspace{0.5cm}
  \underbrace{\underbrace{\bar{L} \otimes \chi_0}_2 \otimes \underbrace{\Delta_0}_3}_{2} \otimes \underbrace{\bar{\Phi} \otimes \nu_R}_{2} , \hspace{0.5cm}
   \underbrace{\underbrace{\bar{L} \otimes \bar{\Phi}}_3 \otimes \underbrace{\chi_0}_1}_{3} \otimes \underbrace{\Delta_0 \otimes \nu_R}_{3} 
    \ee
\be
 \underbrace{\underbrace{\bar{L} \otimes \bar{\Phi}}_3 \otimes \underbrace{\Delta_0}_3}_{1} \otimes \underbrace{\chi_0 \otimes \nu_R}_{1} , \hspace{0.5cm}
  \underbrace{\underbrace{\bar{L} \otimes \Delta_0}_2 \otimes \underbrace{\chi_0}_1}_{2} \otimes \underbrace{\bar{\Phi} \otimes \nu_R}_{2} , \hspace{0.5cm}
   \underbrace{\underbrace{\bar{L} \otimes \Delta_0}_2 \otimes \underbrace{\bar{\Phi}}_2}_{1} \otimes \underbrace{\chi_0 \otimes \nu_R}_{1} 
    \ee
  
  Again note that, owing to their different transformation under
  symmetry forbidding lower dimensional operators, the fields $E_{-1}$
  and $E^\prime_{-1}$ in the second and fifth diagrams of Fig.~\ref{o2t5} must
  be different. 
  Also, comparing these two diagrams one can see that they have the
  same messenger field content. However, these two diagrams belong to
  two distinct UV-complete models, due to the different charges of the
  messenger fields under symmetry forbidding lower dimensional
  operators.  Hence they correspond to different models with the same
  field content.

%%%%%%%%%%%%%%%%%%%%%%%%%%%%%%%%%%%%%%%%%%%%%%%%%%%
\section{Operators Involving Only Doublet ($\Phi$) and Triplet ($\Delta$)} 
\label{sec:doublet-triplet}
%%%%%%%%%%%%%%%%%%%%%%%%%%%%%%%%%%%%%%%%%%%%%%%%%%%

There are several possibilities for dimension-6 operators involving
$SU(2)_L$ doublet and triplet scalars, as listed in Table
\ref{Tab:op}.
As before, an extra symmetry is required in order to ensure that these
operators give the leading contribution to neutrino masses. Again, the
nature of this symmetry can vary from a simple $U(1)_X$ symmetry to
its $Z_n$ subgroup, or more complex symmetries involving non-abelian
groups. 
As discussed in Section \ref{sec:operator-analysis}, depending on the
symmetry required these operators can be classified into two
categories (see Table \ref{Tab:op}) namely operators for which the
lower dimensional operators can be forbidden by $U(1)_X$ or $Z_n$
symmetries and operators for which $U(1)_X$ or $Z_n$ symmetry is not
enough.

For example the operators
$\bar{L} \otimes \bar{\Phi} \otimes \Delta_0 \otimes \Delta_0 \otimes
\nu_R $
and
$\bar{L} \otimes \Phi \otimes \Delta_0 \otimes \Delta_{-2} \otimes
\nu_R $ can both be forbidden by a simple $Z_n$ symmetry.
The minimal consistent one is a $Z_4$ symmetry, under which the
various particles transform as 
\begin{eqnarray} 
\bar{L} \otimes \nu_R \sim z^2, \hspace{2mm}
\Phi \sim 1, \hspace{2mm}
\Delta_i \sim z 
\label{op3-charges}
\end{eqnarray}
where $\Delta_i$, $i = 0, -2$, denote the two different types of
$SU(2)_L$ triplets.
Note that under these charge assignments other operators where
$\Delta_0 \to \bar{\Delta}_0$ are also allowed and could in principle
also contribute to neutrino masses, as long as they have consistent
UV-completions.  
The other operators
$\bar{L} \otimes \bar{\Phi} \otimes \bar{\Delta}_i \otimes \Delta_i
\otimes \nu_R $;
$i = 0, -2$ belong to a different class and for such operators simple
$U(1)_X$ or $Z_n$ symmetries are not enough to forbid all other
dimension-6 and lower dimensional operators. 
Just as the case of
$\bar{L} \otimes \bar{\Phi} \otimes \bar{\Phi} \otimes \Phi \otimes
\nu_R $
operator discussed in Section \ref{sec:OnlySM}, these operators can
also give leading contribution to neutrino masses through a softly
broken $Z_n$ symmetry or for certain non-abelian discrete symmetries,
such as $S_4$.
Alternatively, for this case one may also introduce another copy of
$\Delta_i$, in which case (as before) a simple $Z_n$ symmetry could suffice.

Moving on to possible operator contractions and UV-completions we
first note that the UV-completions of all these operators are very
similar to each other.
To avoid unnecessary repetition we will only discuss in detail the
UV-completion of the operator
$\bar{L} \otimes \bar{\Phi} \otimes \bar{\Delta}_0 \otimes \Delta_0
\otimes \nu_R $.
We have singled out this operator since, amongst all the operators of
this category, it offers the maximum number of possible completions,
see Table \ref{Tab:op}.  
The other operators can be treated in the same manner and we will
comment on some of their salient features as we go along.

There are thirty one different ways of contracting this operator. Six
digramas lie in each of the topologies $T_1$, $T_2$ and $T_3$. One
belongs to $T_4$ topology, while the remaining twelve belong to $T_5$
topology. 

\begin{figure}[!h] 
\centering
 \includegraphics[scale=0.25]{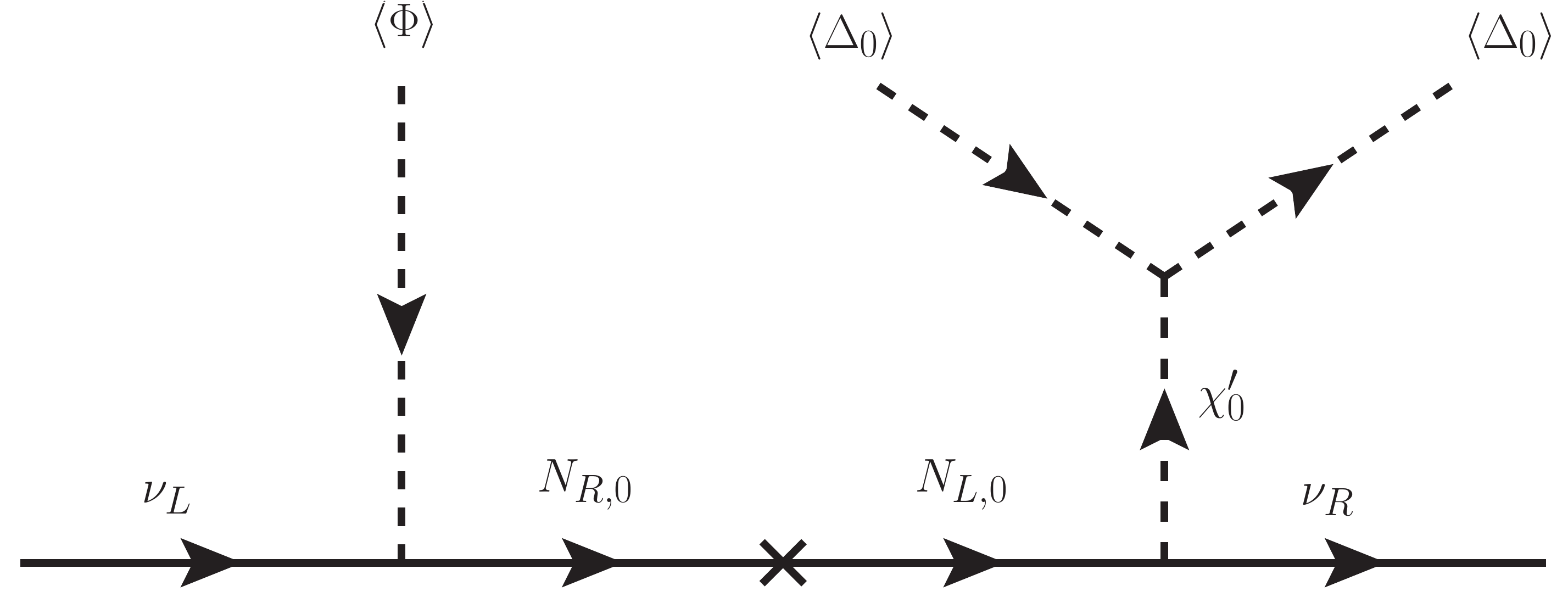}, \hspace{2mm}
  \includegraphics[scale=0.25]{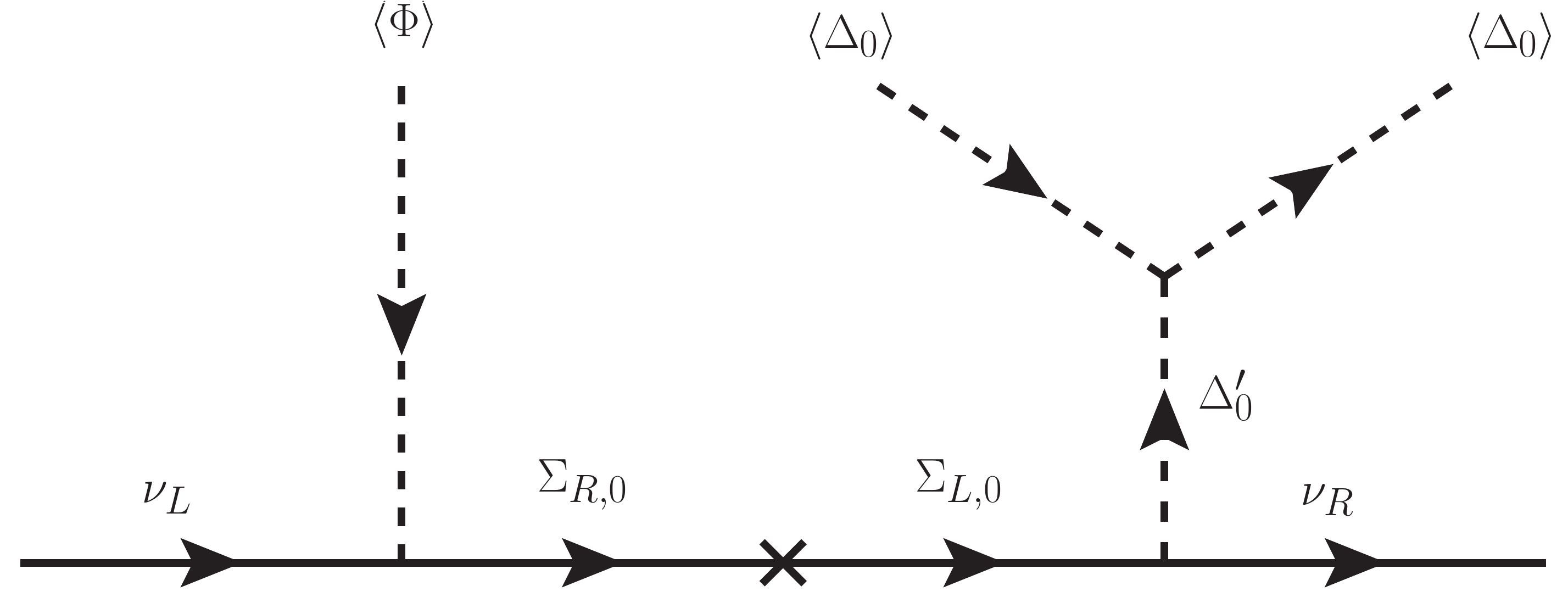}, \hspace{2mm}
   \includegraphics[scale=0.25]{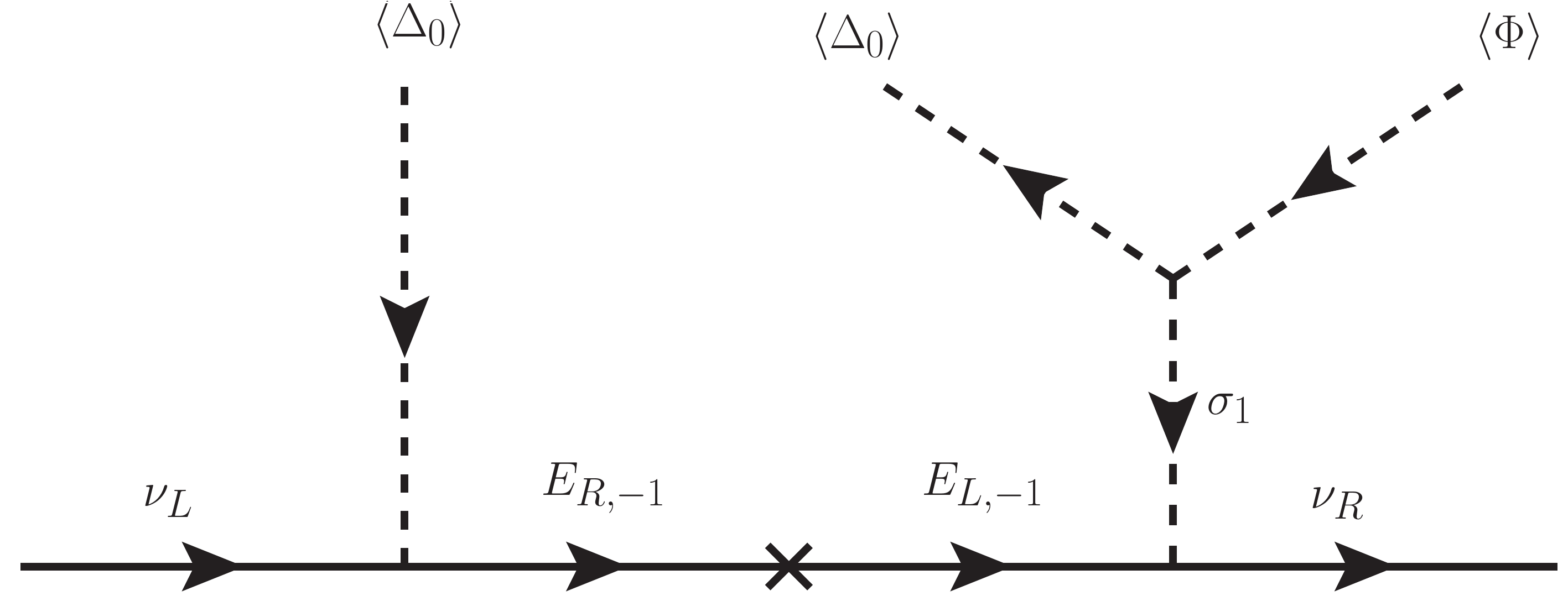}, \hspace{2mm}
    \includegraphics[scale=0.25]{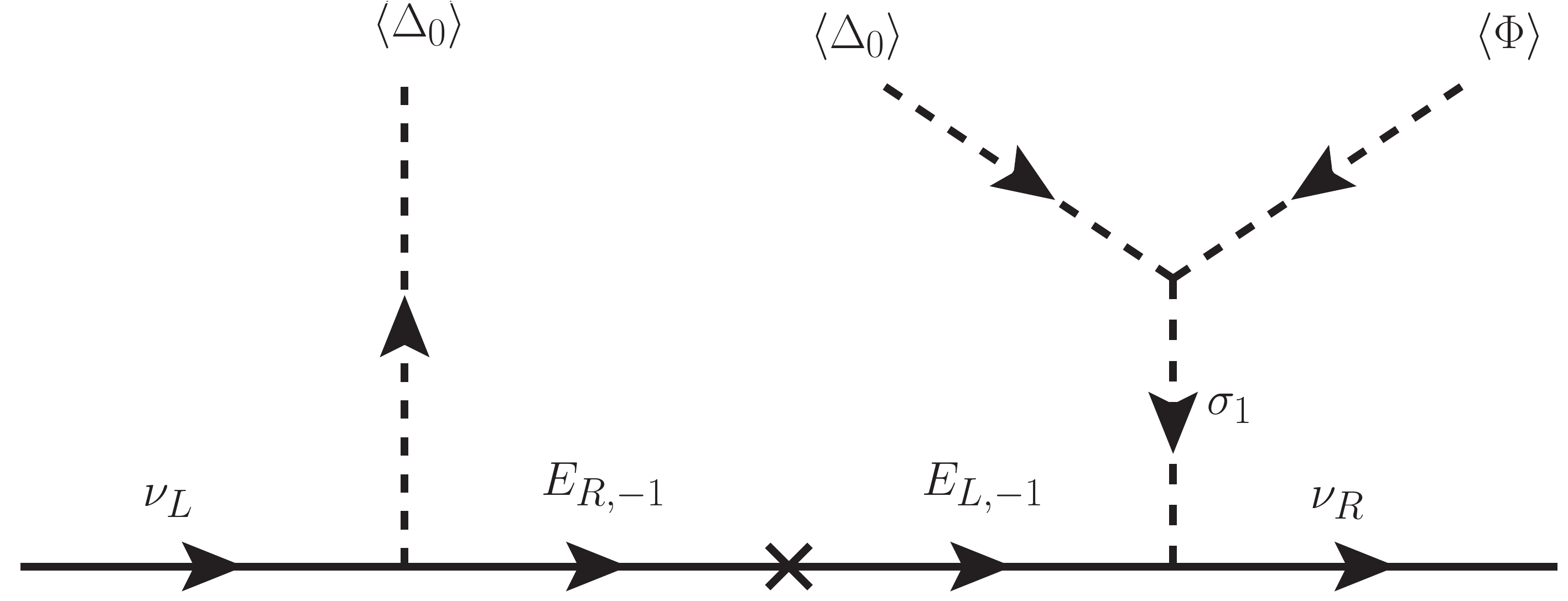}, \hspace{2mm}
     \includegraphics[scale=0.25]{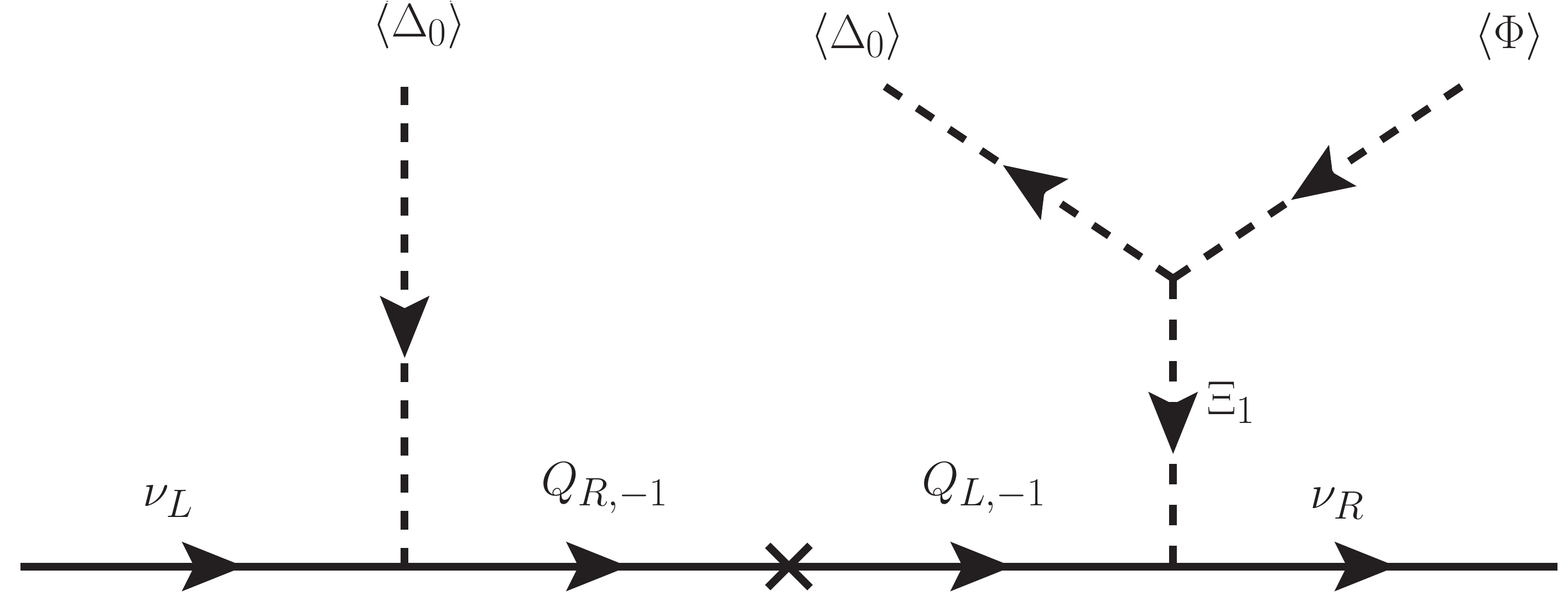}, \hspace{2mm}
      \includegraphics[scale=0.25]{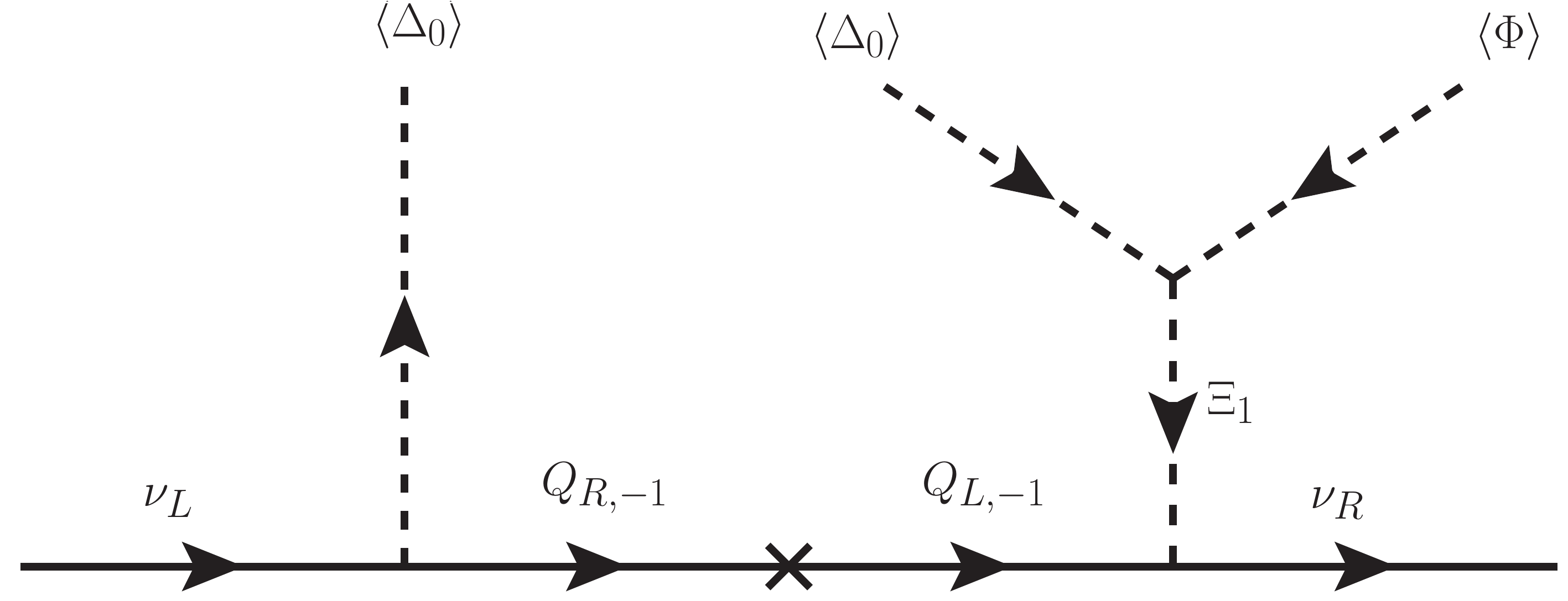}
     \caption{Diagrams showing the $T_1$ topology of the operator $\bar{L} \bar{\Phi}  \bar{\Delta} _0 \Delta_0 \nu_R$.}
      \label{o3t1}
\end{figure}

The six operator contractions are given in \eqref{233t1}, while their
UV-complete diagrams are shown in Figure \ref{o3t1}.
\begin{eqnarray}
&& \underbrace{\underbrace{\bar{L} \otimes \bar{\Phi}}_1 \otimes \underbrace{\bar{\Delta}_0 \otimes \Delta_0}_1}_{1} \otimes \underbrace{\nu_R}_{1} , \hspace{0.5cm}
 \underbrace{\underbrace{\bar{L} \otimes \bar{\Phi}}_3 \otimes \underbrace{\bar{\Delta}_0 \otimes \Delta_0}_3}_{1} \otimes \underbrace{\nu_R}_{1} , \hspace{0.5cm}
 \underbrace{\underbrace{\bar{L} \otimes \bar{\Delta}_0}_2 \otimes \underbrace{\Delta_0 \otimes \bar{\Phi}}_2}_{1} \otimes \underbrace{\nu_R}_{1} , \hspace{0.5cm}
 \underbrace{\underbrace{\bar{L} \otimes \Delta_0}_2 \otimes \underbrace{\bar{\Delta}_0 \otimes \bar{\Phi}}_2}_{1} \otimes \underbrace{\nu_R}_{1} , \nonumber \\
&& \underbrace{\underbrace{\bar{L} \otimes \bar{\Delta}_0}_4 \otimes \underbrace{\Delta_0 \otimes \bar{\Phi}}_4}_{1} \otimes \underbrace{\nu_R}_{1} , \hspace{0.5cm}
\underbrace{\underbrace{\bar{L} \otimes \Delta_0}_4 \otimes \underbrace{\bar{\Delta}_0 \otimes \bar{\Phi}}_4}_{1} \otimes \underbrace{\nu_R}_{1} , \hspace{0.5cm}
\label{233t1}
\end{eqnarray}
Notice that the last two diagrams in Fig.~\ref{o3t1} involve
messengers transforming as quartets of $SU(2)_L$. The fermionic
quartet $Q_{L,R}$ carry $U(1)_Y = -1$, while the scalar quartet $\Xi$
has $U(1)_Y = 1$. 

At this point we comment on the other operators mentioned in Table
\ref{Tab:op}.
For example, for the operator
$\bar{L} \otimes \bar{\Phi} \otimes \bar{\Delta}_{-2} \otimes
\Delta_{-2} \otimes \nu_R$
the third contraction of \eqref{233t1} will be forbidden, since it
involves a contraction of $\bar{L}$ and $\bar{\Delta}_{-2}$ going to a
doublet. 
This contraction implies that the resulting messenger will be an
$SU(2)_L$ doublet with $U(1)_Y = 3$. Hence it will not contribute to
neutrino mass since it has no neutral component. 

For the operator
$\bar{L} \otimes \Phi \otimes \Delta_0 \otimes \Delta_{-2} \otimes
\nu_R $
the first contraction of \eqref{233t1} will have the same problem,
namely the contraction of $\bar{L}$ with $\Phi$ implies an $SU(2)_L$
singlet messenger field with $U(1)_Y = -2$. 
For the operator
$\bar{L} \otimes \bar{\Phi} \otimes \Delta_0 \otimes \Delta_0 \otimes
\nu_R $,
the second contraction of \eqref{233t1} will be forbidden, as in this
case it involves a contraction of two identical triplets going to a
triplet, which is zero.
Moreover, for this case the third and fourth, as well as fifth and
sixth contractions of \eqref{233t1} are identical to each
other. Therefore only one diagram out of each pair should be counted
for this case. 

\begin{figure}[!h] 
\centering
 \includegraphics[scale=0.25]{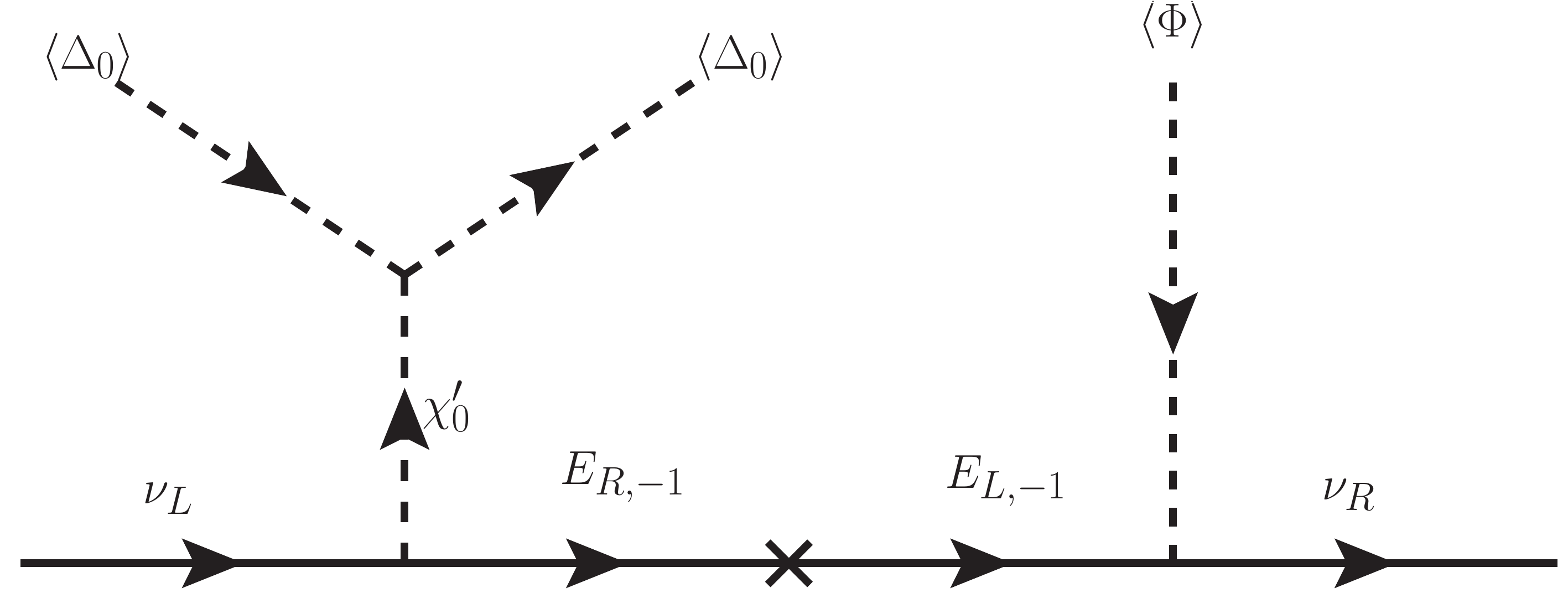}, \hspace{2mm}
  \includegraphics[scale=0.25]{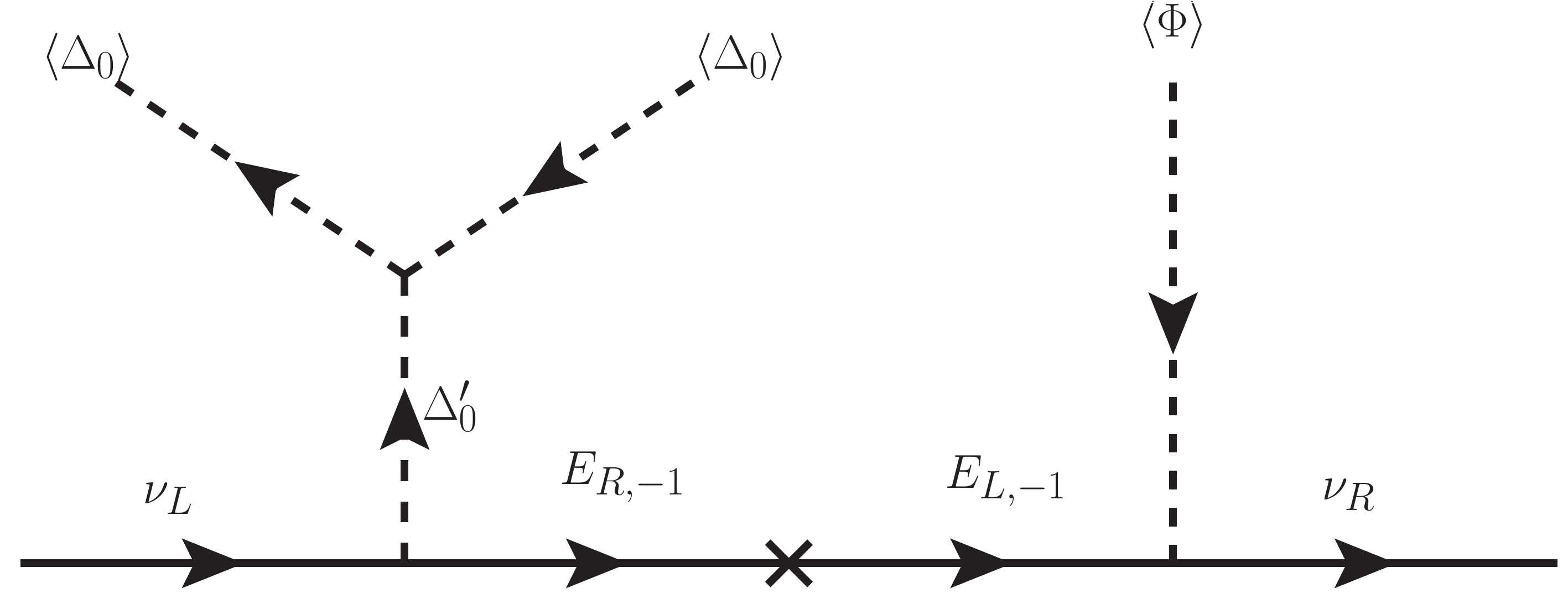}, \hspace{2mm}
   \includegraphics[scale=0.25]{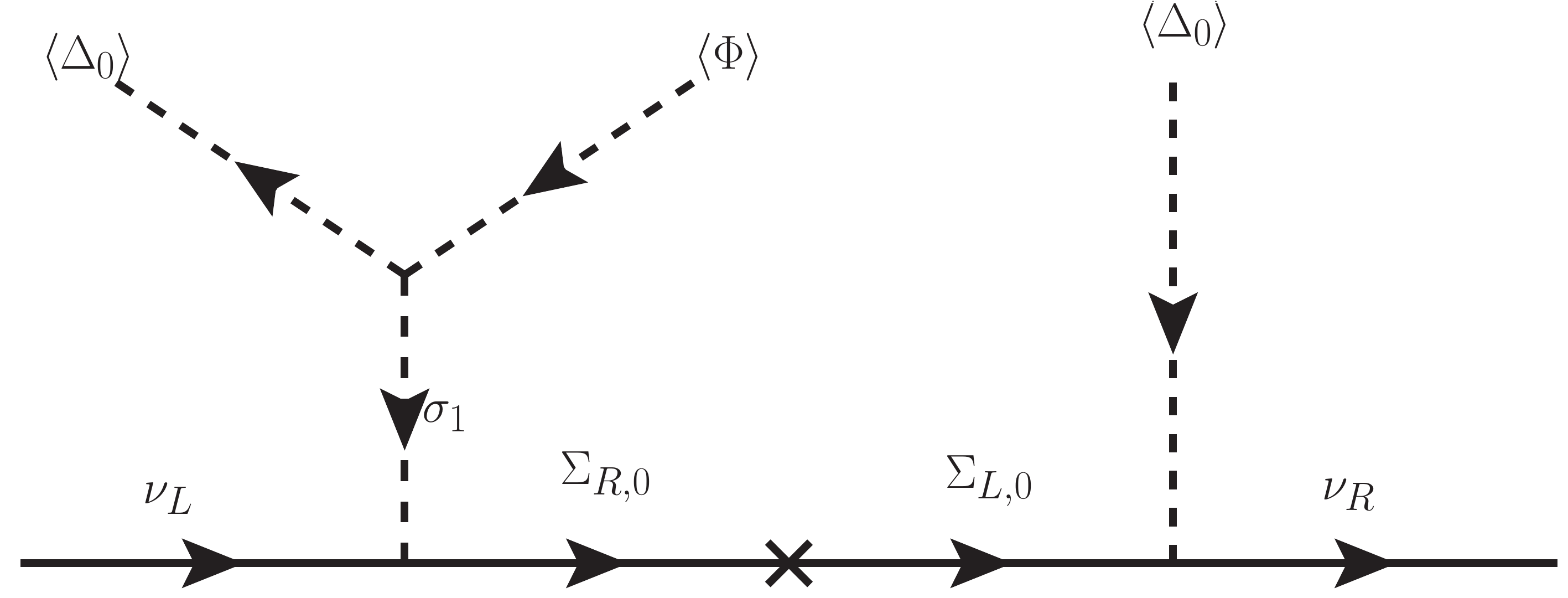}, \hspace{2mm}
    \includegraphics[scale=0.25]{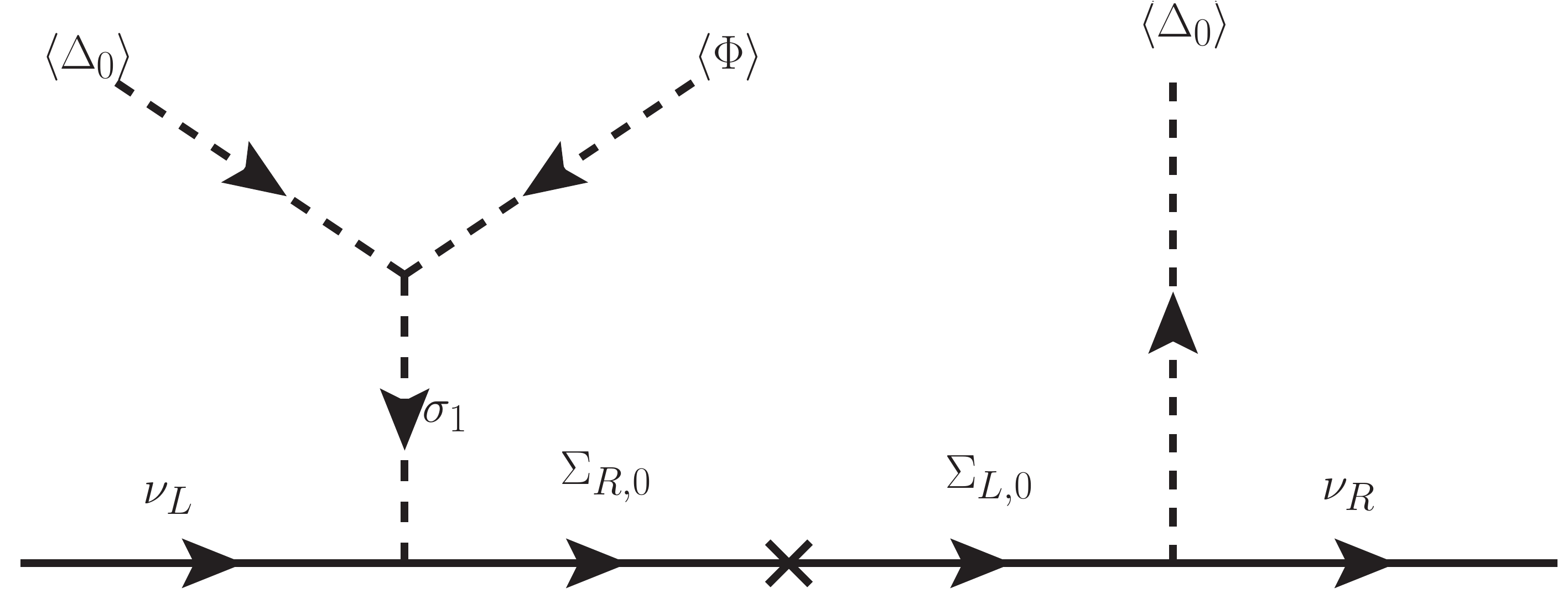}, \hspace{2mm}
     \includegraphics[scale=0.25]{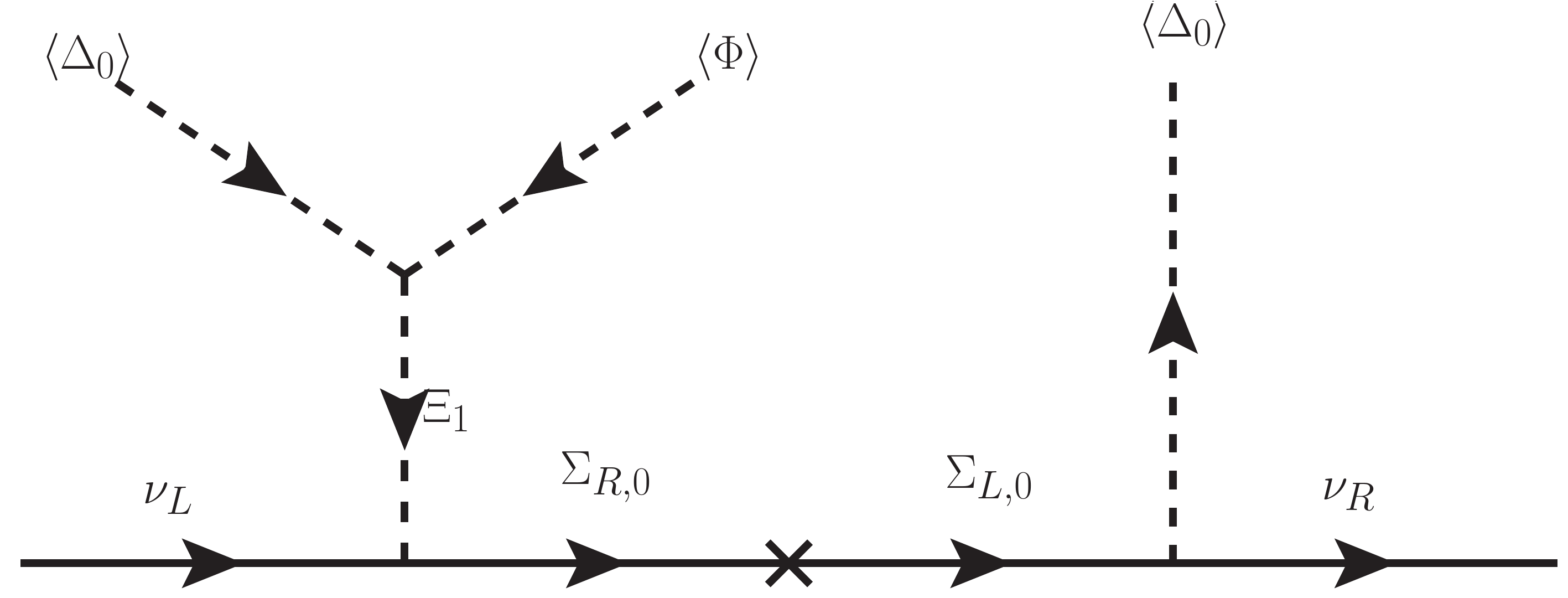}, \hspace{2mm}
      \includegraphics[scale=0.25]{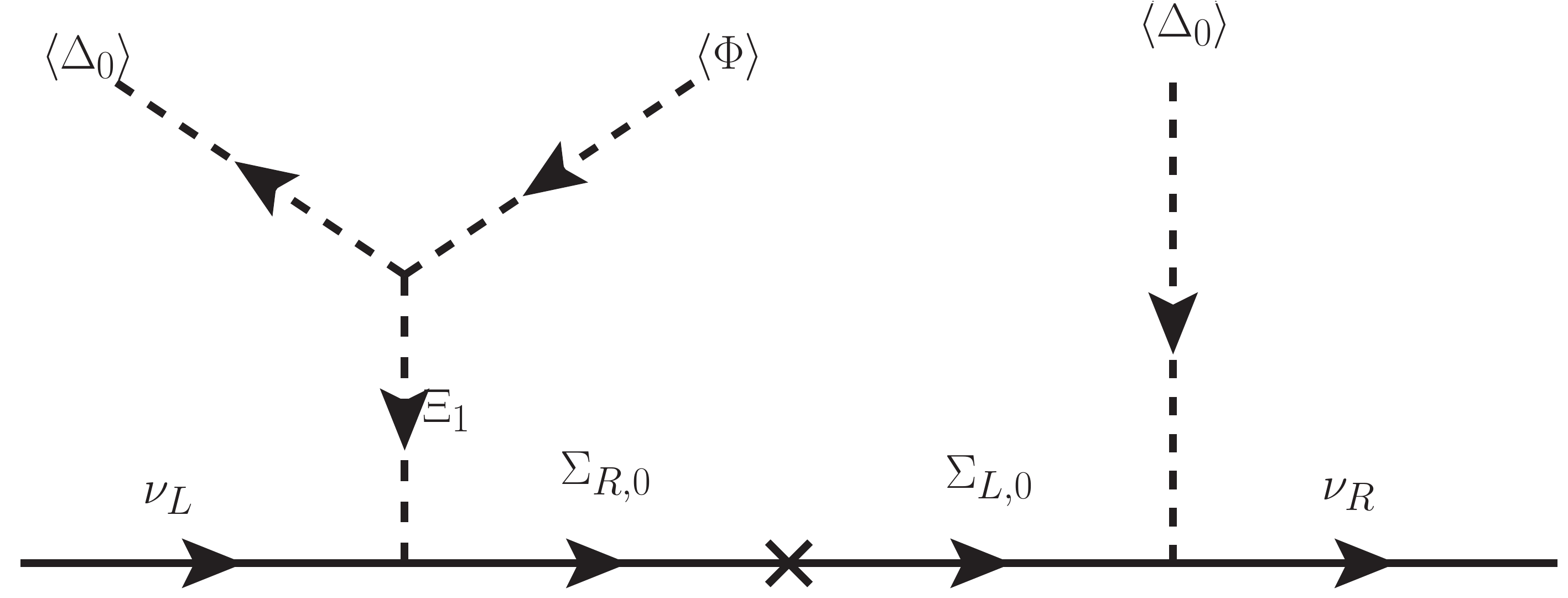}
     \caption{Diagrams showing the $T_2$ topology of the operator $\bar{L} \bar{\Phi}  \bar{\Delta} _0 \Delta_0 \nu_R$.}
      \label{o3t2}
 \end{figure}

The six operator contractions which lead to $T_2$ topologies are given
in \eqref{233t2}. The corresponding diagrams are shown in Figure
\ref{o3t2}. 
\begin{eqnarray}
&& \underbrace{\underbrace{\bar{L}}_2 \otimes \underbrace{\bar{\Delta}_0 \otimes \Delta_0}_1}_{2} \otimes \underbrace{\bar{\Phi} \otimes \nu_R}_{2}, \hspace{0.5cm}
\underbrace{\underbrace{\bar{L}}_2 \otimes \underbrace{\bar{\Delta}_0 \otimes \Delta_0}_3}_{2} \otimes \underbrace{\bar{\Phi} \otimes \nu_R}_{2}, \hspace{0.5cm}
\underbrace{\underbrace{\bar{L}}_2 \otimes \underbrace{\Delta_0 \otimes \bar{\Phi}}_2}_{3} \otimes \underbrace{\bar{\Delta}_0 \otimes \nu_R}_{3}, \hspace{0.5cm}
\underbrace{\underbrace{\bar{L}}_2 \otimes \underbrace{\bar{\Delta}_0 \otimes \bar{\Phi}}_2}_{3} \otimes \underbrace{\Delta_0 \otimes \nu_R}_{3}
\nonumber \\
&&\underbrace{\underbrace{\bar{L}}_2 \otimes \underbrace{\bar{\Delta}_0 \otimes \bar{\Phi}}_4}_{3} \otimes \underbrace{\Delta_0 \otimes \nu_R}_{3} , \hspace{0.5cm}
\underbrace{\underbrace{\bar{L}}_2 \otimes \underbrace{\Delta_0 \otimes \bar{\Phi}}_4}_{3} \otimes \underbrace{\bar{\Delta}_0 \otimes \nu_R}_{3}
\label{233t2}
\end{eqnarray}
 
 Again notice the appearance of the scalar $SU(2)_L$ quartet $\Xi$ in
 the last two diagrams of Fig.~\ref{o3t2}. 
 For the other operators listed in Table \ref{Tab:op}, some of the
 contractions in \eqref{233t2} are forbidden.
 For the operator
 $\bar{L} \otimes \bar{\Phi} \otimes \bar{\Delta}_{-2} \otimes
 \Delta_{-2} \otimes \nu_R$
 the third contraction of \eqref{233t2} is forbidden as it implies a
 messenger doublet with $U(1)_Y = 3$ which does not have a neutral
 component.
 For the operator
 $\bar{L} \otimes \Phi \otimes \Delta_0 \otimes \Delta_{-2} \otimes
 \nu_R $
 the first contraction of \eqref{233t2} leads to a singlet messenger
 with $U(1)_Y = -2$ and hence is forbidden. 
 On the other hand, for the operator
 $\bar{L} \otimes \bar{\Phi} \otimes \Delta_0 \otimes \Delta_0 \otimes
 \nu_R $,
 the second contraction of \eqref{233t2} is forbidden, as it involves
 a contraction of two identical triplets going to a triplet, which is
 zero.
 Also, the third and fourth, as well as the fifth and sixth
 contractions are identical for this case. Therefore only one
   diagram out of each pair should be counted for this case.

\begin{figure}[!h]
 \centering
  \includegraphics[scale=0.25]{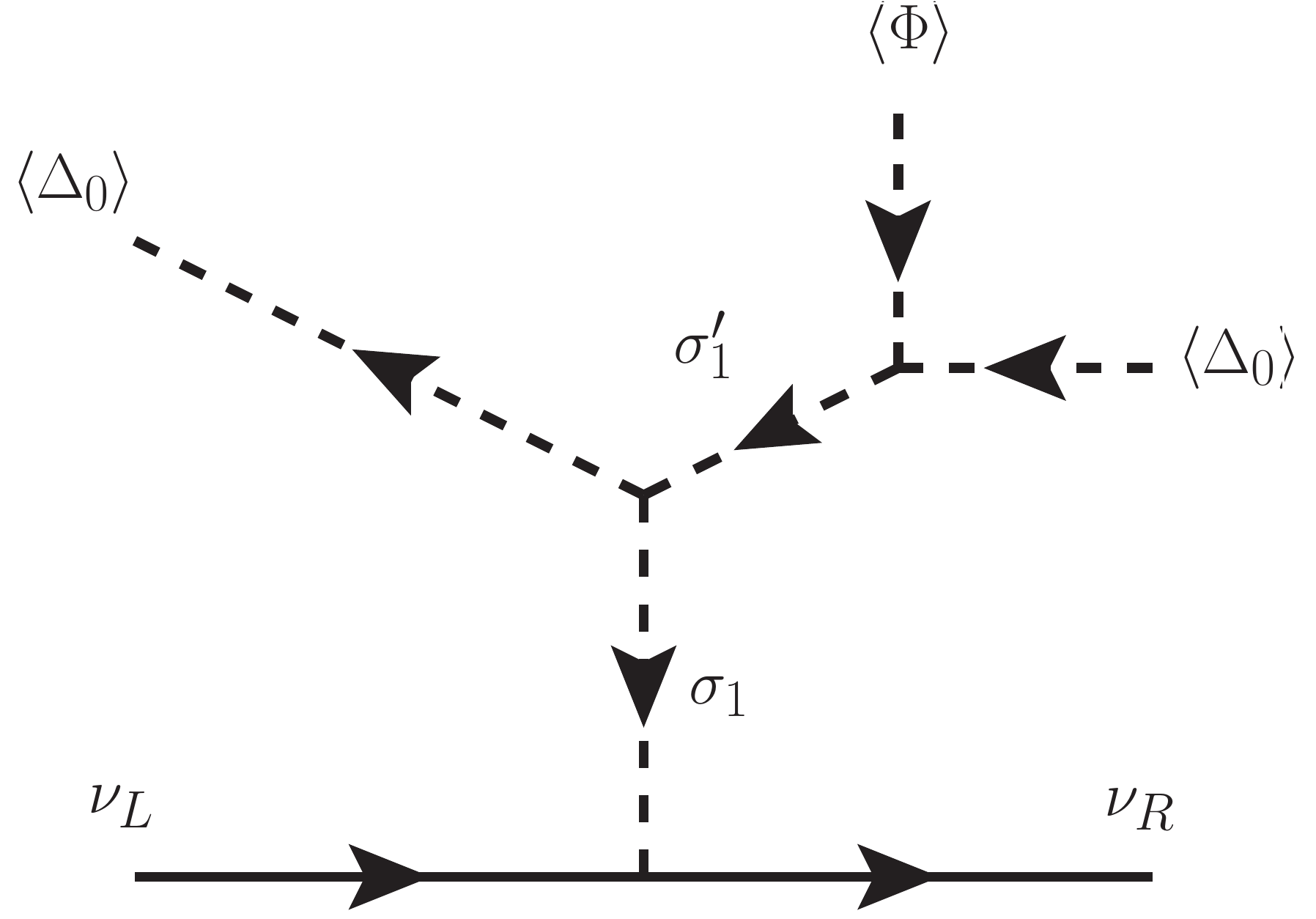}, \hspace{2mm}
   \includegraphics[scale=0.25]{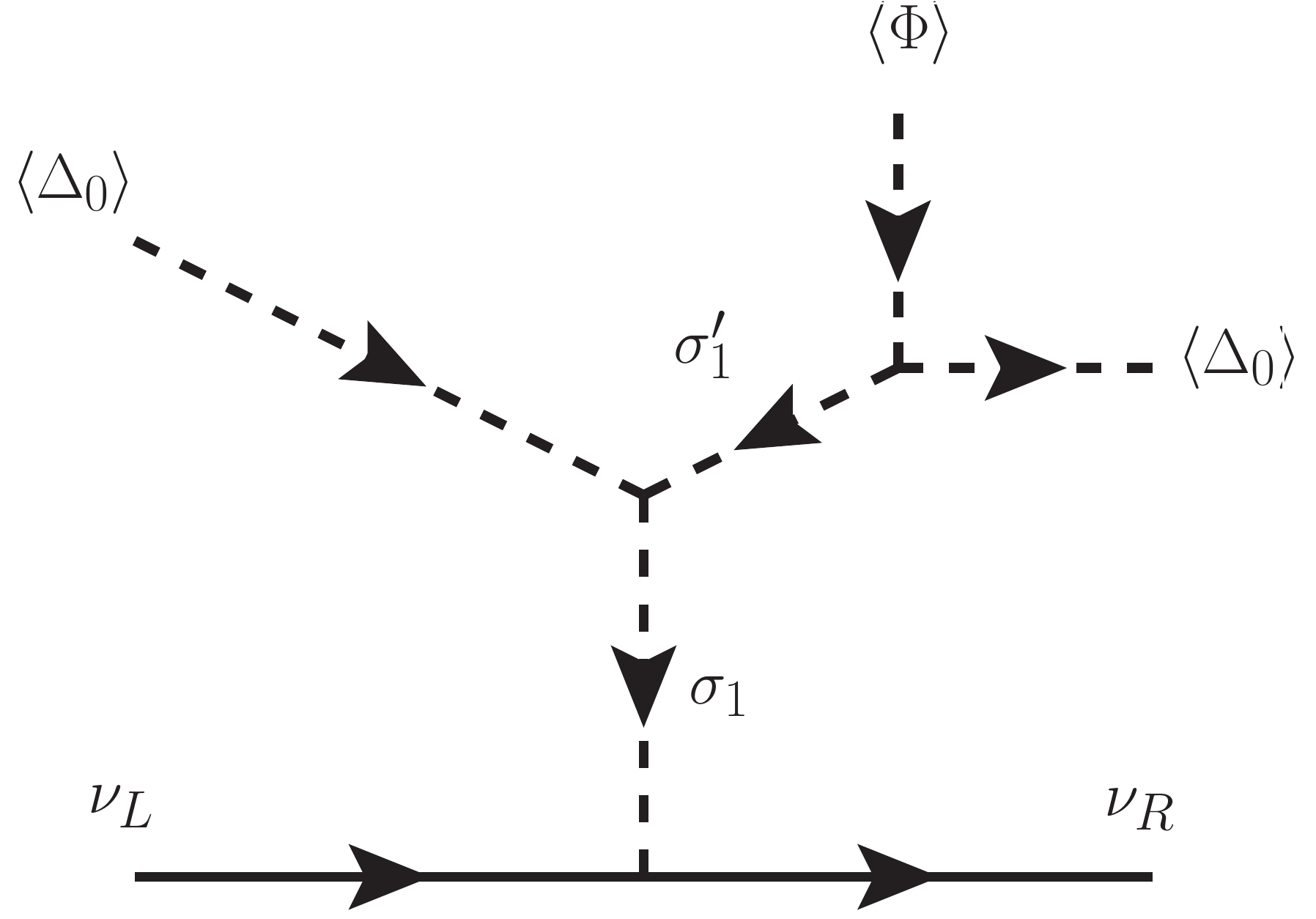}, \hspace{2mm} \\
    \includegraphics[scale=0.25]{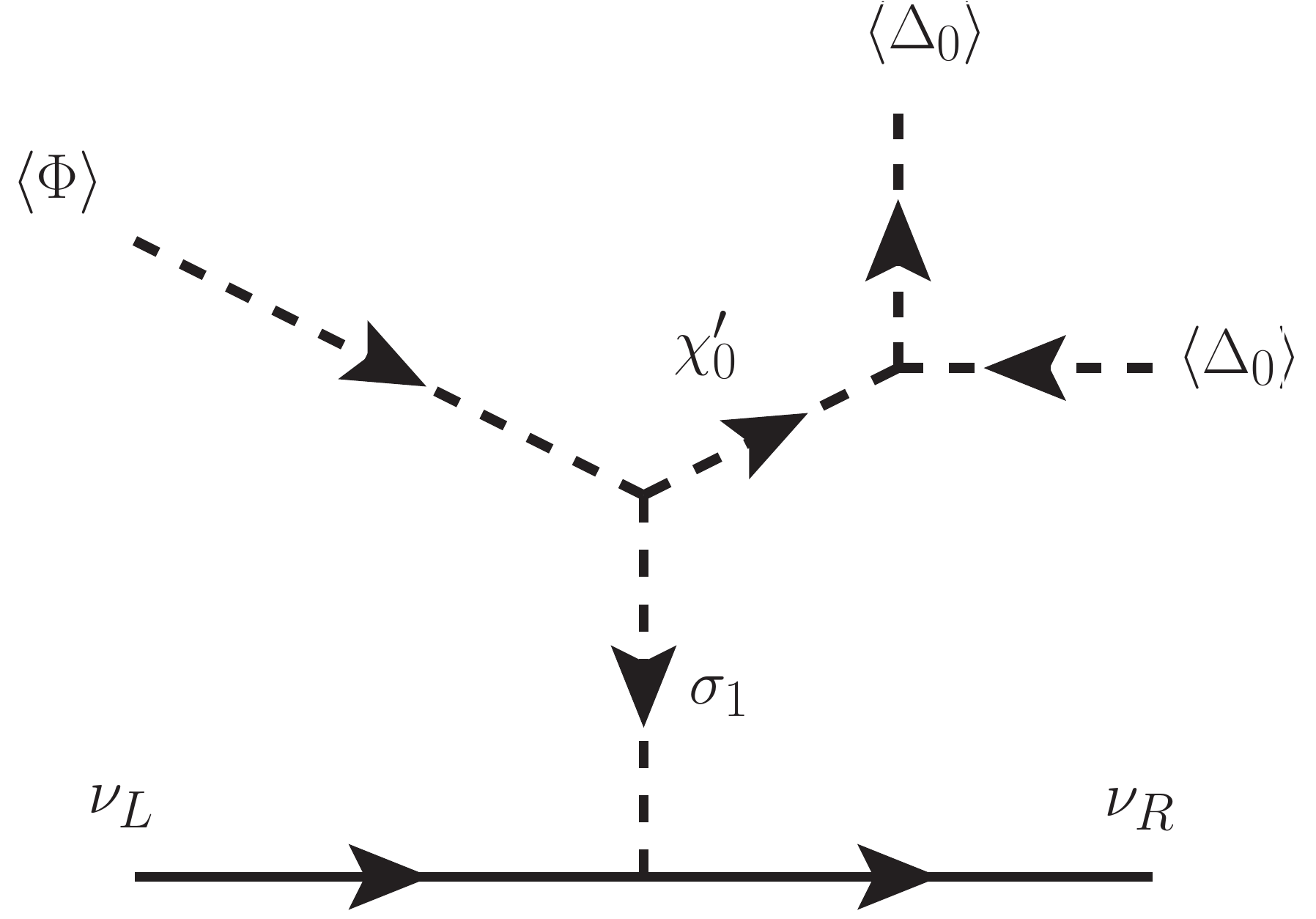}, \hspace{2mm}
     \includegraphics[scale=0.25]{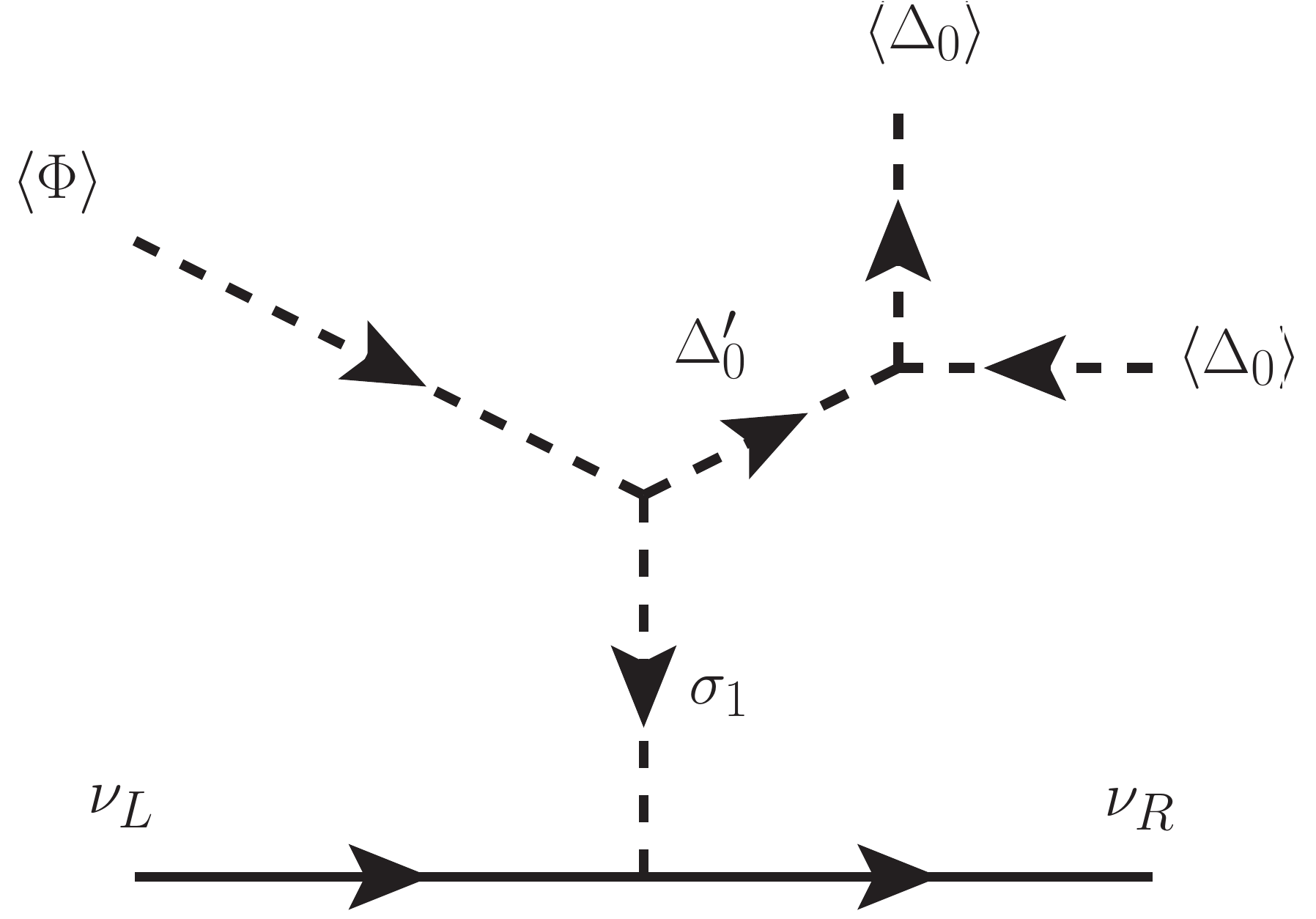}, \hspace{2mm}
      \includegraphics[scale=0.25]{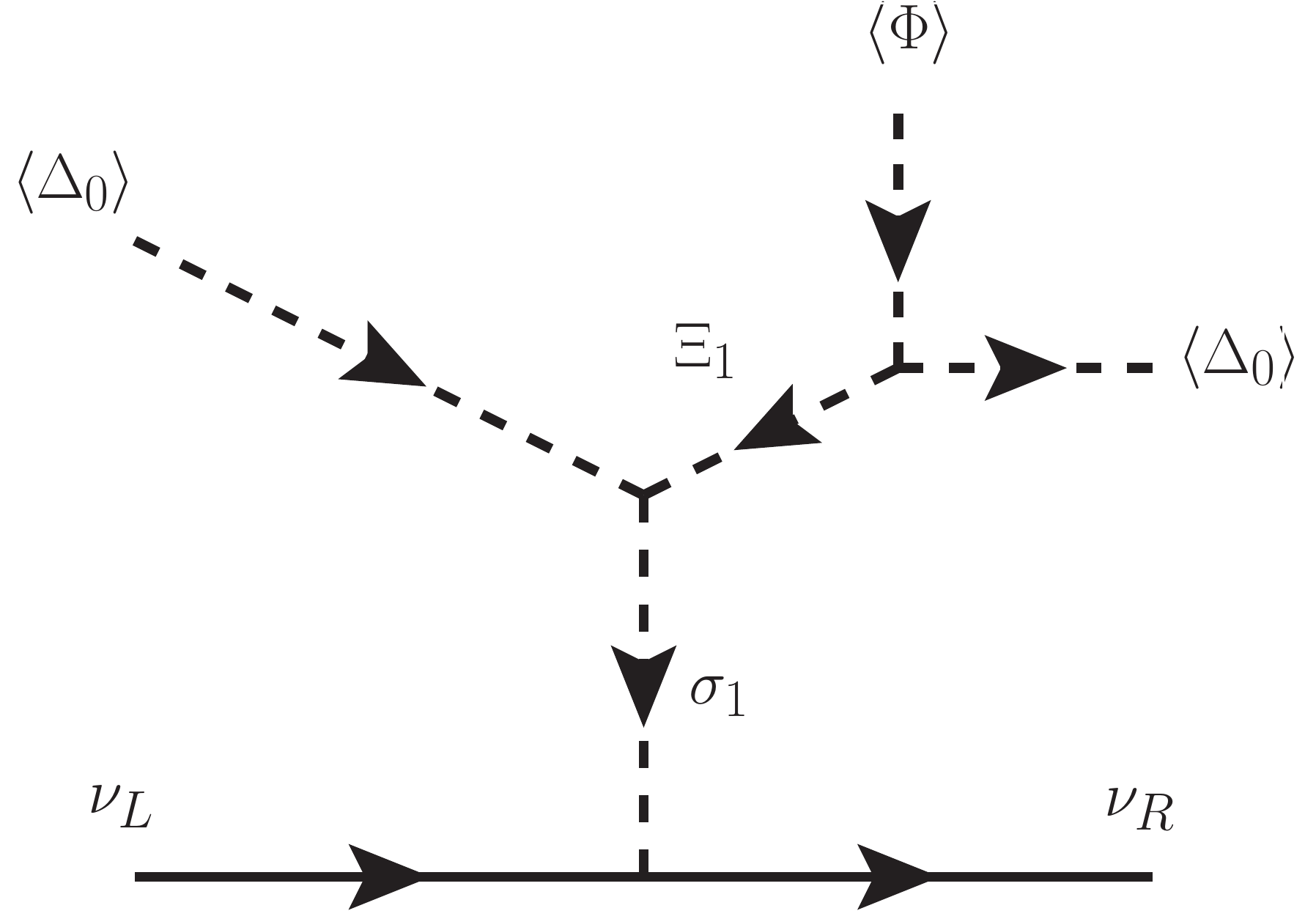}, \hspace{2mm}
       \includegraphics[scale=0.25]{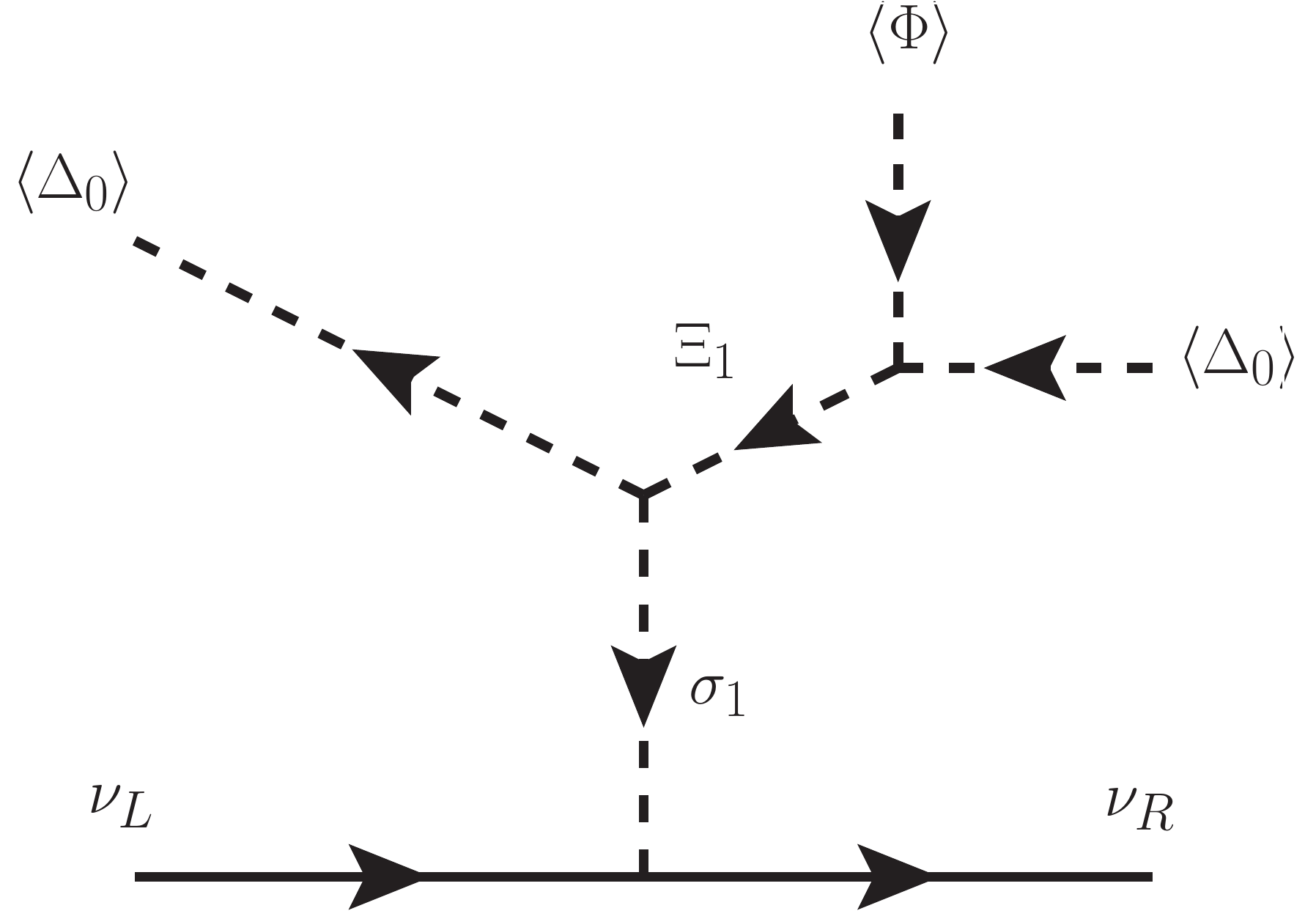}, \hspace{2mm}
        \includegraphics[scale=0.25]{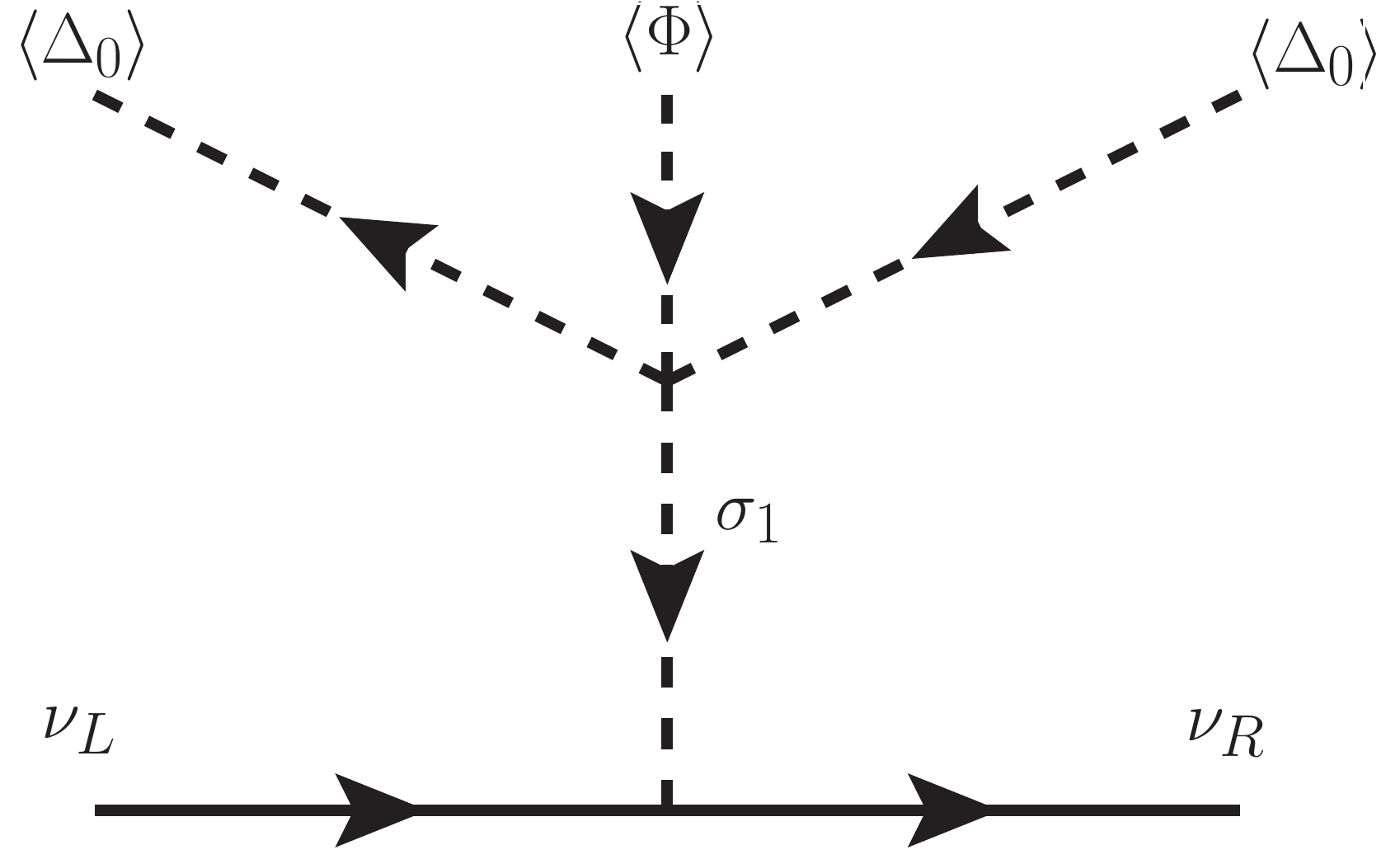}
     \caption{Diagrams showing the $T_3$ and $T_4$ topologies of the operator $\bar{L} \bar{\Phi}  \bar{\Delta} _0 \Delta_0 \nu_R$.}
      \label{o3t3t4}
\end{figure}

 The possible operator contractions leading to the $T_3$ and $T_4$
 topologies are shown in \eqref{233t3-1} and in \eqref{233t3-2}, while
 the corresponding diagrams are shown in Fig.~\ref{o3t3t4}.
\be
 \underbrace{\underbrace{\bar{L} \otimes \nu_R}_2 \otimes \underbrace{\Delta_0}_3}_{2} \otimes \underbrace{\bar{\Phi} \otimes \bar{\Delta}_0}_{2} , \hspace{0.5cm}
 \underbrace{\underbrace{\bar{L} \otimes \nu_R}_2 \otimes \underbrace{\bar{\Delta}_0}_3}_{2} \otimes \underbrace{\bar{\Phi} \otimes \Delta_0}_{2} , \hspace{0.5cm}
 \underbrace{\underbrace{\bar{L} \otimes \nu_R}_2 \otimes \underbrace{\bar{\Phi}}_2}_{1} \otimes \underbrace{\bar{\Delta}_0 \otimes \Delta_0}_{1} , \hspace{0.5cm}
 \underbrace{\underbrace{\bar{L} \otimes \nu_R}_2 \otimes \underbrace{\bar{\Phi}}_2}_{3} \otimes \underbrace{\bar{\Delta}_0 \otimes \Delta_0}_{3} 
\label{233t3-1}
\ee

\be
 \underbrace{\underbrace{\bar{L} \otimes \nu_R}_2 \otimes \underbrace{\bar{\Delta}_0}_3}_{4} \otimes \underbrace{\bar{\Phi} \otimes \Delta_0}_{4} , \hspace{0.5cm}
 \underbrace{\underbrace{\bar{L} \otimes \nu_R}_2 \otimes \underbrace{\Delta_0}_3}_{4} \otimes \underbrace{\bar{\Phi} \otimes \bar{\Delta}_0}_{4} , \hspace{0.5cm}
\underbrace{\bar{L} \otimes \nu_R}_2 \otimes \underbrace{\bar{\Delta}_0 \otimes \bar{\Phi} \otimes \Delta_0}_2 
\label{233t3-2}
\ee

Notice the appearance of the scalar $SU(2)_L$ quartet $\Xi$ in the fifth
and sixth diagrams of Fig.~\ref{o3t3t4} and the fact that $\sigma_1$
and $\sigma_1^\prime$ must be different fields, owing to their different
transformations under symmetries forbidding lower dimensional
operators. 

Concerning other operators in Table \ref{Tab:op}, some of the
contractions of \eqref{233t3-1} and \eqref{233t3-2} are forbidden.
For example, for the operator
$\bar{L} \otimes \bar{\Phi} \otimes \bar{\Delta}_{-2} \otimes
\Delta_{-2} \otimes \nu_R$
the second contraction of \eqref{233t3-1} is again forbidden because
the messenger field will have no neutral component. 
The same happens with the third contraction of \eqref{233t3-1} for the
operator
$\bar{L} \otimes \Phi \otimes \Delta_{-2} \otimes \Delta_{0} \otimes
\nu_R$.
Lastly, for the operator
$\bar{L} \otimes \bar{\Phi} \otimes \Delta_{0} \otimes \Delta_{0}
\otimes \nu_R$ the fourth diagram of \eqref{233t3-1} is forbidden and
only one out of the first and second contraction of \eqref{233t3-1}
and one out of first and second contraction of \eqref{233t3-2} should
be counted. 
 
\begin{figure}[!h] 
\centering
 \includegraphics[scale=0.25]{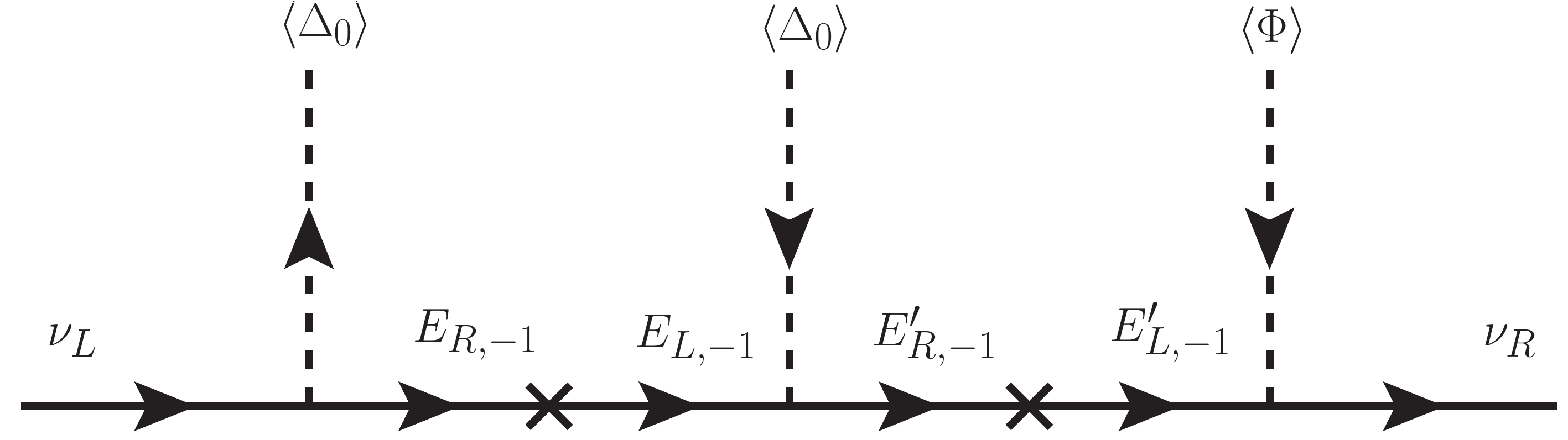}, \hspace{2mm}
  \includegraphics[scale=0.25]{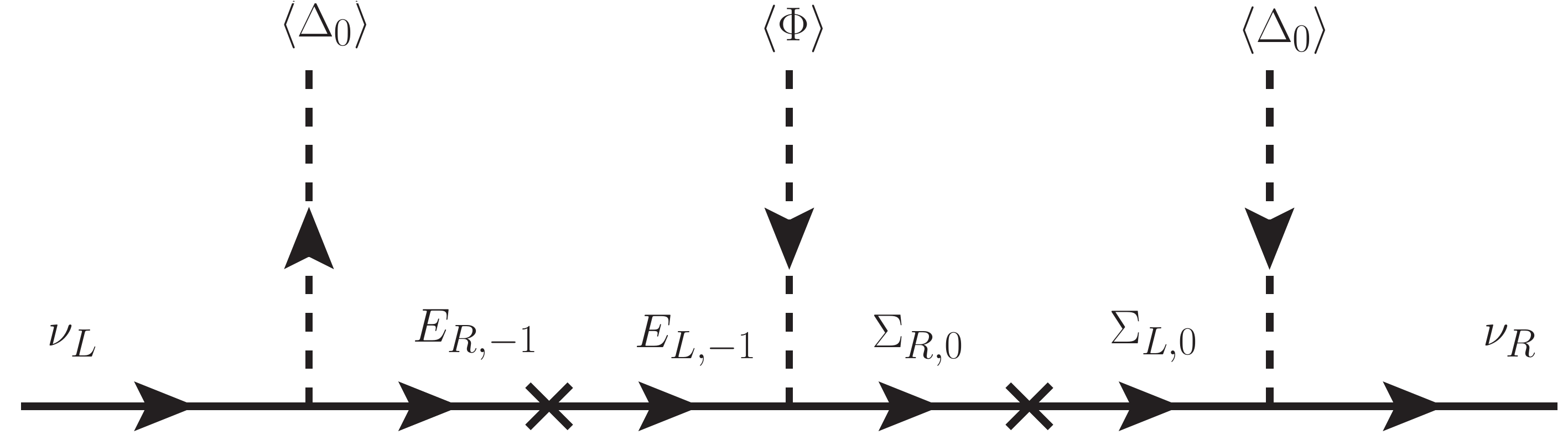}, \hspace{2mm}
   \includegraphics[scale=0.25]{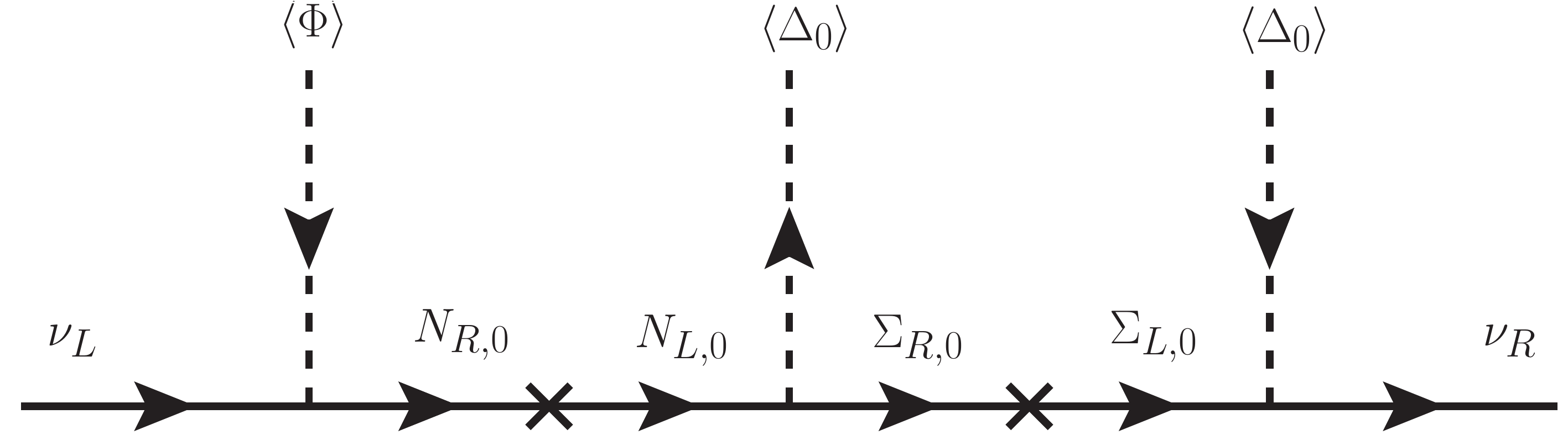}, \hspace{2mm}
    \includegraphics[scale=0.25]{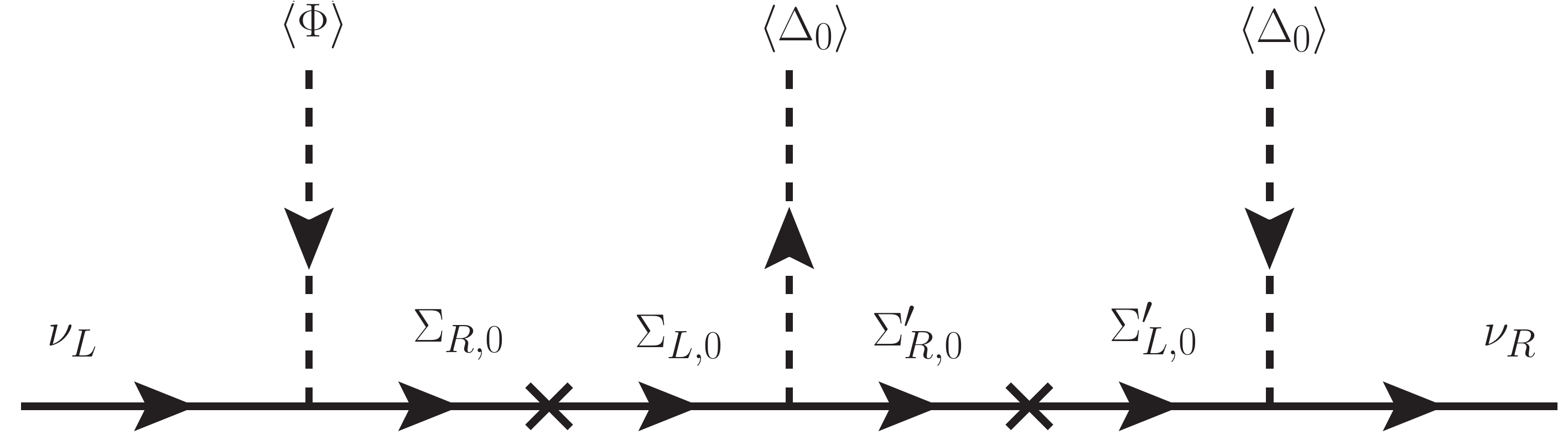}, \hspace{2mm}
     \includegraphics[scale=0.25]{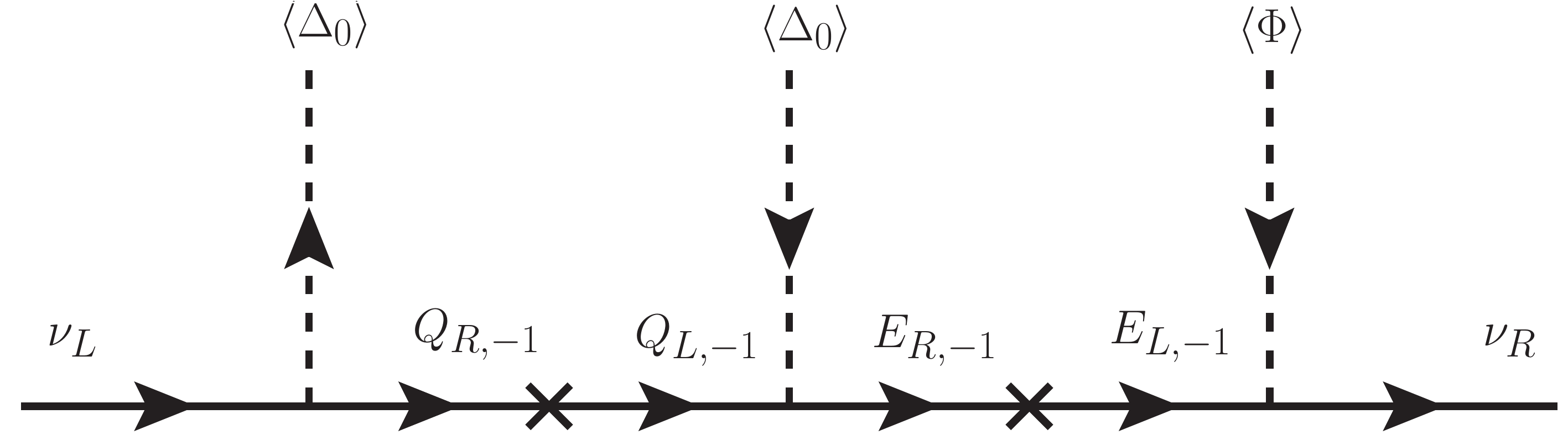}, \hspace{2mm}
      \includegraphics[scale=0.25]{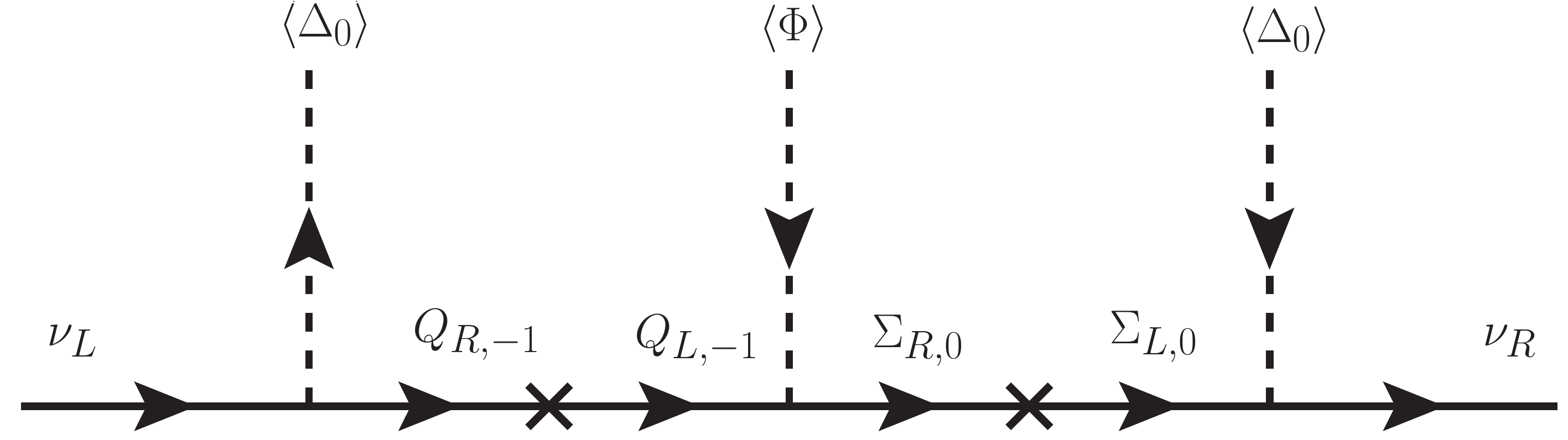}
       \caption{Diagrams showing the $T_5$ topology of the operator $\bar{L} \bar{\Phi}  \bar{\Delta} _0 \Delta_0 \nu_R$.}
        \label{o3t5}
 \end{figure}

\begin{figure}[!h] 
\centering
 \includegraphics[scale=0.25]{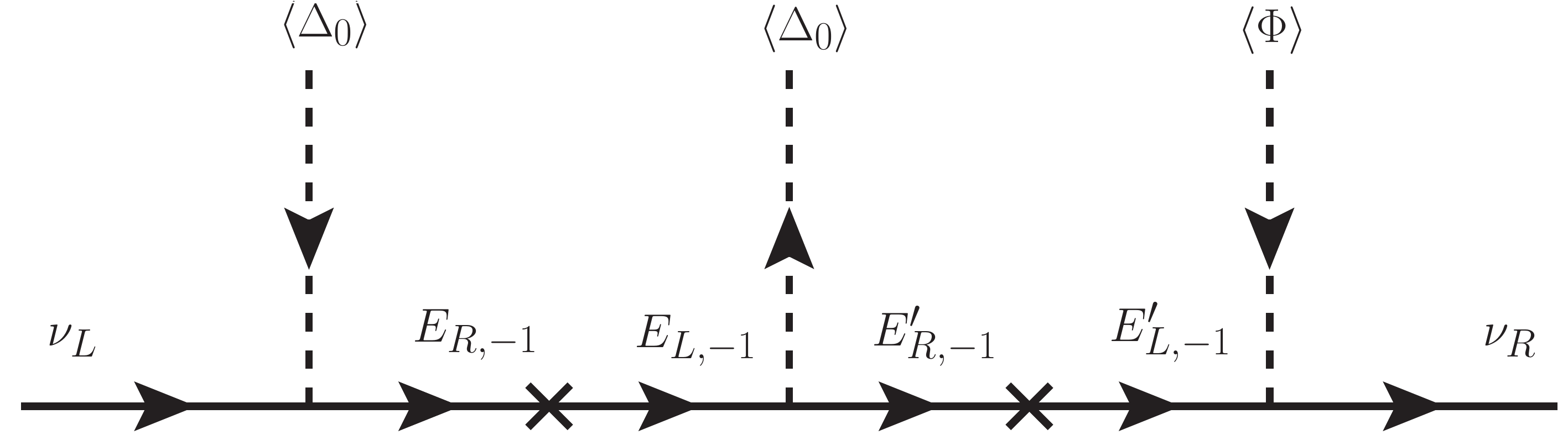}, \hspace{2mm}
  \includegraphics[scale=0.25]{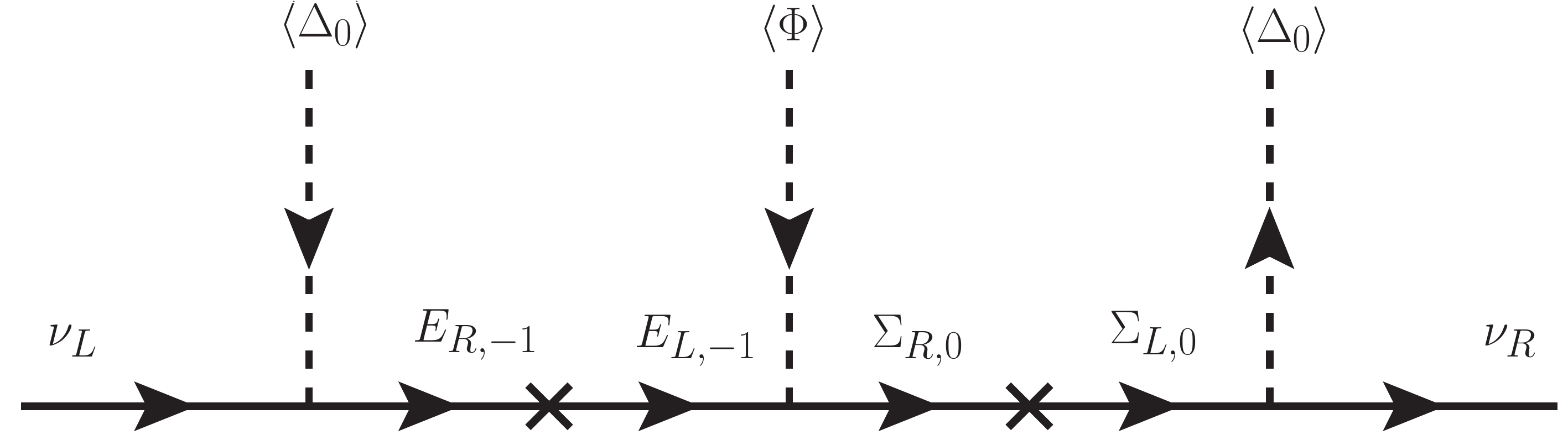}, \hspace{2mm}
   \includegraphics[scale=0.25]{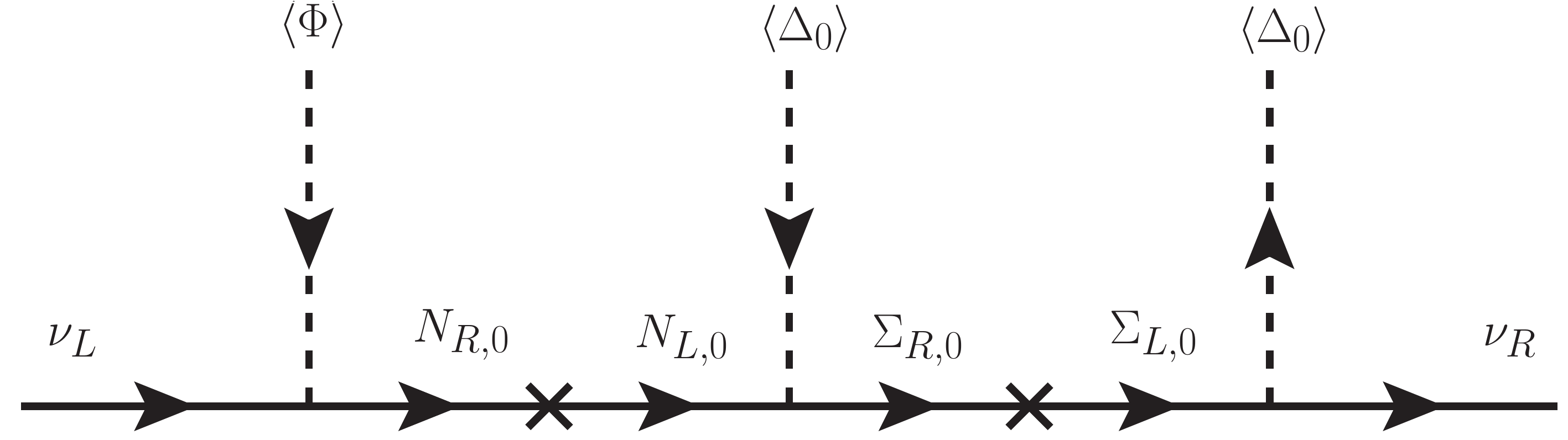}, \hspace{2mm}
    \includegraphics[scale=0.25]{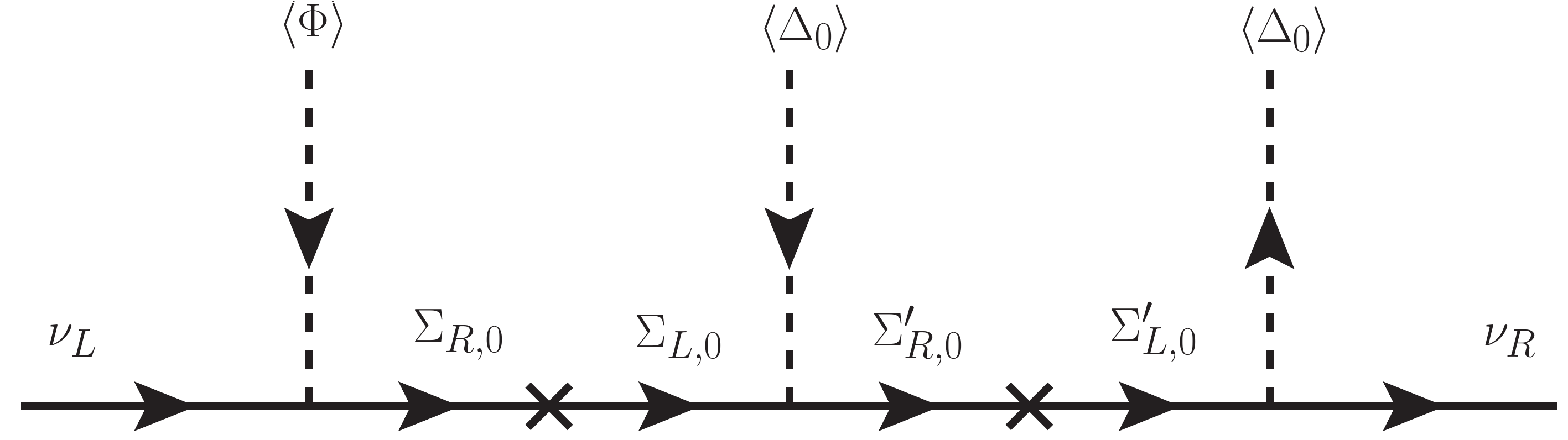}, \hspace{2mm}
     \includegraphics[scale=0.25]{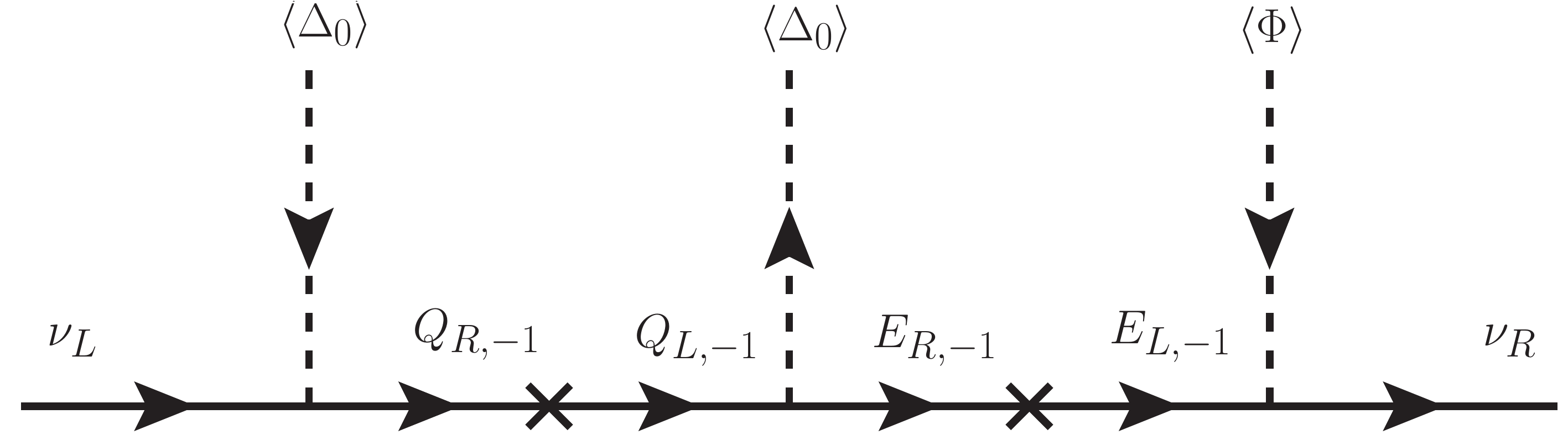}, \hspace{2mm}
      \includegraphics[scale=0.25]{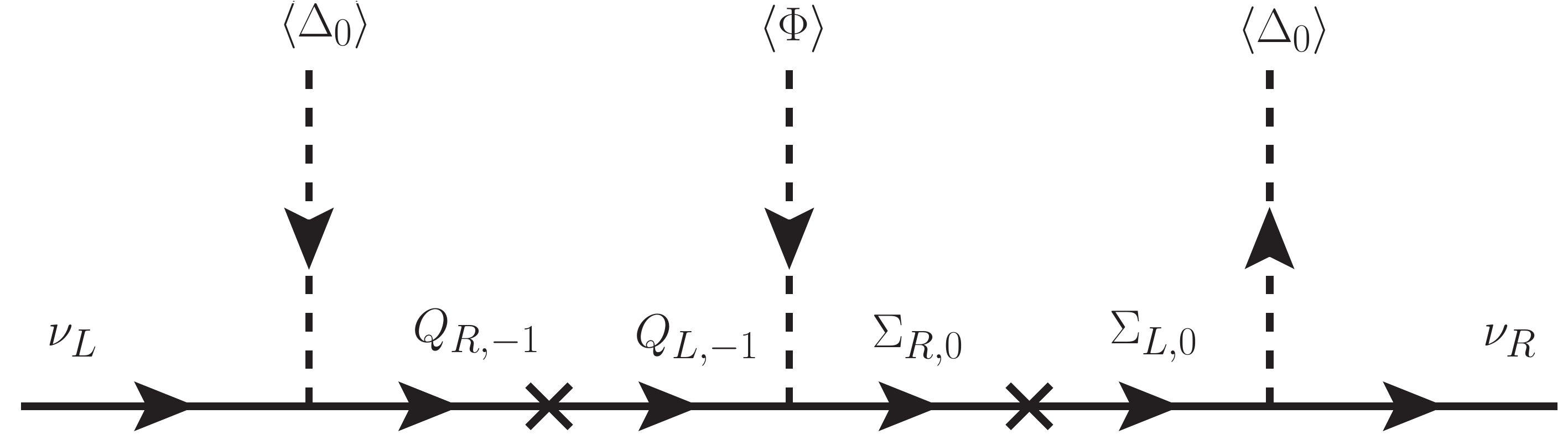}
       \caption{Diagrams showing the $T_5$ topology of the operator $\bar{L} \bar{\Phi}  \bar{\Delta} _0 \Delta_0 \nu_R$.}
        \label{o3t52}
 \end{figure}

Finally, the twelve possible operator contractions leading to the
$T_5$ topology are shown in \eqref{233t5-1}-\eqref{233t5-4}. The
corresponding diagrams are shown in Fig.~\ref{o3t5} and \ref{o3t52}.
\be
  \underbrace{\underbrace{\bar{L} \otimes \Delta_0}_2 \otimes \underbrace{\bar{\Delta}_0}_3}_{2} \otimes \underbrace{\bar{\Phi} \otimes \nu_R}_{2} , \hspace{0.5cm}
   \underbrace{\underbrace{\bar{L} \otimes \Delta_0}_2 \otimes \underbrace{\bar{\Phi}}_2}_{3} \otimes \underbrace{\bar{\Delta}_0 \otimes \nu_R}_{3} , \hspace{0.5cm}
    \underbrace{\underbrace{\bar{L} \otimes \bar{\Phi}}_1 \otimes \underbrace{\Delta_0}_3}_{3} \otimes \underbrace{\bar{\Delta}_0 \otimes \nu_R}_{3} 
  \label{233t5-1}
  \ee
 \be
     \underbrace{\underbrace{\bar{L} \otimes \bar{\Phi}}_3 \otimes \underbrace{\Delta_0}_3}_{3} \otimes \underbrace{\bar{\Delta}_0 \otimes \nu_R}_{3}  , \hspace{0.5cm}
       \underbrace{\underbrace{\bar{L} \otimes \Delta_0}_4 \otimes \underbrace{\bar{\Delta}_0}_3}_{2} \otimes \underbrace{\bar{\Phi} \otimes \nu_R}_{2} , \hspace{0.5cm}
          \underbrace{\underbrace{\bar{L} \otimes \Delta_0}_4 \otimes \underbrace{\bar{\Phi}}_2}_{3} \otimes \underbrace{\bar{\Delta}_0 \otimes \nu_R}_{3}
   \label{233t5-2} 
\ee
\be
  \underbrace{\underbrace{\bar{L} \otimes \bar{\Delta}_0}_2 \otimes \underbrace{\Delta_0}_3}_{2} \otimes \underbrace{\bar{\Phi} \otimes \nu_R}_{2} , \hspace{0.5cm}
   \underbrace{\underbrace{\bar{L} \otimes \bar{\Delta}_0}_2 \otimes \underbrace{\bar{\Phi}}_2}_{3} \otimes \underbrace{\Delta_0 \otimes \nu_R}_{3} , \hspace{0.5cm}
    \underbrace{\underbrace{\bar{L} \otimes \bar{\Phi}}_1 \otimes \underbrace{\bar{\Delta}_0}_3}_{3} \otimes \underbrace{\Delta_0 \otimes \nu_R}_{3} 
  \label{233t5-3}
  \ee
 \be
     \underbrace{\underbrace{\bar{L} \otimes \bar{\Phi}}_3 \otimes \underbrace{\bar{\Delta}_0}_3}_{3} \otimes \underbrace{\Delta_0 \otimes \nu_R}_{3}  , \hspace{0.5cm}
      \underbrace{\underbrace{\bar{L} \otimes \bar{\Delta}_0}_4 \otimes \underbrace{\Delta_0}_3}_{2} \otimes \underbrace{\bar{\Phi} \otimes \nu_R}_{2} , \hspace{0.5cm}
       \underbrace{\underbrace{\bar{L} \otimes \bar{\Delta}_0}_4 \otimes \underbrace{\bar{\Phi}}_2}_{3} \otimes \underbrace{\Delta_0 \otimes \nu_R}_{3}
    \label{233t5-4}
\ee

Notice that the contractions in Fig.~\ref{o3t5} and the ones in
Fig.~\ref{o3t52} differ just in the exchange of
$\Delta \leftrightarrow \bar{\Delta}$. 
While these diagrams involve messengers having similar transformations
under the \sm gauge group, they differ from each other in how they
transform under the symmetry group used to forbid lower dimensional
operators. 
This also implies that the messenger pairs $E_{-1}$ and $E^\prime_{-1}$ as well as
$\Sigma_0$ and $\Sigma^\prime_0$ must be different from each other. Keeping
this in mind we have counted them as different UV-complete models. 

For the operator
$\bar{L} \otimes \bar{\Phi} \otimes \bar{\Delta}_{-2} \otimes
\Delta_{-2} \otimes \nu_R$
the first and second contractions of \eqref{233t5-3} are forbidden due
to messenger fields not having any neutral component. 
For the same reason, the third contractions of \eqref{233t5-1} and
\eqref{233t5-3} are forbidden for the operator
$\bar{L} \otimes \Phi \otimes \Delta_{0} \otimes \Delta_{-2} \otimes
\nu_R$.
For the case of the operator
$\bar{L} \otimes \bar{\Phi} \otimes \Delta_{0} \otimes \Delta_{0}
\otimes \nu_R$
all the contractions in \eqref{233t5-3} and \eqref{233t5-4} are
indistinguishable from those in \eqref{233t5-1} and \eqref{233t5-2}
and hence should not be counted as separate contractions.

%%%%%%%%%%%%%%%%%%%%%%%%%%%%%%%%%%%%%%%%%%%%%%%%%%%%%%%%%%%%%%%%%%%%%%%%%%%%
\section{Discussion and Summary } 
\label{sec:summary-conclusions}
%%%%%%%%%%%%%%%%%%%%%%%%%%%%%%%%%%%%%%%%%%%%%%%%%%%%%%%%%%%%%%%%%%%%%%%%%%%%

As a follow-up to our recent paper in
Ref.~\cite{CentellesChulia:2018gwr}, here we have classified and
analysed the various ways to generate Dirac neutrino mass through the
use of dimension-6 operators.
The UV-completion of such scenarios will require new messenger fields
carrying $SU(2)_L \otimes U(1)_Y$ charges that may be probed at
colliders, since the scale involved may be phenomenologically
accessible.
By using only the Standard Model Higgs doublet in the external legs
one has a unique operator, Eq.~(\ref{op-sm}).
We have shown, however, that the presence of new scalars implies the
existence of many possible field contractions. 
We have described in detail the simplest ones of these, involving
$SU(2)_L$ singlets, doublets and triplets. 
In order to ensure the Dirac nature of neutrinos, as well as the
seesaw origin of their mass (in our case, at the dimension-6 level)
extra symmetries are needed.
They can be realized in several ways, a simple example being lepton
quarticity.
Such symmetries can also be used to provide the stability of dark
matter.
In fact one should emphasize the generality of this connection,
already explained in Ref.~\cite{CentellesChulia:2018gwr} in the
context of the dimension-5 Dirac seesaw scenario. 

%%%%%%%%%%%%%%%%%%%%%%%%%%%%%%%%%%%%%%%%%%%%%%%%%%%%%%%%%%%%%%%%%%%%%%%%%%%
\begin{acknowledgments}

  Work supported by the Spanish MINECO grants FPA2017-85216-P and
  SEV2014-0398, and also PROMETEOII/2014/084 from Generalitat
  Valenciana.  The Feynman diagrams were drawn using Jaxodraw
  \cite{Binosi:2003yf}.

\end{acknowledgments}

%%%%%%%%%%%%%%%%%%%%%%%%%%%%%%%%%%%%%%%%%%%%%%%%%%%%%%%%%%%%%%%%%%%%%%%%

\bibliographystyle{bib_style_T1}
%\bibliography{bibliography} 

\end{document}